\documentclass[journal]{IEEEtran}
\IEEEoverridecommandlockouts
\usepackage{cite}
\usepackage{amsmath,amssymb,amsfonts}
\DeclareMathOperator*{\argmax}{arg\,max}
\DeclareMathOperator*{\argmin}{arg\,min}
\usepackage{algorithm,algpseudocode}
\usepackage{setspace}
\let\Algorithm\algorithm
\renewcommand\algorithm[1][]{\Algorithm[#1]\setstretch{1.0}}   %%% THIS TO CONTROL LINE SPACING IN ALGO TABLES

\usepackage[inline]{enumitem}
\usepackage{graphicx}
\usepackage{textcomp}
\usepackage{xcolor}
\def\BibTeX{{\rm B\kern-.05em{\sc i\kern-.025em b}\kern-.08em
    T\kern-.1667em\lower.7ex\hbox{E}\kern-.125emX}}

\usepackage{lettrine}
\usepackage{afterpage}
\usepackage{tikz}
\usepackage{microtype}
\usepackage[acronym]{glossaries}
\newacronym{3GPP}{3GPP}{3rd Generation Partnership Project}
\newacronym{5G}{5G}{fifth generation}
\newacronym{5G NR}{5G NR}{fifth generation New Radio}
\newacronym{iid}{iid}{independent and identically distributed}
\newacronym{ACF}{ACF}{autocorrelation function}
\newacronym{AGV}{AGV}{automated guided vehicle}
\newacronym{AR}{AR}{auto-regressive}
\newacronym{LR}{LR}{likelihood ratio}
\newacronym{AN}{AN}{access node}
\newacronym{AoA}{AoA}{angle-of-arrival}
\newacronym{AWGN}{AWGN}{additive white Gaussian noise}
\newacronym{BB}{BB}{baseband}
\newacronym{BBU}{BBU}{base band unit}
\newacronym{BS}{BS}{base-station}
\newacronym{BIC}{BIC}{Bayesian information criterion}
\newacronym{MDL}{MDL}{Minimum Description Length}
\newacronym{BLER}{BLER}{block error rate}
\newacronym{BPSK}{BPSK}{binary phase shift keying}
\newacronym{CCE}{CCE}{control channel element}
\newacronym{CDF}{CDF}{cumulative distribution function}
\newacronym{CSI-RS}{CSI-RS}{channel state information reference signal}
\newacronym{CL}{CL}{centroid localization}
\newacronym{CP}{CP}{cyclic prefix}
\newacronym{CPPS}{CPPS}{constant phase \gls{PS}}
\newacronym{CR}{CR}{cognitive radio}
\newacronym{C-RAN}{C-RAN}{centralized radio access network}
\newacronym{CRB}{CRB}{Cram\'er-Rao bound}
\newacronym{CSI}{CSI}{channel state information}
\newacronym{CSIT}{CSIT}{channel state information at transmitter}
\newacronym{CA}{CA}{constant acceleration}
\newacronym{CT}{CT}{constant turn}
\newacronym{CV}{CV}{constant velocity}
\newacronym{D2D}{D2D}{device-to-device}
\newacronym{DBS}{DBS}{different beamwidth sectors}
\newacronym{DCAA}{DCAA}{digitally controlled antenna array}
\newacronym{DCI}{DCI}{downlink control information}
\newacronym{DCPS}{DCPS}{digitally controlled \gls{PS}}
\newacronym{DFU}{DFU}{{}\acrshort{DoA} fusion}
\newacronym{DL}{DL}{downlink}
\newacronym{eMMB}{eMMB}{enhanced mobile broadband}
\newacronym{DMRS}{DMRS}{demodulation reference signal}
\newacronym[firstplural=directions-of-arrival (DoAs)]{DoA}{DoA}{direction-of-arrival}
\newacronym{DoD}{DoD}{direction-of-departure}
\newacronym{DoDs}{DoDs}{directions-of-departure}
\newacronym{DoAs}{DoAs}{directions-of-arrival}
\newacronym{gNB}{gNB}{5G-NR Node B}
\newacronym{DS-CDMA}{DC-CDMA}{direct sequence code division multiple access}
\newacronym{MMA}{MMA}{multi-mode antenna}
\newacronym{E2E}{E2E}{end-to-end}
\newacronym{E911}{E911}{enhanced 911}
\newacronym{EADF}{EADF}{effective aperture distribution function}
\newacronym{EBS}{EBS}{equal beamwidth sectors}
\newacronym{EKF}{EKF}{extended Kalman filter}
\newacronym{ESA}{ESA}{equal sector antenna}
\newacronym{ESPAR}{ESPAR}{electronically steerable parasitic array radiator}
\newacronym{EOTD}{E-OTD}{enhanced observed time difference}
\newacronym{EW}{EW}{equal weighting}
\newacronym{FCC}{FCC}{federal communications commission}
\newacronym{FFT}{FFT}{fast Fourier transform}
\newacronym[firstplural=Fisher information matrices (FIM)]{FIM}{FIM}{Fisher information matrix}
\newacronym{FR1}{FR1}{Frequency Range 1}
\newacronym{FR2}{FR2}{Frequency Range 2}
\newacronym{GMM}{GMM}{Gaussian mixture model}
\newacronym{GNSS}{GNSS}{global navigation satellite system}
\newacronym{GPS}{GPS}{global positioning system}
\newacronym{GSM}{GSM}{global system for mobile communications}
\newacronym{ISAC}{ISAC}{integrated sensing and communications}
\newacronym{ICI}{ICI}{inter-carrier-interference}
\newacronym{ICMP}{ICMP}{internet control message protocol}
\newacronym{IoT}{IoT}{Internet-of-Things}
\newacronym{ISD}{ISD}{inter-site distance}
\newacronym{ISI}{ISI}{inter-symbol-interference}
\newacronym{ITS}{ITS}{intelligent transportation system}
\newacronym{ITU}{ITU}{international telecommunication union}
\newacronym{KF}{KF}{Kalman filter}
\newacronym{LDPC}{LDPC}{low-density parity-check}
\newacronym{LMMSE}{LMMSE}{linear minimum mean-square error}
\newacronym{LoS}{LoS}{line-of-sight}
\newacronym{LS}{LS}{least-squares}
\newacronym{TLS}{TLS}{total least-squares}
\newacronym{NLLS}{NLLS}{nonlinear least-squares}
\newacronym{LTE}{LTE}{long term evolution}
\newacronym{LWA}{LWA}{leaky-wave antenna}
\newacronym{MAC}{MAC}{medium access control}
\newacronym{MAP}{MAP}{maximum a posteriori}
\newacronym{MCS}{MCS}{modulation and coding scheme} 
\newacronym{MEC}{MEC}{multi-access edge computing}     
\newacronym{METIS}{METIS}{Mobile and Wireless Communications Enablers for the Twenty-Twenty Information Society}
\newacronym{MF}{MF}{matched filter}
\newacronym{MGSCM}{MGSCM}{METIS geometry-based stochastic channel model}
\newacronym{MIMO}{MIMO}{multiple-input multiple-output}
\newacronym{MSE}{MSE}{mean-squared error}
\newacronym{ML}{ML}{maximum likelihood}
\newacronym{MLE}{MLE}{maximum likelihood estimator}
\newacronym{MMSE}{MMSE}{minimum mean-square error}
\newacronym{MTC}{MTC}{machine-type communication}
\newacronym{mMTC}{mMTC}{massive \gls{MTC}}
\newacronym{mmW}{mmWave}{millimeter-wave}
\newacronym{mmWave}{mmWave}{millimeter-wave}
\newacronym{MU-MIMO}{MU-MIMO}{multiuser multiple-input-multiple-output}
\newacronym{MU-MISO}{MU-MISO}{multiuser multiple-input-single-output}
\newacronym{MVUE}{MVUE}{minimum variance unbiased estimator}
\newacronym{NLoS}{NLoS}{non-line-of-sight}
\newacronym{NR}{NR}{new radio}
\newacronym{ppm}{ppm}{parts per million}
\newacronym{OFDM}{OFDM}{orthogonal frequency-division multiplexing}
\newacronym{OFDMA}{OFDMA}{orthogonal frequency-division multiple access}
\newacronym{ON}{ON}{observing node}
\newacronym{OTDoA}{OTDoA}{observed time-difference-of-arrival}
\newacronym{PHY}{PHY}{physical}
\newacronym{PDF}{PDF}{probability density function}
\newacronym{PDoA}{PDoA}{phase difference-of-arrival}
\newacronym{PDCCH}{PDCCH}{physical downlink control channel}
\newacronym{QPSK}{QPSK}{quadrature phase shift keying}
\newacronym{NR-PDCCH}{NR-PDCCH}{\acrshort{NR} \acrlong{PDCCH}}
\newacronym{REG}{REG}{resource element group}
\newacronym{REM}{REM}{radio environment map}
\newacronym{RAT}{RAT}{radio access technology}
\newacronym{RAN}{RAN}{radio access network}
\newacronym{RToA}{RToA}{round-trip time of arrival}
\newacronym{RZF}{RZF}{regularized zero-forcing}
\newacronym{PPMCC}{PPMCC}{Pearson product-moment correlation coefficient}
\newacronym{PU}{PU}{primary user}
\newacronym{PS}{PS}{phase shifter}
\newacronym{PW}{PW}{power weighting}
\newacronym{RMSE}{RMSE}{root-mean-squared error}
\newacronym{RRMSE}{RRMSE}{relative root-mean-squared error}
\newacronym[firstplural=reference signals (RSs)]{RS}{RS}{reference signal}
\newacronym{RSS}{RSS}{received signal strength}
\newacronym{RRH}{RRH}{remote radio head}
\newacronym{RRM}{RRM}{radio resource management}
\newacronym{RF}{RF}{radio frequency}
\newacronym{RFI}{RFI}{radio frequency interference}
\newacronym[firstplural=receivers (RX)]{RX}{RX}{receiver}
\newacronym{SBS}{SBS}{switched-beam system}
\newacronym{SDE}{SDE}{sector-pair \acrshort{DoA} estimation}
\newacronym{SIMO}{SIMO}{single-input-multiple-output}
\newacronym{SINR}{SINR}{signal-to-interference-plus-noise ratio}
\newacronym{SLS}{SLS}{simplified least squares}
\newacronym{SNR}{SNR}{signal-to-noise ratio}
\newacronym{SSP}{SSP}{side-sector suppression}
\newacronym{SSL}{SSL}{sector selection}
\newacronym{STD}{STD}{standard deviation}
\newacronym{SU}{SU}{secondary user}
\newacronym{SZF}{SZF}{suppressed zero-forcing}
\newacronym{TCP}{TCP}{transmission control protocol}
\newacronym{TDD}{TDD}{time division duplex}
\newacronym{TDoA}{TDoA}{time difference of arrival}
\newacronym{TN}{TN}{target node}
\newacronym[firstplural=times of arrival (ToA)]{ToA}{ToA}{time of arrival}
\newacronym{ToF}{ToF}{time of flight}
\newacronym{TRP}{TRP}{transmission/reception point}
\newacronym{TSLS}{TSLS}{three-stage SLS}
\newacronym{TTI}{TTI}{transmission time interval}
\newacronym[firstplural=transmitters (TX)]{TX}{TX}{transmitter}
\newacronym{UCA}{UCA}{uniform circular array}
\newacronym{UAV}{UAV}{unmanned aerial vehicle}
\newacronym{UDN}{UDN}{ultra-dense network}
\newacronym{UE}{UE}{user equipment}
\newacronym{UKF}{UKF}{unscented Kalman filter}
\newacronym{UL}{UL}{uplink}
\newacronym{ULA}{ULA}{uniform linear array}
\newacronym{URLLC}{URLLC}{ultra-reliable low latency communication}
\newacronym{UTDoA}{U-TDoA}{uplink-time difference of arrival}
\newacronym{UTD}{UTD}{the uniform theory of diffraction}
\newacronym{UMi}{UMi}{urban micro}
\newacronym{UN}{UN}{user node}
\newacronym{V2V}{V2V}{vehicle-to-vehicle}
\newacronym{VW}{VW}{variance weighting}
\newacronym{WCL}{WCL}{weighted centroid localization}
\newacronym{WLAN}{WLAN}{wireless local area network}
\newacronym{wrt}{wrt}{with respect to}
\newacronym{WSS}{WSS}{wide sense stationary}
\newacronym{ZF}{ZF}{zero-forcing}
\newacronym{BRSRP}{BRSRP}{beam reference signal received power}
\newacronym{RSRP}{RSRP}{reference signal received power}
\newacronym{pdf}{pdf}{probability density function}
\newacronym{PF}{PF}{particle filter}
\newacronym{PRS}{PRS}{positioning reference signal}
\newacronym{NLLF}{NLLF}{negative log likelihood function}
\newacronym{LLF}{LLF}{log-likelihood function}
\newacronym{ARU}{ARU}{analog radio unit}
\newacronym{DF}{DF}{direction finding}
\newacronym{SS}{SS}{synchronization signal}
\newacronym{HPBW}{HPBW}{half power beamwidth}
\newacronym{RSSI}{RSSI}{received signal strength indicator}
\newacronym{CWNV}{CWNV}{constant white-noise velocity}
\newacronym{CWNA}{CWNA}{constant white-noise acceleration}
\newacronym{UPA}{UPA}{uniform planar array}
\newacronym{DM-RS}{DM-RS}{demodulation reference signal}
\newacronym{PDSCH}{PDSCH}{physical downlink shared channel}
\newacronym{PUSCH}{PUSCH}{physical uplink shared channel}
\newacronym{SSB}{SSB}{synchronization signal block}
\newacronym{SRS}{SRS}{sounding reference signal}
\newacronym{SCS}{SCS}{subcarrier spacing}
\newacronym{AoD}{AoD}{angle-of-departure}
\newacronym{SLAM}{SLAM}{simultaneous localization and mapping}
\newacronym{AGC}{AGC}{automitic gain control}
\newacronym{FOV}{FoV}{field-of-view}
\newacronym{VA}{VA}{virtual anchor}
\newacronym{PHD}{PHD}{probability hypothesis density}
\newacronym{COTS}{COTS}{commercial-off-the-shelf}
\newacronym{RFS}{RFS}{random finite set}
\newacronym{PPP}{PPP}{Poisson point process}
\newacronym{RBPF}{RBPF}{Rao-Blackwellized particle filter}
\newacronym{RB}{RB}{Rao-Blackwellized}
\newacronym{MC}{MC}{Monte Carlo}
\newacronym{GM}{GM}{Gaussian mixture}
\newacronym{DA}{DA}{data association}
\newacronym{IPL}{IPL}{iterated posterior linearization}
\newacronym{SLR}{SLR}{statistical linear regression}
\newacronym{OID}{OID}{optimal importance density}
\newacronym{ESS}{ESS}{effective sample size}
\newacronym{SIR}{SIR}{sequential importance resampling}
\newacronym{SIS}{SIS}{sequential importance sampling}
\newacronym{RHS}{RHS}{right-hand side}
\newacronym{IEKF}{IEKF}{iterated extended Kalman filter}
\newacronym{SVD}{SVD}{singular value decomposition}
\newacronym{SP}{SP}{scattering point}
\newacronym{RTT}{RTT}{round-trip time}
\newacronym{CFAR}{CFAR}{constant false alarm rate}
\newacronym{GOSPA}{GOSPA}{generalized optimal sub-pattern assignment}
\newacronym{DS}{DS}{diffuse scattering}
\newacronym{SR}{SR}{specular reflection}
\newacronym{D}{D}{diffraction}
\newacronym[firstplural=sidelobe false detections (SLFD)]{SLFD}{SLFD}{sidelobe false detection}
\newacronym{XR}{XR}{extended reality}

\ifCLASSOPTIONcompsoc
\usepackage[caption=false,font=normalsize,labelfont=sf,textfont=sf]{subfig}
\else
\usepackage[caption=false,font=footnotesize]{subfig}
\fi

\usepackage[font = footnotesize]{caption} 
\usepackage{amsthm} % Include the amsthm package

\newtheorem{lemma}{Lemma}

\usepackage{scalerel}
\usepackage{tikz}
\usetikzlibrary{svg.path}
\definecolor{orcidlogocol}{HTML}{A6CE39}
\tikzset{
  orcidlogo/.pic={
    \fill[orcidlogocol] svg{M256,128c0,70.7-57.3,128-128,128C57.3,256,0,198.7,0,128C0,57.3,57.3,0,128,0C198.7,0,256,57.3,256,128z};
    \fill[white] svg{M86.3,186.2H70.9V79.1h15.4v48.4V186.2z}
                 svg{M108.9,79.1h41.6c39.6,0,57,28.3,57,53.6c0,27.5-21.5,53.6-56.8,53.6h-41.8V79.1z M124.3,172.4h24.5c34.9,0,42.9-26.5,42.9-39.7c0-21.5-13.7-39.7-43.7-39.7h-23.7V172.4z}
                 svg{M88.7,56.8c0,5.5-4.5,10.1-10.1,10.1c-5.6,0-10.1-4.6-10.1-10.1c0-5.6,4.5-10.1,10.1-10.1C84.2,46.7,88.7,51.3,88.7,56.8z};
  }
}

\newcommand\orcidicon[1]{\href{https://orcid.org/#1}{\mbox{\scalerel*{
\begin{tikzpicture}[yscale=-1,transform shape]
\pic{orcidlogo};
\end{tikzpicture}
}{|}}}}
\usepackage{hyperref} %<--- Load after everything else

\newcommand{\N}[0]{\ensuremath{\mathcal{N}}}

\newcommand{\appropto}[0]{%
    \ensuremath{%
        \mathrel{%
            \vcenter{%
                \offinterlineskip\halign{%
                    \hfil$##$\cr%
                    \propto\cr\noalign{\kern2pt}%
                    \sim\cr\noalign{\kern-2pt}%
                }%
            }%
        }%
    }%
}

\renewcommand{\vec}[1]{\ensuremath{{\mathbf{#1}}}}

\newcommand{\vb}[0]{\vec{b}}

\newcommand{\vh}[0]{\vec{h}}
\newcommand{\vm}[0]{\vec{m}}
\newcommand{\vp}[0]{\vec{p}}

\newcommand{\vs}[0]{\vec{s}}

\newcommand{\vx}[0]{\vec{x}}

\newcommand{\vz}[0]{\vec{z}}
\newcommand{\vA}[0]{\vec{A}}

\newcommand{\vH}[0]{\vec{J}}  %modified to J without changing the command otherwise, MV

\newcommand{\vP}[0]{\vec{P}}

\newcommand{\vR}[0]{\vec{R}}
\newcommand{\vQ}[0]{\vec{Q}}

\newcommand{\vW}[0]{\vec{W}}
\newcommand{\vecsymbol}[1]{\ensuremath{\boldsymbol{#1}}}
\newcommand{\vmu}[0]{\vecsymbol{\mu}}
\newcommand{\vSigma}[0]{\vecsymbol{\Sigma}}

\algnewcommand{\algorithmicsubalgorithm}{\textit{Line search}}
\algdef{SE}[SUBALG]{SubAlgorithm}{EndSubAlgorithm}{\algorithmicsubalgorithm}{\algorithmicend\ \algorithmicsubalgorithm}

\makeatletter
\let\OldStatex\Statex
\renewcommand{\Statex}[1][3]{%
  \setlength\@tempdima{\algorithmicindent}%
  \OldStatex\hskip\dimexpr#1\@tempdima\relax}
\makeatother

\newcommand{\blueline}{\tikz[baseline]{\draw[blue,solid,line width = 1.0pt](0,0.8mm) -- (5.3mm,0.8mm)}}
\newcommand{\bluetriangle}{\tikz[baseline]{\draw[blue,solid,line width = 0.5pt] 
(0.0mm,1.6mm) -- (2.0mm,1.6mm) --  (1.0mm,0.0mm) -- (0.0mm,1.6mm)}}
\newcommand{\blackcross}{\tikz[baseline]{\draw[black,solid,line width = 0.5pt] 
(0.0mm,0.0mm) -- (1.6mm,1.6mm) (0.0mm,1.6mm) -- (1.6mm,0.0mm)}}
\newcommand{\redplus}{\tikz[baseline]{\draw[red,solid,line width = 0.5pt]
(0mm,0.8mm) -- (1.6mm,0.8mm)  (0.8mm,1.6mm) -- (0.8mm,0mm) }}

\definecolor{matlab_blue}{rgb}{0.00, 0.45, 0.74}
\definecolor{matlab_red}{rgb}{0.85,0.33,0.10}
\definecolor{matlab_yellow}{rgb}{0.93,0.69,0.13}

\newcommand{\matlabbluebar}{\tikz[baseline]{\draw[matlab_blue,solid,line width = 1.0pt,fill=matlab_blue](0,0) -- (0,2mm) -- (4mm,2mm) -- (4mm,0) -- (0,0)}}
\newcommand{\matlabredbar}{\tikz[baseline]{\draw[matlab_red,solid,line width = 1.0pt,fill=matlab_red](0,0) -- (0,2mm) -- (4mm,2mm) -- (4mm,0) -- (0,0)}}

\definecolor{supone_red}{rgb}{0.5000,0.0000,0.0000}
\definecolor{suptwo_red}{rgb}{0.6562,0.0000,0.0000}
\definecolor{cfarone_yel}{rgb}{1.0000,0.4531,0.0000}
\definecolor{cfartwo_yel}{rgb}{1.0000,0.8438,0.0000}
\definecolor{svdone_blu}{rgb}{0.0000, 0.5469,1.0000}
\definecolor{svdtwo_blu}{rgb}{0.0000, 0.0000,0.5156}
\newcommand{\suponeline}{\tikz[baseline]{\draw[supone_red,dashed,line width = 1.0pt](0,0.8mm) -- (5.3mm,0.8mm)}}
\newcommand{\suptwoline}{\tikz[baseline]{\draw[suptwo_red,solid,line width = 1.0pt](0,0.8mm) -- (5.3mm,0.8mm)}}
\newcommand{\cfaroneline}{\tikz[baseline]{\draw[cfarone_yel,dashed,line width = 1.0pt](0,0.8mm) -- (5.3mm,0.8mm)}}
\newcommand{\cfartwoline}{\tikz[baseline]{\draw[cfartwo_yel,solid,line width = 1.0pt](0,0.8mm) -- (5.3mm,0.8mm)}}
\newcommand{\svdoneline}{\tikz[baseline]{\draw[svdone_blu,dashed,line width = 1.0pt](0,0.8mm) -- (5.3mm,0.8mm)}}
\newcommand{\svdtwoline}{\tikz[baseline]{\draw[svdtwo_blu,dashdotted,line width = 1.0pt](0,0.8mm) -- (5.3mm,0.8mm)}}
\newcommand{\svdthreeline}{\tikz[baseline]{\draw[black,solid,line width = 1.0pt](0,0.8mm) -- (5.3mm,0.8mm)}}
\newcommand{\expectation}{ \mathbb{E}  }

\newcommand{\Abold}{ \mathbf{A}  } 
\newcommand{\abold}{ \mathbf{a}  }

\newcommand{\diag}{ \textrm{diag}  }
\newcommand{\atan}{ \textrm{atan}  }
\newcommand{\imagunit}{\mathrm{i}}

\newcommand{\cc}{ \mathbf{c} }
\newcommand{\bb}{ \mathbf{b} }

\newcommand{\Ncross}{ N_{\mathrm{c}}  }
\newcommand{\Ndirect}{ N_{\mathrm{d}}  }
\newcommand{\Sdirect}{ S_{\mathrm{d}}  }
\newcommand{\Scross}{ S_{\mathrm{c}}  }

\newcommand{\blue}[1]{{\color{blue}#1}}

\renewcommand{\algorithmicrequire}{\textbf{Input: }}
\renewcommand{\algorithmicensure}{\textbf{Output: }}

\usepackage{etoolbox}
\makeatletter 
\pretocmd\@bibitem{\color{black}\csname keycolor#1\endcsname}{}{\fail}
\newcommand\citecolor[1]{\@namedef{keycolor#1}{\color{red}}}
\makeatother

\begin{document}
\bstctlcite{IEEEexample:BSTcontrol}

\title{Millimeter-wave Radio SLAM: End-to-End Processing Methods and Experimental Validation}

\author{\IEEEauthorblockN{
Elizaveta Rastorgueva-Foi \orcidicon{0000-0002-2576-7078},
Ossi Kaltiokallio
\orcidicon{0000-0002-9336-7703}, \IEEEmembership{Member, IEEE},
Yu Ge 
\orcidicon{0000-0003-1747-2664}, \IEEEmembership{Student Member, IEEE},
{Matias Turunen},
Jukka Talvitie \orcidicon{0000-0001-7685-7666}, \IEEEmembership{Member, IEEE},
Bo Tan \orcidicon{0000-0002-6855-6270}, \IEEEmembership{Member, IEEE},
Musa Furkan Keskin \orcidicon{0000-0002-7718-8377}, \IEEEmembership{Member, IEEE},
Henk Wymeersch
\orcidicon{0000-0002-1298-6159}, 
\IEEEmembership{Fellow, IEEE},
and
Mikko Valkama \orcidicon{0000-0003-0361-0800}, \IEEEmembership{Fellow, IEEE}}
\vspace{-0mm}
%\thanks{This paper is supported by the Vinnova B5GPOS Project under Grant 2022-01640.}
%\thanks{{This work was partially supported by the Academy of Finland \red{and by XXX and ZZZ (..or omitted in full)}.}} %(grants \#323244 \#315858, \#319994, \#328214, \#338224, \#341489, and \#346622), and by XXX} {(TO BE UPDATED)}}
\thanks{E. Rastorgueva-Foi, O. Kaltiokallio, {M. Turunen}, J. Talvitie, B. Tan and M. Valkama are with %the Department of Electrical Engineering, 
Tampere University, Finland.}
\thanks{Y. Ge, M.~F.~Keskin and H. Wymeersch are with %the Department of Electrical Engineering, 
Chalmers University of Technology, Sweden.}
\thanks{This manuscript contains 60\,GHz measurement data in the form of raw I/Q samples as well as corresponding channel parameter estimates. {The data is available openly at IEEE DataPort, DOI \blue{\url{https://dx.doi.org/10.21227/xskh-dk87}}}}
}

\maketitle
\begin{abstract}
In this article, we address the timely topic of cellular bistatic simultaneous localization and mapping (SLAM) with specific focus on {end-to-end processing solutions, from raw I/Q samples, via channel parameter estimation to user equipment (UE) and landmark location information} in millimeter-wave (mmWave) networks{, with minimal prior knowledge}. Firstly, we propose a new multipath channel parameter estimation solution that operates directly with beam reference signal received power (BRSRP) measurements, alleviating the need to know the true antenna beampatterns or the underlying beamforming weights. Additionally, the method has built-in robustness against unavoidable antenna sidelobes. Secondly, we propose new snapshot SLAM algorithms that have increased robustness and identifiability compared to prior art, in practical built environments with complex clutter and multi-bounce propagation scenarios{, and do not rely on any a priori motion model}. The performance of the proposed methods is assessed at the 60\,GHz mmWave band, via both realistic ray-tracing evaluations as well as true experimental measurements, in an indoor environment. A wide set of offered results demonstrate the improved performance, compared to the relevant prior art, in terms of the channel parameter estimation as well as the end-to-end SLAM performance. Finally, the article provides the measured 60\,GHz data openly available for the research community, facilitating results reproducibility as well as further algorithm development.
\end{abstract}
\vspace{-1mm}

% ############################################################################
% INTRODUCTION
% ############################################################################

\vspace{-4mm}
\section{Introduction}
\vspace{-0mm}
\label{sec:Introduction}
\lettrine[lines=2]{\textbf{W}} \enskip hile the primary purpose of mobile cellular networks is to provide efficient connectivity services, the ability to extract location information and situational awareness of the surrounding environment is also receiving increasing interest \cite{pmn_wcm_2023,NLoS_HPBW_COMMAG_2021,ITU-R-M2516,Tataria_IEEEProc_2021}. The related notion of \gls{ISAC} refers to extending the situational awareness from ordinary \gls{UE} positioning to the ability to sense also various passive objects in the environment, through cellular radio-based measurements and corresponding signal processing \cite{Liu2022,pmn_wcm_2023}. Such knowledge of the \gls{UE} locations and surrounding environment can be harnessed in numerous ways, for example in different \gls{XR} use cases \cite{Jukka_XR_JSTSP_2023}, vehicular applications \cite{koivisto2017}, or industrial systems \cite{Yi_MUPOSAC}. In general, the prospects for high-accuracy situational awareness are known to improve \cite{journal_78,Que2023} when the networks are expanding towards the \gls{mmW} frequency bands. The current \gls{5G NR} specifications support already operating bands up to 71\,GHz \cite{3GPPTS38104}, while further extensions towards the sub-THz regime are expected in the 6G era \cite{6g_netw_2020}.

\begin{figure}[t!]
    \centering
    \includegraphics[width=0.95\columnwidth]{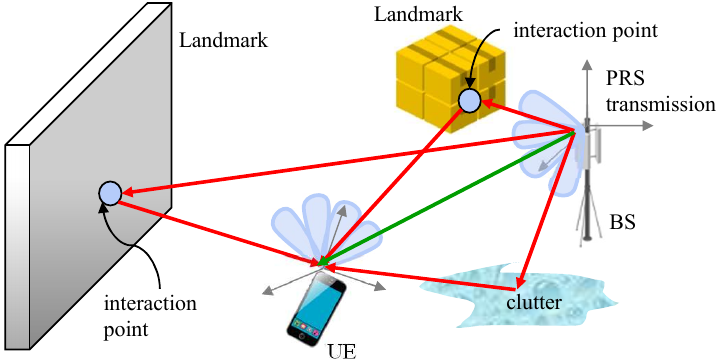}
    \vspace{-0mm}
    \caption{Illustration of the bistatic cellular SLAM paradigm where UE is jointly estimating its own state as well as those of the environment landmarks.}
    \vspace{-0mm}
    \label{fig:LizaPaperScenario}
\end{figure}

Bistatic cellular \gls{SLAM} is one of the prominent ISAC applications where the coordinates of both the \gls{UE} as well as those of the environment scattering points -- commonly referred to as the landmarks -- are all jointly estimated, based on either \gls{UL} or \gls{DL} reference signals and known \gls{BS} locations \cite{yu_jsac_2022,kim2020,Yang2023}. 
{This is illustrated conceptually in Fig.~\ref{fig:LizaPaperScenario}}. Complete end-to-end solutions for \gls{SLAM} comprise \emph{estimators for spatio-temporal channel parameters}, often in the form of \gls{ToA}, \gls{AoA}, and/or \gls{AoD} of the involved propagation paths, combined with the actual \emph{SLAM filters or snapshot estimators} that process the channel parameter estimates into the corresponding estimates of the \gls{UE} and landmark locations \cite{shahmansoori2018}. This is also the main scope of this article{, in contrast to many other studies that focus either exclusively on channel parameter estimation in the context of wireless communication or on cellular SLAM, without considering the challenges of obtaining the channel parameters}. We have a specific focus on implementation-feasible yet high-performing end-to-end processing solutions that can operate with minimal knowledge of the involved antenna system beampatterns and facilitate \gls{mmW} SLAM in practical complex built environments, particularly indoors, while operating on downlink \gls{PRS} standardized already for the existing \gls{5G NR} networks. Additionally, we emphasize research reproducibility and provide measured \gls{mmW} I/Q and channel data openly for the research community, while use the same measured data to evaluate and benchmark the proposed methods against the relevant prior art.

\vspace{-4mm}
\subsection{Prior Art}
The available literature on \emph{channel parameter estimation} is generally-speaking wide, however, a vast majority of works is carried out under the assumptions of known antenna steering vectors, beamforming weights, and thereon beampatterns. Such methods are described, e.g., in \cite{Salmi2009, Jost2012, Alkhateeb2014,Qin2019,Tulino2022,SanchezFernandez2021,Zhu2017,Boljanovic2021,Boljanovic2023}. To this end, \cite{Salmi2009} and \cite{Jost2012} present Bayesian channel estimation algorithms based on different variants of Kalman filters, while \cite{Alkhateeb2014, Tulino2022,SanchezFernandez2021} harness the \gls{mmW} channel sparsity through compressed sensing (CS) methods. The work in \cite{Qin2019}, in turn, proposes an \gls{AoA} estimation method using virtual subarrays which, however, requires a special beamformer or antenna pattern design, similar to \cite{Zhu2017}. Furthermore, \cite{Boljanovic2021,Boljanovic2023} propose joint \gls{AoD} and \gls{AoA} estimation methods with specific frequency-dependent codebooks in true-time-delay array context. Opposed to these previous works, an angle estimation algorithm building on the \gls{BRSRP} measurements is described in the recent work in \cite{Yang2023}. The method builds on thresholding and successive cancellation principle, operating on \gls{AoA}-\gls{AoD} power map, however, is lacking, e.g., explicit treatment of antenna sidelobes whose impact can be substantial in \gls{mmW} systems. This is particularly so in \gls{SLAM} context where the multipaths and their dynamic range {are one of the key factors.}
For clarity, we also note that works like \cite{Gu2022} exist that deploy machine-learning to the channel parameters estimation -- however, such works are still commonly at their early phases.

In the context of the available \gls{SLAM} methods, we first note that both snapshot approaches \cite{shahmansoori2018,wymeersch2018,wen2021,fascista2021} as well as sequential filtering based solutions \cite{kaltiokallio2021,yu_jsac_2022,kim2020,Yang2023,kaltiokallio2022spawc} exist. Both research directions are relevant and the preferred choice depends on the overall system and application scenario.
Snapshot \gls{SLAM} is fundamentally important as it serves as a baseline for what can be done with radio signals alone, without any movement models, while a snapshot method can also be used as input to filtering \cite{wymeersch2018}. On the other hand, filtering based SLAM methods process the observations sequentially over time and are expected to remove false detections and improve the accuracy \cite{kim2020}. Both \gls{SLAM} methods also have their limitations. A major drawback of snapshot \gls{SLAM} is that it is not always applicable, since the measurements may not be sufficient to solve the \gls{SLAM} problem \cite{shahmansoori2018}. For example, the \gls{UE} cannot be localized with the \gls{LoS} alone if the clocks of the \gls{UE} and \gls{BS} are not synchronized \cite{wymeersch2018} and one must resort to filtering based techniques to solve the problem over time. On the other hand, filtering methods always require a snapshot algorithm for initialization when prior information is not available. Another disadvantage of filtering methods is that they require solving a complex data association problem which increases computational overhead of the algorithms \cite{kim2020,kaltiokallio2022spawc}. It is to be noted that low complexity alternatives exist \cite{kaltiokallio2021,yu_jsac_2022}, however, at the expense of reduced accuracy. {Finally, different from mono-static sensing where \gls{TX} and \gls{RX} entities are directly mutually synchronized \cite{Liu2022},\cite{guidi2016,BaqueroBarneto2022}, an important practical aspect and challenge in bi-static SLAM is the ability to estimate and track the time-varying clock-bias between the \gls{UE} and the network elements. Examples of existing works where clock parameters are incorporated as part of the overall state estimation or tracking problem include, e.g., \cite{wymeersch2018,kim2020}.}

\vspace{-2mm}
\subsection{Novelty and Contributions}
Compared to the available methods and literature, this article provides the following contributions.
Firstly, as opposed to the vast majority of existing literature in \cite{Salmi2009, Jost2012, Alkhateeb2014,Qin2019,Tulino2022,SanchezFernandez2021,Zhu2017,Boljanovic2021,Boljanovic2023}, we focus on enabling accurate multipath \gls{AoA}/\gls{AoD} estimation with \gls{BRSRP} measurements only{, \emph{without knowledge of the complex antenna patterns or the underlying steering vectors and beamforming weights}.} This is motivated by the fact that in real networks, only the beam indices and corresponding nominal beam directions are commonly available \cite{Dwivedi2021}. Furthermore, various practical aspects such as errors in the antenna element spacings as well as mutual coupling create anyway varying levels of uncertainties in the true antenna patterns -- in particular in \gls{mmW} networks  \cite{schmid2013effects} where analog/RF beamforming dominates and the design and implementation of antenna elements and beamforming units are tedious. Additionally, the proposed channel parameter estimator that builds on the \gls{SVD} of the \gls{AoA}-\gls{AoD} \gls{BRSRP} map is shown to have built-in robustness against the antenna sidelobes which is a clear additional benefit compared to the prior art in \cite{Yang2023}. The proposed method is also compatible with \gls{5G NR} \gls{PRS} signal structure and beam-sweaping procedures defined in \cite{3GPPTS38211}.

{Secondly, we focus on advancing the SLAM in three directions: first, through increased \emph{robustness and tolerance} of snapshot SLAM, compared to the prior art in \cite{shahmansoori2018,wen2021,fascista2021}, against practical measurement imperfections or outliers; second,  by improving the SLAM problem \emph{identifiability}{; and third, to perform SLAM \emph{without knowledge of the user motion model}.} }These aspects have high importance in real \gls{mmW} deployment environments, where the amount of the available \gls{LoS}/\gls{NLoS} measurements can easily vary from measurement location to another \cite{heath2016overview}. Additionally, despite the advances in channel parameter estimation, measurement outliers are commonly occurring, e.g., due to the clutter, multi-bounce phenomenon and antenna sidelobes \cite{Liu2022}. Specifically, we improve the snapshot \gls{SLAM} identifiability via including an appropriate regularization term into the objective function that embeddes the prior information to the processing system. Additionally, a robust cost function is introduced to handle outliers originating, e.g., from false detections or clutter.

Finally, the openly available \gls{mmW} SLAM measurement data is scarce -- or almost non-existing -- hence, we bridge this important gap and provide 60\,GHz indoor measurement data building on \gls{5G NR} standard-compliant beam-based \gls{PRS} transmissions. We also utilize the measurement data to assess the achievable end-to-end performance of the methods proposed in this article, while benchmarking against the prior-art.

Thus, to summarize, the novelty and contributions of the article can be shortly stated as follows:
\begin{itemize}
    \item {\textbf{End-to-end robust snapshot SLAM approach:} We develop and demonstrate an approach for end-to-end snapshot SLAM, under minimal knowledge of the array parameters and the user motion model. The end-to-end approach comprises a novel channel parameter estimation method and a novel snapshot SLAM method. }
    \item {\textbf{Robust channel parameter estimation:}} We propose a new propagation path \gls{AoA}/\gls{AoD} estimation method utilizing standardized \gls{BRSRP} measurements, alleviating the need for knowledge of the underlying antenna system beampatterns while offering controlled robustness against antenna sidelobes;
    \item {\textbf{Robust snapshot SLAM:}}   We propose a new snapshot \gls{SLAM} algorithm that offers increased system identifiability and improved robustness against measurement outliers compared to prior-art;
    \item {\textbf{Realistic performance evaluation:}} We evaluate and benchmark the performance of the proposed methods in a realistic indoor environment through accurate ray-tracing as well as experimental measurement campaign, both carried out at the 60\,GHz band;
    \item {\textbf{Open-source data-set:}} We release the complete 60\,GHz I/Q measurement data set as well as the corresponding processed channel parameter data set, together with supportive scripts for their utilization in any follow-up research; 
\end{itemize}

The rest of this article is organized as follows:
Section~\ref{sec:SystemModel} describes the basic assumptions, the problem geometry and the corresponding fundamental system model. The proposed channel parameter estimation methods are described in Section~\ref{sec:ChannelEstimation}, while the proposed snapshot \gls{SLAM} method is provided in Section~\ref{sec:SLAM}. The considered indoor evaluation environment, ray-tracing assumptions and the actual 60\,GHz measurement setup and data are all described in Section~\ref{sec:environments}. The complete end-to-end performance results and corresponding comparisons against benchmark methods are reported and analyzed in Section~\ref{sec:Results}. Finally, the conclusions are drawn in~\ref{sec:Conclusions}, {while selected modeling details are provided in the Appendix.}

\textit{Notations:}
Vectors are denoted by bold lowercase letters (i.e., $\abold$), bold uppercase letters are used for matrices (i.e., $\Abold$) and scalars are denoted by normal font (i.e., $a$). The operators $(\cdot)^\top$, $(\cdot)^*$, $(\cdot)^\mathrm{H}$, $(\cdot)^\dagger$, $\expectation \{ \cdot \}$, $|\cdot|$ and $\|\cdot\|$ denote the transpose, conjugate, Hermitian transpose, pseudoinverse, expectation, absolute value and Euclidian norm, respectively. Finally, $\imagunit$ denotes the imaginary unit for which $\imagunit^2=-1$.

% ############################################################################
% SYSTEM MODEL
% 
\vspace{-1mm}
\section{System Model}
\label{sec:SystemModel}
\vspace{-0.0mm}
\subsection{Basic Assumptions and {System Geometry}}
In this work, we consider \gls{OFDM} based \gls{mmW} cellular systems where \glspl{BS} are regularly broadcasting beamformed downlink reference signals, that allow \glspl{UE} to estimate the multipath channel parameters for localization, sensing and mapping purposes. Concrete example in \gls{5G NR} context is the \gls{PRS} \cite{3GPPTS38211}, however, also e.g. the \gls{SS} burst can in practice be utilized, though offering lower bandwidth compared to \gls{PRS}. Additionally, since time- and angle-based measurements allow for a paradigm shift from classical multi-\gls{BS} localization and \gls{SLAM} approaches \cite{koivisto2017} towards single-\gls{BS} solutions \cite{liu2019}, 
we also focus on the single-\gls{BS} scenario. However, the proposed channel parameter estimator is applicable also in multi-\gls{BS} scenarios when the corresponding \glspl{PRS} are properly orthogonalized.

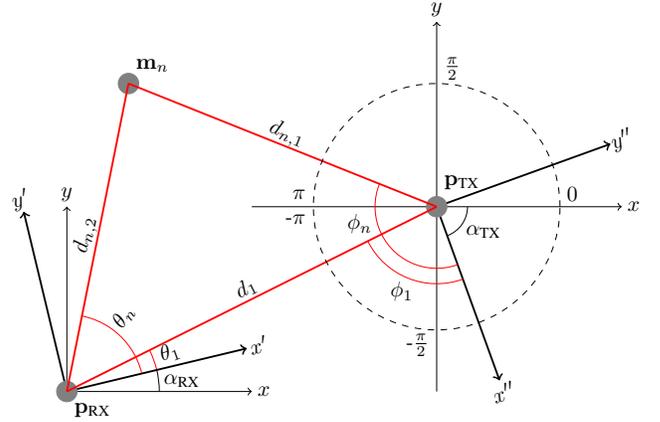
\begin{figure}[!t]
\centering
     \resizebox{88mm}{57mm}{
\begin{tikzpicture}

\draw[->] (0,0) -- (0,3);
\draw[->] (0,0) -- (3,0);
\node[xshift=2mm,yshift=0mm] at (3,0) {$x$};
\node[xshift=0mm,yshift=2mm] at (0,3) {$y$};
\draw[black,line width=0.3mm,->] (0,0) -- (2.92,0.70);
\draw[black,line width=0.3mm,->] (0,0) -- (-0.70,2.92);
\node[rotate=15,xshift=2mm,yshift=0mm] at (2.92,0.70) {$x'$};
\node[rotate=15,xshift=0mm,yshift=2mm] at (-0.70,2.92) {$y'$};

\node[xshift=2mm,yshift=2mm] at (8,3) {$0$};
\node[right,xshift=0mm,yshift=2.5mm] at (6,5) {$\tfrac{\pi}{2}$};
\node[left,xshift=0mm,yshift=2mm] at (4,3) {$\pi$};
\node[left,xshift=0mm,yshift=-2mm] at (4,3) {-$\pi$};
\node[left,xshift=0mm,yshift=-2.5mm] at (6,1) {-$\tfrac{\pi}{2}$};

\draw[->] (3,3) -- (9,3);
\draw[->] (6,0) -- (6,6);
\node[xshift=2mm,yshift=0mm] at (9,3) {$x$};
\node[xshift=0mm,yshift=2mm] at (6,6) {$y$};
\draw[black,line width=0.3mm,->] (6,3) -- (7.02,0.18); 
\draw[black,line width=0.3mm,->] (6,3) -- (8.82,4.02); 
\node[rotate=20,xshift=0mm,yshift=-2mm] at (7.02,0.18) {$x''$};
\node[rotate=20,xshift=2mm,yshift=0mm] at (8.82,4.02) {$y''$};

\node[fill,draw,circle,color=gray,minimum size=1mm] at (6,3) (a) {};
\node[fill,draw,circle,color=gray,minimum size=1mm] at (0,0) (b) {};
\node[fill,draw,circle,color=gray,minimum size=1mm] at (1,5) (c) {};

\draw[red, line width=0.3mm,-] (6,3) -- (0,0);
\draw[red, line width=0.3mm,-] (6,3) -- (1,5);
\draw[red, line width=0.3mm,-] (0,0) -- (1,5);

\draw[black,dashed] (8,3) arc [start angle=0, delta angle=360, radius=2];
\draw[black] (6.5,3) arc [start angle=0, delta angle=-70, radius=0.5];
\draw[black] (1.5,0) arc [start angle=0, delta angle=14, radius=1.5];
\draw[red] (6.3420,    2.0603) arc [start angle=290, delta angle=-131.8014, radius=1];
\draw[red] (6.4275,    1.8254) arc [start angle=290, delta angle=-83.4349, radius=1.25];
\draw[red] (1.4554,    0.3629) arc [start angle=14, delta angle=12.5651, radius=1.5];
\draw[red] (1.2129,    0.3024) arc [start angle=14, delta angle=64.6901, radius=1.25];
\draw[] (a) node[right,xshift=0mm,yshift=4mm] {$\mathbf{p}_\textrm{TX}$};
\draw[] (b) node[right,xshift=0mm,yshift=-3mm] {$\mathbf{p}_\textrm{RX}$};
\draw[] (c) node[right,xshift=0mm,yshift=3mm] {$\mathbf{m}_n$};

\node[right,xshift=-0.5mm,yshift=1.75mm] at (1.5,    0) {$\alpha_\textrm{RX}$};
\node[right,xshift=-1.5mm,yshift=-4.00mm] at (6.5,3) {$\alpha_\textrm{TX}$};
\node[right,xshift=-13mm,yshift=-2.00mm] at (6.4275,    1.8254) {$\phi_1$};
\node[right,xshift=-0.5mm,yshift=1.75mm,rotate=20] at (1.4554,    0.3629) {$\theta_1$};
\node[rotate=26.5651,yshift=2mm] at (3,1.5) {$d_1$};
\node[right,xshift=-20mm,yshift=9.00mm] at (6.4275,    1.8254) {$\phi_n$};
\node[right,xshift=-7mm,yshift=6mm,rotate=56.5] at (1.4554,    0.3629) {$\theta_n$};
\node[rotate=-21.8014,yshift=2mm] at (3.5,4) {$d_{n,1}$};
\node[rotate=78.6901,yshift=2mm] at (0.5,2.5) {$d_{n,2}$};
\end{tikzpicture}
}
\vspace{-5mm}
\caption{Problem geometry for \gls{LoS} and $n$th \gls{NLoS} propagation path with angles  expressed in the local frames of the \gls{TX} (BS) and \gls{RX} (UE). {$\alpha_\textrm{TX}$ and $\alpha_\textrm{RX}$ represent the location orientations relative to the global coordinate system.}}
\label{fig:problem_geometry}
\vspace{-1mm}
\end{figure}

To this end, we consider \gls{TX} and \gls{RX} entities equipped with \glspl{UPA} with vertical times horizontal dimensions of the form  ${N}^{v}_\textrm{TX}\times {N}^{h}_\textrm{TX}$ and ${N}^{v}_\textrm{RX}\times {N}_\textrm{RX}^{h}$, respectively. The antenna elements are separated by a half-wavelength  distance $d = \lambda/2$, where $\lambda = c/f_\text{c}$ denotes the wavelength at carrier frequency $f_\text{c}$, and $c$ is the speed of light. The total number of antenna elements in \gls{TX} and \gls{RX} are ${N}_\textrm{TX} = {N}^{v}_\textrm{TX}\times {N}^{h}_\textrm{TX}$ and ${N}_\textrm{RX} = {N}^{v}_\textrm{RX}\times{N}_\textrm{RX}^{h}$, respectively. 
{The locations (3D coordinates) of the involved TX and RX entities are denoted by $\vp_\textrm{TX}^\textrm{3D} = [\vp_\textrm{TX}^\top,\, z_\textrm{TX}]^\top$ and $\vp_\textrm{RX}^\textrm{3D} = [\vp_\textrm{RX}^\top,\, z_\textrm{RX}]^\top$, respectively, where $\vp_\textrm{TX}$ and $\vp_\textrm{RX}$ are the corresponding {horizontal} 2D coordinates, {while $z_\textrm{TX}$ and $z_\textrm{RX}$ denote the respective vertical coordinates} -- all in \emph{global coordinate system}. Additionally, the 3D location of an arbitrary single-bounce landmark is denoted by $\vm_n^\textrm{3D} = [\vm_n^\top,\, z_n]^\top${,} where $\vm_n$ refers to the respective {horizontal} 2D coordinates, {$z_n$ is a vertical coordinate}, while the subscript $n$ serves as a landmark or path index. Furthermore, we denote the 3D \gls{AoD} as $\boldsymbol{\psi}^\top_{\text{TX},n}=[\phi_n, \phi_n^{\text{EL}}]$, where $\phi_n$ and $\phi_n^{\text{EL}}$ are the azimuth and elevation \glspl{AoD}, respectively, Similarly, the 3D \gls{AoA} is denoted by $\boldsymbol{\psi}^\top_{\text{RX},n} = [\theta_n, \theta_n^{\text{EL}}]$. All the involved angles are expressed in \emph{local coordinate systems}, i.e., relative to the local orientations of the TX and RX entities.

Finally, as our main emphasis is on indoor \gls{mmW} systems, we eventually focus mostly on 2D (azimuth) estimation and SLAM algorithms. The corresponding 2D system geometry is illustrated in Fig.~\ref{fig:problem_geometry}. However, the basic system and received signal models are provided in 3D for generality.  
}

%%%%%%%%%%%%%%%%%%%%%%%%
% DRAWING FIGURE 2
%%%%%%%%%%%%%%%%%%%%%%%%

\vspace{-3mm}
\subsection{Received Signal Model}
{We assume that coarse timing information is established between \gls{TX} and \gls{RX} entities, through, e.g., correlation with known \gls{PRS} sequences as discussed further in Section~\ref{sec:ChannelEstimation}.}
Now, under multipath radio propagation environment with $N$ propagation paths, the received signal at $k^\text{th}$ subcarrier and $m^\text{th}$ OFDM symbol, using the $i^\text{th}$ \gls{TX} beam and $j^\text{th}$ \gls{RX} beam, can be represented as
\begin{equation} \label{eq:rx_symbol_vec}
    y_{k,m}^{i,j} = \mathbf{w}_{\text{RX},j}^\mathrm{H} \left( \mathbf {H}_{k,m} \mathbf{w}^{*}_{\text{TX},i} x_{k,m}^{i,j} + \mathbf{n}_{k,m}^{i,j} \right), 
\end{equation}
where $\mathbf{w}_{\text{TX},i} \in \mathbb{C}^{\text{N}_{\textrm{TX}}}$ and $\mathbf{w}_{\text{RX},j} \in \mathbb{C}^{\text{N}_{\textrm{RX}}}$ are the \gls{TX} and \gls{RX} beamformers, $x_{k,m} \in \mathbb{C}$ with $|x_{k,m}|=1\,\forall k,m$ is the transmitted \gls{PRS} sample, and $\mathbf{n}_{k,m}^{i,j} \in \mathbb{C}^{\text{N}_{RX}}$ denotes the antenna element wise \gls{AWGN} at the \gls{RX}. Furthermore, $\mathbf {H}_{k,m} \in \mathbb{C}^{\text{N}_{\textrm{RX}} \times \text{N}_{\textrm{TX}}}$ is the effective spatial multipath channel matrix defined as \cite[eq.~(4)]{heath2016overview} 
\begin{align} \label{eq:channel}
\begin{split}
    \mathbf{H}_{k,m} = \sum_{n=1}^{N} & \xi_n e^{-\imagunit2\pi k \Delta f {\tau_{\mathrm{f},n}} } e^{\imagunit 2\pi m T_{\text{sym}} f_{\text{D},n}} \\ 
    & \times \mathbf{a}_{\text{RX}}(\boldsymbol{\psi}_{\text{RX},n}) \mathbf {a}_{\text{TX}}(\boldsymbol{\psi}_{\text{TX},n})^\top,
\end{split}
\end{align}
{where $\Delta f$ is the \gls{SCS}, $T_\text{sym}$ denotes the \gls{OFDM} symbol duration, and $\xi_n$, $\tau_{\mathrm{f},n}$ and $f_{\text{D},n}$ are the complex path coefficient, the propagation delay with respect to the beginning of the received \gls{OFDM} symbol, and the Doppler frequency for the $n^\text{th}$ propagation path, respectively. The true physical propagation delay for path $n$ is denoted by $\tau_n = \tau_{\text{c}} + \tau_{\mathrm{f},n}$, with $d_n = c\tau_n$ denoting the corresponding physical distance, where $\tau_{\text{c}}$ refers to the delay between the received \gls{OFDM} symbol and transmit time. Considering now a receiver with an unknown clock bias $b_\textrm{UE}$, the corresponding biased propagation delay $\tau^b_n$ can be expressed as 
\begin{equation}
    \tau^b_n = \tau_n - b_\textrm{UE} = \tau_{\text{c}} + \tau_{\mathrm{f},n} - b_\textrm{UE}.
\end{equation}
Estimation of $\tau_{\text{c}}$, subject to unknown bias $b_\textrm{UE}$, can be considered part of regular \gls{OFDM} symbol synchronization, for example, through time-based correlation, as discussed further in Section \ref{sec:toa_estimation}.} 
Additionally, $\mathbf{a}_{\text{TX}}(\boldsymbol{\psi}_{\text{TX},n})\!\in\! \mathbb{C}^{N_{\text{TX}}}$ and $\mathbf{a}_{\text{RX}}(\boldsymbol{\psi}_{\text{RX},n}) \!\in\! \mathbb{C}^{N_{\text{RX}}}$ denote the  \gls{TX} and RX steering vectors, respectively. The exact way how the \gls{PRS} sequences are mapped to the physical resources (OFDM symbols and the underlying subcarriers), in \gls{5G NR} context, is described in \cite{3GPPTS38211}.

Now, by combining \eqref{eq:rx_symbol_vec} and \eqref{eq:channel}, the received signal model can be re-expressed as 
\begin{align} \label{eq:rx_symbol2}
\begin{split}
    y_{k,m}^{i,j} = \Big\{ \sum_{n=1}^{N} & \xi_n e^{-\imagunit2\pi k \Delta f {\tau_{\mathrm{f},n} }} e^{\imagunit 2\pi m T_{\text{sym}} f_{\text{D},n}} x_{k,m}^{i,j} \\  &  \times G_{\text{TX},i}(\boldsymbol{\psi}_{\text{TX},n}) G_{\text{RX},j}(\boldsymbol{\psi}_{\text{RX},n}) \Big\} + \tilde{n}_{k,m}^{i,j}, 
\end{split}
\end{align}
where $\tilde{n}_{k,m}^{i,j} =  \mathbf{w}_{\text{RX},j}^\mathrm{H}\mathbf{n}_{k,m}^{i,j}$ denotes beamformed noise, 
while $G_{\text{TX},i}(\boldsymbol{\psi}_{\text{TX},n}) = \mathbf{a}_{\text{TX}}(\boldsymbol{\psi}_{\text{TX},n})\!{^\top}\! \mathbf{w}^{*}_{\text{TX},i}$, and $G_{\text{RX},j}(\boldsymbol{\psi}_{\text{RX},n}) = \mathbf{w}_{\text{RX},j}^\mathrm{H} \mathbf{a}_{\text{RX}}(\boldsymbol{\psi}_{\text{RX},n})$. Importantly, the expression in \eqref{eq:rx_symbol2} applies to arbitrary antenna systems with $G_{\text{TX},i}(\boldsymbol{\psi}_{\text{TX},n})\!\in\! \mathbb{C}$ and $G_{\text{RX},j}(\boldsymbol{\psi}_{\text{RX},n})\!\in\! \mathbb{C}$ 
denoting the corresponding angular responses for the $i^\text{th}$ and  $j^\text{th}$ beams at the \gls{TX} and \gls{RX} sides.

{The fundamental technical problems considered in the article are (\emph{i}) to estimate the involved path angles and delays, with received PRS samples, and (\emph{ii}) to estimate the UE location and the locations of the landmarks, with angle and delay estimates as the inputs. These are addressed in Sections III and IV, respectively, where Section IV also details the relations between the locations and the involved path angles and delays.}

% ############################################################################
% PROPOSED PARAM ESTIMATION METHODS
% ############################################################################
\vspace{-3mm}
\section{Channel Parameter Estimation Methods}
\label{sec:ChannelEstimation}

In this section, we provide a detailed description of the proposed \gls{AoA}/\gls{AoD} estimation method, along with an explanation of the \gls{ToA} estimation approach we have considered. To ensure clarity and simplify the presentation, we focus on describing these methods in the 2D/azimuth domain. This approach is particularly relevant in indoor scenarios, where the floor and ceiling impose constraints on the extent of the vertical direction.
{We further assume that directional mmWave beams are deployed, such that the TX and RX beam indices $i$ and $j$ have physical correspondence to the \gls{TX} and \gls{RX} beamforming angles, respectively. Such directional beams are considered in large majority of the mmWave systems research, particularly when analog/RF beamforming is assumed.}

\vspace{-3mm}
\subsection{BRSRP Measurements}

Taking \gls{RSRP} measurements is generally a standard procedure in \gls{5G NR}~{\cite{3GPPTS38133}}, e.g., for paging and beam alignment purposes.
To this end, for each TX-RX beam pair $(i,j)$, let us define the corresponding \gls{BRSRP} as a beam-based \gls{RSRP} measurement of the considered \gls{RS}. Specifically, this is defined as
\begin{equation} \label{eq:brsrp}
\beta_{i,j}=\frac{1}{{N}_\text{RS}}\sum_{(k,m) \in \mathcal{M}_\text{RS}} \left\vert y_{k,m}^{i,j} \right\vert^{2},
\end{equation}
where $\mathcal{M}_\text{RS}$ is a set of all \gls{RS} symbols and subcarriers mapped to the \gls{OFDM} resource grid with cardinality $\vert \mathcal{M}_\text{RS} \vert = N_\text{RS}$.

\begin{lemma}\label{lemma_approx}
Let $\mathbf{b}_{n} \in \mathbb{C}^{K}$ and $\mathbf{c}_{n} \in \mathbb{C}^{M}$ denote the frequency-domain and time-domain steering vectors of the $n^{\text{th}}$ path, respectively, where $K$ is the number of active RS subcarriers, and $M$ is the number of RS \gls{OFDM} symbols. Moreover, the $k^{\text{th}}$ element of $\mathbf{b}_{n}$ is defined as $\mathbf{b}_{n}[k] = e^{-\imagunit2\pi k \Delta f \tau_{\mathrm{f},n}}$, and the $m^{\text{th}}$ element of $\mathbf{c}_{n}$ as $\mathbf{c}_{n}[m] = e^{\imagunit 2\pi m T_{\text{sym}} f_{\text{D},n}}$.
Suppose that the paths in \eqref{eq:channel} are non-overlapping in either delay or Doppler, and that $K$ and/or $M$ are/is sufficiently large, i.e.,
\begin{align}\label{eq_delayDoppler_separation}
    \frac{\bb_{n_1}^\mathrm{H} \bb_{n_2}}{{K}} \approx 0 ~~~ {\rm{or}} ~~~ \frac{\cc_{n_1}^\mathrm{H} \cc_{n_2}}{{M}} \approx 0
\end{align}
for any $n_1 \neq n_2$. Then, the BRSRP measurement in \eqref{eq:brsrp} can be approximated as
\begin{equation} \label{eq:power_angle} %\vspace{-2mm}
\!\! \beta_{i,j} \! \approx \!\sum_{n=1}^{N} \! \vert \xi_n \vert^2 \vert G_{\mathrm{TX},i} (\boldsymbol{\psi}_{\mathrm{TX},n}) \vert^2 \vert G_{\mathrm{RX},j} (\boldsymbol{\psi}_{\mathrm{RX},n}) \vert^2 + {\sigma_{\mathrm{noise}}^2}
\end{equation}
where {$\sigma_{\mathrm{noise}}^2$ is the variance of $\tilde{n}_{k,m}^{i,j}$}.

\end{lemma}
%\vspace{-2mm}
\begin{proof}
    Please see {the Appendix.} 
\end{proof}
\vspace{-1mm}

%%%%%%%%%%%%%%%%%%%%

\vspace{-1mm}
\subsection{Proposed SVD-based AoD and AoA Extraction} \label{sec:angle_extraction}
Considering the PRS beam-sweeping procedure with $L_\textrm{TX}$ \gls{TX} beams and $L_\textrm{RX}$ \gls{RX} beams, {comprising the corresponding \gls{TX} and \gls{RX} beamforming angles $\{\Phi_{i}\}_{i=1}^{L_{\text{TX}}}$ and  $\{\Theta_{j}\}_{j=1}^{L_{\text{RX}}}$}, respectively, a total of $L_\textrm{TX} \times L_\textrm{RX}$ directional \gls{BRSRP} measurements are available at the \gls{RX}. The corresponding \gls{BRSRP} matrix $\mathbf{B} \in \mathbb{R}^{L_\textrm{TX} \times L_\textrm{RX}} $ is defined as
\begin{equation} \label{eq:brsrp_matrix}
\mathbf{B} = 
\begin{pmatrix}
\beta_{1,1}&\beta_{1,2}&\cdots&\beta_{1,L_\textrm{RX}} \\
\beta_{2,1}&\beta_{2,2}&\cdots&\beta_{2,L_\textrm{RX}} \\
\vdots  & \vdots  & \ddots & \vdots  \\
\beta_{L_\textrm{TX},1}&\beta_{L_\textrm{TX},2}&\cdots&\beta_{L_\textrm{TX},L_\textrm{RX}} \\
\end{pmatrix}.
\end{equation}
The matrix $\mathbf{B}$ thus represents the spatial channel in the angular domain, in the form of  an \gls{AoD}-\gls{AoA} power map. A concrete example is visualized in Fig.~\ref{fig:example_map}, building on measurement arrangements described in Section~\ref{sec:environments}. 

\subsubsection{{Technical Premise and Intuition}} Taking into account the fairly narrow beamwidths and the mmWave channel sparsity, the {\glspl{AoD} $\{\phi_n\}_{n=1}^{N}$ and \glspl{AoA} $\{\theta_n\}_{n=1}^{N}$ for $N$} propagation paths can be extracted by processing the power map $\mathbf{B}$. For this purpose, based on \eqref{eq:brsrp} and the approximation in \eqref{eq:power_angle}, the angular power map $\mathbf{B}$ can be first represented as
\begin{equation}
    \mathbf{B} = \sum_{n=1}^{N} \vert \xi_n \vert^2 \mathbf{g}_{\text{TX}}(\boldsymbol{\psi}_{\text{TX},n}) \mathbf{g}_{\text{RX}}(\boldsymbol{\psi}_{\text{RX},n})^\top + \bar{\mathbf{N}},
    \label{eq:power_map_B}
\end{equation}
where $\mathbf{g}_{\text{TX}}(\boldsymbol{\psi}_{\text{TX},n})$ and $\mathbf{g}_{\text{RX}}(\boldsymbol{\psi}_{\text{RX},n})$ denote the \gls{TX} and \gls{RX} beam gain vectors, with $\mathbf{g}_{\text{TX}}(\boldsymbol{\psi}_{\text{TX},n})[i] = \vert G_{\text{TX},i}(\boldsymbol{\psi}_{\text{TX},n}) \vert^2$ and  $\mathbf{g}_{\text{RX}}(\boldsymbol{\psi}_{\text{RX},n})[j] = \vert G_{\text{RX},j}(\boldsymbol{\psi}_{\text{RX},n}) \vert^2$, respectively. Furthermore, $\bar{\mathbf{N}}$ is a noise matrix whose element at the $i^{\text{th}}$ row and $j^{\text{th}}$ column is defined as $\bar{\mathbf{N}}[i,j] = 
{(1/N_\mathrm{RS}) \! \sum_{(k,m) \in \mathcal{M}_\mathrm{RS}} \! \vert \tilde{n}_{k,m}^{i,j} \vert^2}$.

\begin{figure}
  \begin{center}
  \vspace{-1mm}
  \includegraphics[width=3.2in,trim={0.3cm 1.9cm 1.0cm 2.65cm},clip]{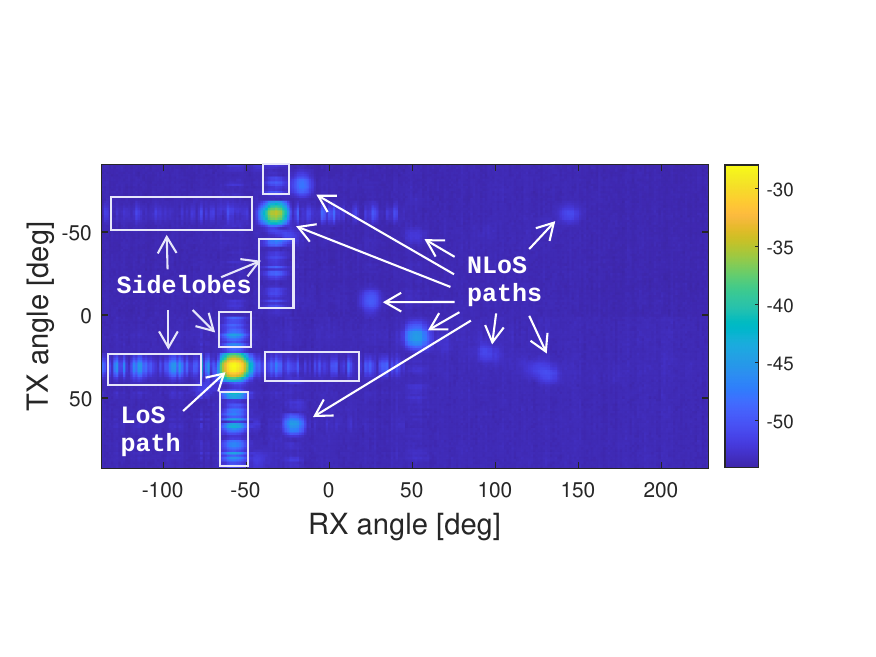}
  \vspace{-1mm}
  \caption{Example visualization of measured 60\,GHz BRSRP data, with the \gls{LoS} path and other landmarks being clearly visible. Antenna sidelobes are creating notable power spreading along the two axes.
  }
  \vspace{-5mm}
  \label{fig:example_map}
  \end{center}
\end{figure}

Inspired by the structure of $\mathbf{B}$ in \eqref{eq:power_map_B}, let us consider its singular value decomposition (SVD), expressed as $\mathbf{B} = \mathbf{U} \mathbf{\Lambda} \mathbf{V}^\top$,  
where $\mathbf{U} \in \mathbb{R}^{L_\textrm{TX} \times L_\textrm{TX}}$, $\mathbf{V} \in \mathbb{R}^{L_\textrm{RX} \times L_\textrm{RX}}$ are unitary matrices, and $\mathbf{\Lambda} \in \mathbb{R}^{L_\textrm{TX} \times L_\textrm{RX}}$ is a rectangular diagonal matrix. In the main diagonal of $\mathbf{\Lambda}$, singular values of $\mathbf{B}$, $[\sigma_1,\sigma_2,...,,\sigma_R] > 0${,} are sorted in the descending order, where $R = \text{rank}(\mathbf{B}) \leq \text{min}(L_\textrm{TX},L_\textrm{RX})$.
The SVD can also be expressed as
\begin{equation} \label{eq:svd_rank1_sum}
    \mathbf{B} = \sum_{r=1}^{R} \sigma_r \mathbf{u}_r \mathbf{v}_r^\top = \sum_{r=1}^{R} \mathbf{B}^{(1)}_r,
\end{equation}
where $\mathbf{u}_r$ and $\mathbf{v}_r$ are the $r^{\text{th}}$ columns of $\mathbf{U}$ and $\mathbf{V}$, respectively. Importantly, {based on \eqref{eq:power_map_B}}, $\mathbf{u}_r$ and $\mathbf{v}_r$ belong to the corresponding spans of $\{\mathbf{g}_{\text{TX}}(\boldsymbol{\psi}_{\text{TX},n})\}_{n=1}^{N}$ and $\{\mathbf{g}_{\text{RX}}(\boldsymbol{\psi}_{\text{RX},n})\}_{n=1}^{N}$. Thus, the $r^{\text{th}}$ singular vectors can be expressed as
\begin{equation} \label{eq:uv_resp}
\mathbf{u}_r \!=\!\! \sum_{n=1}^{N}\! q_{r,n}^\text{u}\mathbf{g}_{\text{TX}}(\boldsymbol{\psi}_{\text{TX},n}) \text { and } \mathbf{v}_r \!=\!\! \sum_{n=1}^{N}\! q_{r,n}^\text{v}\mathbf{g}_{\text{RX}}(\boldsymbol{\psi}_{\text{RX},n}),    
\end{equation} 
where $q_{r,n}^\text{u}$ and $q_{r,n}^\text{v}$ are the linear combination weights of the spanning set vectors $\mathbf{g}_{\text{TX}}(\boldsymbol{\psi}_{\text{TX},n})$ and $\mathbf{g}_{\text{RX}}(\boldsymbol{\psi}_{\text{RX},n})$, respectively. 
In addition, $\mathbf{B}^{(1)}_r$ in \eqref{eq:svd_rank1_sum} is a rank-1 matrix associated with the $r^{\text{th}}$ singular value and the related basis vectors $\mathbf{u}_r$ and $\mathbf{v}_r$. 
Furthermore, in the case of a sparse beamformed channel where all paths are well separated{, combined with directional beamforming at the TX and RX, the following holds \cite{bajwa2010compressed,heath2016overview}}
\begin{align}
\begin{split}
    \mathbf{g}_{\text{TX}}(\boldsymbol{\psi}_{\text{TX},n})^\top \mathbf{g}_{\text{TX}}(\boldsymbol{\psi}_{\text{TX},n'}) &\approx \delta_{n-n'} \text{ and} \\
    \mathbf{g}_{\text{RX}}(\boldsymbol{\psi}_{\text{RX},n})^\top \mathbf{g}_{\text{RX}}(\boldsymbol{\psi}_{\text{RX},n'}) &\approx \delta_{n-n'},
\end{split}
\end{align}
where $\delta_{n-n'}$ is a unit impulse function. In such case, each SVD term in \eqref{eq:svd_rank1_sum} corresponds essentially to one propagation path, and thus $\mathbf{B}^{(1)}_r$ is a rank-1 approximation of the power map stemming from the corresponding path. 
 
Interestingly, based on \eqref{eq:power_map_B}--\eqref{eq:uv_resp} and the corresponding fact that the {singular vectors of the BRSRP matrix are in the span of the beam gain vectors, the SVD-based low-rank modeling approach can also inherently manage different unknown beam patterns, and is thus resilient to sidelobes and other beam pattern fluctuations}. Specifically, the antenna sidelobes illustrated also in {Fig.~\ref{fig:example_map}},  can be decoupled from the peaks in $\mathbf{B}$ by using different low-rank approximations with increasing rank and analyzing the resulting power increment formed at each step.  To this end, a pair of subsequent low-rank approximations of $\mathbf{B}$, denoted as $\mathbf{B}_{K-1}$ and $\mathbf{B}_K$, with %$k-1,
$K < R$, can be expressed as
\begin{equation} \label{eq:rank_K_approx}
    \mathbf{B}_K =  \sum_{r=1}^{K} \sigma_r \mathbf{u}_r \mathbf{v}_r^\top = \sum_{r=1}^{K} \mathbf{B}^{(1)}_r = \mathbf{B}_{K-1} + \mathbf{B}^{(1)}_K,  \vspace{-1mm}
\end{equation}
where the rank-1 matrix $\mathbf{B}^{(1)}_K=\sigma_K \mathbf{u}_K \mathbf{v}_K^\top$ represents a recovered \emph{BRSRP map increment} between the two levels of approximation accuracy. 

\subsubsection{{Overall Approach and Refinements}}Due to the relation of singular vectors to antenna response in \eqref{eq:uv_resp} and the relation of the singular vectors and singular matrix in \eqref{eq:svd_rank1_sum}, the \gls{AoD} and \gls{AoA} estimates can be extracted recursively by processing rank-1 matrices  $\mathbf{B}^{(1)}_r$ for separate singular values, while assuming only known \gls{TX} beam angles $\{\Phi_{i}\}_{i=1}^{L_{\text{TX}}}$ and \gls{RX} beam angles $\{\Theta_{j}\}_{j=1}^{L_{\text{RX}}}$. 
{A suitable rank value for the low-rank approximation in \eqref{eq:rank_K_approx} can be determined in various manner so that noise-dominated singular matrices are omitted from the approximation. In this article, we choose the approximation rank by ensuring that the low-rank approximation recovers a desired portion of the original total measured power, as described in Algorithm \ref{alg:angle_extraction}. Additionally, suitable rank selection can also help 
emphasizing \gls{LoS} and single-bounce paths, over the corresponding multi-bounce paths \cite{BaqueroBarneto2022} that are typically of low power.}

\begin{algorithm}[t]
\small
\caption{Proposed AoD-AoA Extraction Algorithm}
\label{alg:angle_extraction}
\algorithmicrequire{\gls{TX} beam angles $\{\Phi_{i}\}_{i=1}^{L_{\text{TX}}}$, \gls{RX} beam angles $\{\Theta_{j}\}_{j=1}^{L_{\text{RX}}}$, power measurement matrix $\mathbf{B} \in \mathbb{R}^{L_\text{TX} \times L_\text{RX}}$, desired power ratio $p$, and power threshold $\beta_\text{th}$}\\
\algorithmicensure{Number of estimated paths $\hat{N}$, \gls{AoD} estimates $\{ \hat{\phi}_n \}_{n=1}^{\hat{N}}$, and \gls{AoA} estimates $\{ \hat{\theta}_n \}_{n=1}^{\hat{N}}$}
\begin{algorithmic}[1]
    \State Compute SVD of \gls{BRSRP} matrix:  $\mathbf{B} = \mathbf{U} \mathbf{\Lambda} \mathbf{V}^\top$
    \State Compute the total power  $p_{\text{TOT}}=\sum_{r=1}^{R} \sigma_r^2$, where $\sigma_r$ is the $r^{\text{th}}$ singular value, and $R = \text{min}(L_\textrm{TX},L_\textrm{RX})$
    
    \State Set $r = 1$, $p_r = 0\%$, $\mathcal{M}_{\text{est}}=\{\}$
    \While {$p_r < p$}
    \State \textit{Compute a rank-1 matrix:} $\mathbf{B}^{(1)}_r = \sigma_r \mathbf{u}_r \mathbf{v}_r^\top$
    \State \textit{Find the indices of the element with maximum value:} 
    \Statex[2] $(\hat{i}_r,\hat{j}_r) = \argmax_{i,j}\mathbf{B}^{(1)}_r[i,j]$,
    \Statex[2] where $\mathbf{B}^{(1)}_r[i,j]$ is the matrix $\mathbf{B}^{(1)}_r$ element at the \Statex[2] $i^{\text{th}}$ row and $j^{\text{th}}$ column.
    \State \textit{Save the angles and power as a 3D set element:} 
    \Statex[2] $\mathcal{M}_{\text{est}} \leftarrow \mathcal{M}_{\text{est}} \cup \{(\Phi_{\hat{i}_r},\Theta_{\hat{j}_r},\beta_{\hat{i}_r,\hat{j}_r})\}$ 
    \State {\textit{Compute power ratio:} $p_r \leftarrow \frac{1}{p_{\text{TOT}}}\sum_{\tilde{r}=1}^{r} \sigma_{\tilde{r}}^2 \times 100\%$}
    \State \textit{Set } $r \leftarrow r + 1$
    \EndWhile, $K=r$
    %\Statex % Add vertical space
    \vspace{1mm}
    \State \textbf{Power thresholding} -- choose only the components with power exceeding a given threshold: 
    \Statex[2] 
    $\mathcal{M}_{\text{est}} \leftarrow \{ \mathcal{M}_{\text{est}} \, \vert \, \text{elements with } \beta_{\hat{i}_r,\hat{j}_r} \geq \beta_\text{th} \}$
    %\Statex % Add vertical space
    \State \textbf{Clustering} -- Perform clustering of elements in $\mathcal{M}_{\text{est}}$ to obtain $\hat{N} \leq K$ clusters $\mathcal{M}_{\text{est}}^n$, $n=1,\dots,\hat{N}$.
    For each cluster, angle estimates are obtained by computing a power-weighted cluster mean as $\bar{{\phi}}_n = (\sum_{s} \beta_{{\hat{i}_s},{\hat{j}_s}} {\Phi}_{\hat{i}_s})/\sum_{s} \beta_{{\hat{i}_s},{\hat{j}_s}}$ and $\bar{{\theta}}_n = (\sum_{s} \beta_{{\hat{i}_s},{\hat{j}_s}} {\Theta}_{\hat{i}_s})/\sum_{s} \beta_{{\hat{i}_s},{\hat{j}_s}}$, where the index $s$ is used to select elements from the specific cluster. 
    
    \State \textbf{Polynomial fitting} -- a subset of the power matrix $\mathbf{B}$ around each cluster mean is fitted with a $2^{\text{nd}}$ degree polynomial surface. For the $n^{\text{th}}$ cluster, the final \gls{AoD} and \gls{AoA} estimates are found in closed form as a {vertex of the local polynomial surface as $\hat{\phi}_n = (c_5 c_3 - 2 c_6 c_2)/(4 c_6 c_4 - c_5^2)$, $\hat{\theta}_n = (c_5 c_2 - 2 c_4 c_3)/(4 c_6 c_4 - c_5^2)$, where $c_l, l=1...6$ are the coefficients of the polynomial surface.} 
    
\end{algorithmic}
\end{algorithm} %\vspace{-2mm}

The complete proposed procedure for obtaining \gls{AoD} estimates $\{ \hat{\phi}_n\}_{n=1}^{\hat{N}}$ and \gls{AoA} estimates $\{ \hat{\theta}_n \}_{n=1}^{\hat{N}}$ for $\hat{N}$ estimated paths, is stated and described in Algorithm \ref{alg:angle_extraction}. Besides the fundamental processing of rank-1 matrices for extracting angle information, three additional refinements are shown and applied, namely, \textit{power thresholding}, \textit{clustering}, and \textit{polynomial fitting}. With power thresholding, we omit peaks which are close to noise level, and thus mitigate noise-related estimation errors. {One feasible approach to assign a value for the related power threshold $\beta_\text{th}$ is to set it at or marginally higher than the prevailing noise floor. Concrete examples are provided along the numerical results, in Section~\ref{sec:Results}.} The clustering step, in turn, considers the fact that powerful peaks create large residuals and require more than one rank-1 singular matrix to be properly represented. Assuming spatially sparse channel, we expect such residual peaks to occur only in vicinity of powerful peaks that are spatially separated from each other, thus creating clusters in \gls{AoD}-\gls{AoA} domain. After the clustering step, {where any existing clustering method \cite{aggarwal2013data} can be adopted,} $\hat{N}$ clusters are obtained, whose cluster means represent the coarsely estimated \gls{AoD} and \gls{AoA} angles. Lastly, to enhance the angle estimation precision, each peak associated with a coarsely estimated \gls{AoD} and \gls{AoA} is fitted with a parabolic surface using a weighted least squares method. Fitting is performed within a local window of size approximately representing the beamwidth. The final \gls{AoD} estimates $\{ \hat{\phi}_n \}_{n=1}^{\hat{N}}$ and \gls{AoA} estimates $\{ \hat{\theta}_n \}_{n=1}^{\hat{N}}$ can be then found in closed form as a vertex of the parabolic surface for each peak. 

{Finally, we note that if paths share the same \gls{AoD}, but have different \glspl{AoA}, or vice versa, multiple paths can be represented by one rank-1 singular matrix. Thus, one could argue that the \gls{AoD}-\gls{AoA} extraction algorithm would benefit from multi-peak selection, however, with such measurement geometry, the weaker paths can be assumed as undesired multi-bounce paths in the considered SLAM setting.}

{\subsubsection{Complexity Assessment} The fundamental complexity order of the proposed SVD-based angle estimation method is $\mathcal{O}(\text{min}(L_\textrm{TX} L_\textrm{RX}^2,L_\textrm{TX}^2 L_\textrm{RX}) + \text{min}(L_\textrm{TX}, L_\textrm{RX})^2 K + L_\textrm{TX} L_\textrm{RX} K)$,  
excluding further refinements of thresholding, clustering and polynomial fitting. The corresponding complexities of the method in \cite{Yang2023} as well as that of a classical cell-averaging \gls{CFAR} detector \cite{Richards2005_book}, being used as benchmark methods along the numerical results, read $\mathcal{O}(\text{min}(L_\textrm{TX},L_\textrm{RX})^3 + N_{\text{peak}} L_\textrm{TX} L_\textrm{RX})$, and $\mathcal{O}(L_\textrm{TX} L_\textrm{RX} N_{\text{TB}}^2)$, respectively, where $N_{\text{peak}}$ is a maximum number of support squares \cite{Yang2023} and $N_{\text{TB}}$ is a size of the \gls{CFAR} training window. For good performance, $N_{\text{TB}}^2$ is comparable to or larger than $L$ (i.e., $L_\textrm{TX}$ or $L_\textrm{RX}$), and thus all three methods have similar complexity order of $\mathcal{O}(L^3)$.}

\vspace{-5mm}
\subsection{ToA Estimation} \label{sec:toa_estimation}
In general, the \gls{SS} burst \cite{3GPPTS38211,3GPPTS38213} allows to establish the basic frame and OFDM symbol synchronization, however, \gls{PRS} allows for larger bandwidth and thus facilitates more accurate symbol time estimation and particularly the fine \gls{ToA} estimation. We thus next shortly address the \gls{ToA} estimation, building on \gls{PRS} and the corresponding \gls{PRS} IDs \cite{3GPPTS38211,3GPPTS38213} -- both known at the \gls{UE}. The coarse \gls{ToA} estimation is carried out using time-domain I/Q signals, already before the angle estimation phase, while the fine \gls{ToA} estimation or refinement is carried out in frequency-domain only for the identified path angles. These together provide the estimates for the pathwise \glspl{ToA}.

\subsubsection{Coarse ToA Estimation in Time Domain}
First, sample-wise time delay estimation and beam ID search is performed by maximising the cross-correlation between the received waveform and potential reference waveforms in time domain. For sampling rate of $F_s$, we denote the $q^{\text{th}}$ time domain signal sample for the received signal and the PRS with PRS ID $\nu$ as $Y^{i,j}(q)$ and $X^{i,j}_\nu(q)$, respectively. Now, the estimated PRS ID $\hat{\nu}_{i,j}$ and sample-wise delay $\widehat{\Delta q}_{i,j}$ can be obtained as
\begin{equation}
    (\hat{\nu}_{i,j},\widehat{\Delta q}_{i,j}) = \argmax_{\nu,\Delta q}\sum_{q=0}^{N_\text{s}-1}|Y^{i,j}[q+\Delta q] X^{i,j}_\nu[q]^*|^2.
    \label{eq:ToA_coarse}
\end{equation}
Furthermore, the coarse delay estimate reads then 
\begin{equation}
    \hat{\tau}^{i,j}_\text{c} = \widehat{\Delta q}_{i,j} / F_s,
    \label{eq:ToA_coarse2}
\end{equation}
which defines the delay between the received \gls{OFDM} symbol and the transmit time with respect to the receiver clock -- and is thus subject to bias.

\subsubsection{Fine ToA Estimation in Frequency Domain}
The fine time delay estimation is performed per estimated path, using the frequency domain samples $y_{k,m}^{i,j}$ with $(k,m) \in \mathcal{M}_{RS}$, corresponding to  
the beam pair $(i,j)$ whose beam angles are closest to the estimated \gls{AoD} and \gls{AoA}.  
Such fine \gls{ToA} estimate for path $n$, determined with respect to the beginning of the \gls{OFDM} symbol, can be obtained as \cite[Ch.~3.2]{Sand2014_book} 
\begin{equation} \label{eq:fractional_toa}
    \hat{\tau}_{\text{f},n} = \argmax_{\tau} \left\vert \sum_{(k,m) \in \mathcal{M}_{RS}} (x_{k,m}^{\hat{i}_n,\hat{j}_n})^* y_{k,m}^{\hat{i}_n,\hat{j}_n} e^{j 2 \pi k \Delta f  \tau} \right\vert,
\end{equation}
where $\hat{i}_n = \argmin_i(\vert \Phi_{i}-\hat{\phi}_n\vert)$ and $\hat{j}_n = \argmin_j(\vert \Theta_{j}-\hat{\theta}_n\vert)$ are the \gls{TX} and \gls{RX} beam indices associated with the $n^{\text{th}}$ estimated path, respectively. In practice,  
\eqref{eq:fractional_toa} can be solved using an optimization algorithm, {interpolated IFFT}, or performing a brute force search over suitable propagation delays. 

Finally, the complete {(biased) \gls{ToA} estimates}  
are obtained as
\begin{equation} \label{eq:finetoa}
    {\hat{\tau}^{b}_{n}} = \hat{\tau}^{\hat{i}_n,\hat{j}_n}_\text{c} + \hat{\tau}_{\text{f},n}
\end{equation}
for path indices $n=1,\dots,\hat{N}$, {which correspond to (biased) path distance estimates of $c\hat{\tau}^b_n$.}
As noted in \cite{Omri2019,3GPPTS38213}, 
the coarse \gls{ToA} estimate can also be determined wrt. the frame start time.  
In such case, the possible excess time between the beginning of the frame and the actual RS transmission time can be taken into account in the coarse \gls{ToA} estimate, while the path-wise fine \gls{ToA} estimates remains intact. With unsynchronized \gls{TX} and \gls{RX} clocks, the underlying bias of \gls{ToA} estimate in \eqref{eq:finetoa} can be estimated and dealt with {in} the considered \gls{SLAM} scenario, as shown in Section \ref{sec:SLAM}.

% ############################################################################
% PROPOSED SLAM METHODS
% ############################################################################
\vspace{-2mm}
\section{Proposed Snapshot SLAM Method}
\label{sec:SLAM}

Next, we describe the proposed snapshot \gls{SLAM} method, building on the previously described \gls{AoA}, \gls{AoD} and \gls{ToA} estimates. 
The fundamental problem geometry is illustrated in Fig.~\ref{fig:problem_geometry}, while in the following, for clarity, we explicitly refer to \gls{BS} and \gls{UE} as the \gls{TX} and \gls{RX} entities, respectively.

\vspace{-3mm}
\subsection{Problem Formulation}
It is assumed that the \gls{BS} position and orientation, denoted with $[\vp_\textrm{BS}^\top, \, \alpha_\textrm{BS}]^\top$, are known while the unknown \gls{UE} state is represented using the 2D position, heading and clock bias (cast in meters, {$B_\textrm{UE}=c\hspace{0.3mm}b_\textrm{UE}$}) as $\vs = [\vp_\textrm{UE}^\top, \, \alpha_\textrm{UE}, \, B_\textrm{UE}]^\top$. Furthermore, the $n$th single bounce propagation path is represented using the 2D interaction point or landmark $\vm_n = [x_n,\, y_n]^\top$, $n=2,3, ..., \hat{N}$, where $\hat{N}$ denotes the number of the available \gls{AoA}, \gls{AoD} and \gls{ToA} estimates in a given measurement location. For notational convenience, the \gls{LoS} path index -- if existing -- is $n=1$.

In addition, let $\vx_n = [\vs^\top, \, \vm_n^\top]^\top$ denote the joint state of the \gls{UE} and the $n$th landmark, $\vm = [\vm_2^\top,\ldots,\vm_{\hat{N}}^\top]^\top$ denotes the map which is the joint state of the $\hat{N}-1$ landmarks, and the unknown 
joint state of the \gls{UE} and map is $\vx = [\vs^\top, \vm^\top]^\top$. Now, the estimation problem can be defined as
\vspace{-1mm}
\begin{equation}\label{eq:slam_posterior}
    p(\vx \mid \vz) \propto p(\vx) \prod_{n=1}^{\hat{N}} p(\vz_n \mid \vx_n),
\end{equation}
where $p(\vx) = \N(\vx \mid \vmu, \vSigma)$ is the prior for $\vx$ obtained for example using external sensors or a previous estimate and $p(\vz_n \mid \vx_n)$ is the likelihood of the $n$th measurement. It is to be noted that typically snapshot \gls{SLAM} implies that no prior information exists since the primary interest is analyzing what can be done with radio signals alone \cite{shahmansoori2018,wen2021} whereas in this article, we evaluate the system performance both with and without the {\gls{UE}} prior. {No prior knowledge on the map is required, but for a comprehensive treatment of the problem, the solution is presented considering both the UE and map priors.} The measurements are defined as $\vz_n = [{c\hat{\tau}^{b}_n}, \, \hat{\phi}_n, \, \hat{\theta}_n]^\top${,} in which the delay estimates are converted to meters for convenience. An estimate for $\vx$ can be obtained by maximizing \eqref{eq:slam_posterior}, mathematically given by
\begin{equation}\label{eq:slam_estimate}
\hat{\vx} = \underset{\vx}{\arg \max} \, p(\vx \mid \vz).
\end{equation}

Since the \gls{UE} does not known whether the \gls{LoS} exists or not, \eqref{eq:slam_estimate} is solved under \gls{NLoS} only and under \gls{LoS}+\gls{NLoS}, separately. Propagation paths with distance to within one meter of the shortest path and power within $3 \text{ dB}$ to the path with maximum power are considered as candidate \gls{LoS} signals and $N_\textrm{LoS}$ denotes the number of \gls{LoS} candidates. Furthermore, for each candidate, \eqref{eq:slam_estimate} is solved with and without a prior (see Section \ref{sec:gn_initialization}). This will give $2 N_\textrm{LoS}+1$ solutions to \eqref{eq:slam_estimate} with different costs and the solution with lowest cost measured in terms of \eqref{eq:robust_rwls_objective_function} can be selected as the estimate.

\vspace{-4mm} 
\subsection{Measurement Models}
Assuming that the measurement noise is zero-mean Gaussian, which is a common assumption is bistatic SLAM \cite{kim2020,kaltiokallio2022spawc}, the likelihood function is Gaussian 
\begin{equation}\label{eq:slam_likelihood}
    p(\vz_n \mid \vx_n) = \N(\vz_n \mid \vh_n(\vx), \vR_n),
\end{equation}
with mean $\vh_n(\vx)$ and covariance $\vR_n$. Building on the geometry in Fig.~\ref{fig:problem_geometry}, the mean is given by
\begin{equation}\label{eq:slam_measurement_model}
    \vh_n(\vx) = \begin{bmatrix}
    d {-} B_\textrm{UE} \\
    \atan2(-\delta_{1,y},-\delta_{1,x}) - \alpha_\textrm{BS} \\
    \atan2(\delta_{2,y},\delta_{2,x}) - \alpha_\textrm{UE}
    \end{bmatrix}.
\end{equation}
For the \gls{LoS} path $(n=1)$, the parameters are defined as: $d = \lVert \vp_\textrm{BS} - \vp_\textrm{UE} \rVert$ and $[\delta_{1,x}, \, \delta_{1,y}]^\top = [\delta_{2,x}, \, \delta_{2,y}]^\top = \vp_\textrm{BS} - \vp_\textrm{UE}$.
Respectively for the $n$th \gls{NLoS} path, the parameters are defined as: $d = \lVert \vp_\textrm{BS} - \vm_n \rVert + \lVert \vm_n - \vp_\textrm{UE} \rVert$, $[\delta_{1,x}, \, \delta_{1,y}]^\top = \vp_\textrm{BS} - \vm_n${,} and $[\delta_{2,x}, \, \delta_{2,y}]^\top = \vm_n - \vp_\textrm{UE}$.

\vspace{-4mm}
\subsection{Regularized Robust Least Squares Estimator}
Maximizing the posterior as given in \eqref{eq:slam_estimate}, is equivalent to
\begin{equation}
    \hat{\vx} = \underset{\vx}{\arg \min} \, L(\vx),
\end{equation}
{in which $L(\vx)$ is the objective function we wish to minimize.} In this article, we utilize the following objective function
\begin{equation}\label{eq:robust_rwls_objective_function}
    L(\vx) = (\vx - \vmu)^\top \vSigma^{-1} (\vx - \vmu) + \sum_{n=1}^{\hat{N}} f(q_n(\vx)),
\end{equation}
where the first term is a regularization term that encodes the prior information and the second term encodes the evidence provided by the measurements. In the second term, $f(\cdot)$ is a robust cost function which we will define later and 
\begin{equation}\label{eq:quadratic_error}
    q_n(\vx) = (\vz_n - \vh_n(\vx))^\top \vR_n^{-1} (\vz_n - \vh_n(\vx)).
\end{equation}
defines a quadratic error.

The Gauss-Newton algorithm can be utilized to iteratively solve \eqref{eq:robust_rwls_objective_function} and the method is based on approximating $\vh_n(\vx)$ using a first order Taylor series expansion \cite{nocedal1996}, given by
\begin{equation}\label{eq:approximate_nonlinear_function}
    \vh_n(\vx) \approx \vh_n(\hat{\vx}^{(j)}) + \vH_n(\hat{\vx}^{(j)})(\vx - \hat{\vx}^{(j)}),
\end{equation}
where $\hat{\vx}^{(j)}$ is the estimate of $\vx$ at the $j$th iteration and $\vH_n(\hat{\vx}^{(j)}) = \nabla_{\vx} \vh_n(\vx) \vert_{\vx = \hat{\vx}^{(j)}}$ is the Jacobian. The parameter update of the Gauss-Newton algorithm can be derived by plugging \eqref{eq:approximate_nonlinear_function} into \eqref{eq:robust_rwls_objective_function}, setting the gradient of $L(\vx)$ to zero and solving for $\vx$, expressed as 
\begin{align}\label{eq:robust_rwls_objective_function_minimum}
    \frac{\partial L(\vx)}{\partial \vx} &= \frac{\partial}{\partial \vx}  (\vx - \vmu)^\top \vSigma^{-1} (\vx - \vmu) +  \sum_{n=1}^{\hat{N}} \frac{\partial f}{\partial q_n} \frac{\partial q_n}{\partial \vx}   \\
    & \approx 2 \vSigma^{-1} (\vx - \vmu) -2 \sum_{n=1}^{\hat{N}} \vH_n^\top(\hat{\vx}^{(j)}) \tilde{\vR}_n(\hat{\vx}^{(j)})^{-1} \nonumber \\
    & \quad \times \bigl(\vz_n - \vh_n(\hat{\vx}^{(j)}) - \vH_n(\hat{\vx}^{(j)})(\vx - \hat{\vx}^{(j)})\bigr) = 0, \nonumber
\end{align}
in which 
\begin{equation}
    \tilde{\vR}_n(\hat{\vx}^{(j)})^{-1} = \frac{\partial f}{\partial q_n} \biggr{|}_{q_n = q_n(\hat{\vx}^{(j)})} \vR_n^{-1}.
\end{equation}
{In order to minimize $L(\vx)$, w}e then set the next estimate $\hat{\vx}^{(j+1)}$ to be equal to the minimum, which gives
\begin{align}\label{eq:robust_rwls_update}
    \hat{\vx}^{(j+1)} &= \hat{\vx}^{(j)} + \vA^{-1}\vb, \quad \text{where} \\
    \vA &= \vSigma^{-1} + \sum_{n=1}^{\hat{N}} \vH_n^\top(\hat{\vx}^{(j)}) \tilde{\vR}_n(\hat{\vx}^{(j)})^{-1} \vH_n(\hat{\vx}^{(j)}), \nonumber\\
    \vb &= \vSigma^{-1} (\vmu - \hat{\vx}^{(j)} ) \nonumber \\
    & \quad + \sum_{n=1}^{\hat{N}} \vH_n^\top(\hat{\vx}^{(j)}) \tilde{\vR}_n(\hat{\vx}^{(j)})^{-1} (\vz_n - \vh_n(\hat{\vx}^{(j)})). \nonumber
\end{align}

In general, there are many possible robust cost functions that reduce the weight of components with large errors so that they have a smaller influence to the solution due to a reduced gradient \cite{mactavish2015}. In this article, we utilize the Cauchy cost function, $f(q_n(\vx)) = \log (1 + q_n(\vx))$, such that the second term of the objective function in \eqref{eq:robust_rwls_objective_function} becomes
 \begin{equation}
     \sum_{n=1}^{\hat{N}} \log \left( 1 \mkern-4mu + \mkern-2mu (\vz_n - \vh_n(\vx))^\top \vR_n^{-1} (\vz_n - \vh_n(\vx)) \right) \mkern-4mu,
 \end{equation}
and the gradient is given by \eqref{eq:robust_rwls_objective_function_minimum} in which
\begin{equation}
    \tilde{\vR}_n(\hat{\vx}^{(j)})^{-1} = \frac{1}{1 + q_n(\hat{\vx}^{(j)})} \vR_n^{-1}.
 \end{equation}
Thus, the new covariance matrix $\tilde{\vR}_n$ is just an inflated version of the original covariance matrix $\vR_n$, given by 
\begin{equation}\label{eq:inflated_covariance_matrix}
    \tilde{\vR}_n(\hat{\vx}^{(j)}) \mkern-4mu = \mkern-4mu \left( 1 + (\vz_n \mkern-4mu - \mkern-2mu \vh_n(\hat{\vx}^{(j)}))^\top \vR_n^{-1} (\vz_n \mkern-4mu - \mkern-2mu \vh_n(\hat{\vx}^{(j)})) \right) \vR_n
\end{equation}
and it gets bigger as the quadratic error increases. As a consequence, cost terms that are very large and potential outliers are given less trust to diminish their impact. % to the solution is diminished.

In practice, taking a full step according to \eqref{eq:robust_rwls_update} might be too large with respect to the neighborhood for which the Taylor series approximation in \eqref{eq:approximate_nonlinear_function} is valid. To avoid this, a scaled Gauss–Newton step can be done instead, proportional to the direction given by the local approximation. In this article, we use backtracking line search \cite{boyd_vandenberghe_2004} to compute the step length. The resulting algorithm is summarized in Algorithm \ref{alg:gauss_newton_with_line_search} {and initialization of $\hat{\vx}^{(0)}$ is presented in the following section.}

\begin{algorithm}[t]
\small
\caption{Proposed Gauss-Newton algorithm}
\label{alg:gauss_newton_with_line_search}
\algorithmicrequire{Initial parameter guess $\hat{\vx}^{(0)}$ and measurements $\vz$} \\
\algorithmicensure{Parameter estimate $\hat{\vx}$ and covariance $\vP$}
\begin{algorithmic}[1]
    \State Set $j \leftarrow 0$
    \Repeat
    \State \textit{Calculate the update direction as given in \eqref{eq:robust_rwls_update}:}
    \Statex[2] $\Delta\hat{\vx}^{(j+1)} = \vA^{-1}\vb$
    %\Statex % Add vertical space
    \vspace{1mm}

    \State \textit{Compute step length $\gamma$ using Algorithm 9.2 in} \cite{boyd_vandenberghe_2004}
    
    %\Statex
    \vspace{1mm}
    \State \textit{Update parameter estimate:}
    \Statex[2] $\hat{\vx}^{(j+1)} = \hat{\vx}^{(j)} + \gamma \Delta\hat{\vx}^{(j+1)}$ 

    \State Set $j \leftarrow j + 1$

    \Until{Converged}
    \State $\hat{\vx} = \hat{\vx}^{(j)}$
    \State $\vP = (\vSigma^{-1} + \sum_{n=1}^{\hat{N}} \vH_n^\top(\hat{\vx}^{(j)}) \tilde{\vR}_n(\hat{\vx}^{(j)})^{-1} \vH_n(\hat{\vx}^{(j)}))^{-1}$ 
    
\end{algorithmic}
\end{algorithm}

\vspace{-4mm}
\subsection{Initialization}\label{sec:gn_initialization}
A major disadvantage of the Gauss–Newton approach is that the linearization of the measurement model is local. In highly nonlinear problems, this means that the solution can converge to a local minima and therefore, initialization of the algorithm is very important. To this end, let us define the prior mean and inverse of the covariance matrix as
\begin{align}
    \vSigma^{-1} &= \textrm{blkdiag}(\vSigma_{\vs\vs}^{-1},\vSigma_{\vm_2\vm_2}^{-1},\ldots,\vSigma_{\vm_{\hat{N}}\vm_{\hat{N}}}^{-1}), \\
    \vmu &= [\vmu_\vs^\top, \, \vmu_{\vm_2}^\top, \, \ldots, \, \vmu_{\vm_{\hat{N}}}^\top]^\top,
\end{align}
where $\vmu_{\vs}$, $\vSigma_{\vs\vs}$, $\vmu_{\vm_n}$ and $\vSigma_{\vm_n\vm_n}$ denote the mean and covariance of the \gls{UE} and $n$th map entry, respectively. {Algorithm \ref{alg:gauss_newton_with_line_search} is initialized using the prior mean, that is, $\hat{\vx}^{(0)} = \vmu$.} 

{Snapshot \gls{SLAM} typically implies that no prior information exists $(\vSigma^{-1} = \mathbf{0})$ \cite{shahmansoori2018,wen2021}, whereas in this article, we evaluate the system performance both with and without the prior. If prior information is not available, the \gls{LoS} must exist so that we can first compute a prior for the \gls{UE} state as described in Section \ref{sec:ue_los_initialization} which can then be used to initialize the landmarks as presented in \ref{sec:landmark_initialization}. If a prior for the \gls{UE} exists, the landmarks can be directly initialized using the prior as described in Section \ref{sec:landmark_initialization}. If a prior for the \gls{UE} and map exists, Algorithm \ref{alg:gauss_newton_with_line_search} can be directly initialized using the prior but in this article it is always assumed that no prior knowledge of the map is available, that is, $\vSigma_{\vm_n\vm_n}^{-1} = \mathbf{0}, \, n = 2,\ldots,\hat{N}$.}

\subsubsection{UE Initialization Without Prior Information}\label{sec:ue_los_initialization}
{The challenge in initializing the UE state using the \gls{LoS} measurement is the unknown clock bias $B_\textrm{UE}$. One can consider different trial values of $B_\textrm{UE}$ over a range of $[{B}_\textrm{UE,min}, \, {B}_\textrm{UE,max}]$\footnote{{We can define the following inequality, $d_\textrm{min} \leq c \hat{\tau}^b_1 - B_{UE} \leq d_\textrm{max}$, in which $d_\textrm{min}$ and $d_\textrm{max}$ define the minimum and maximum propagation distances of the \gls{LoS} path. From the inequality, the bias range can be computed as $B_\textrm{UE,min} = c \hat{\tau}^b_1 - d_\textrm{max}$ and $B_\textrm{UE,max} = c \hat{\tau}^b_1 - d_\textrm{min}$.}} and for each trial value $\hat{B}_\textrm{UE}$, we form an augmented measurement $\check{\vz} = [\hat{\tau}^b_1, \, \hat{\phi}_1, \, \hat{\theta}_1,\, \hat{B}_\textrm{UE}]^\top $ and covariance $\check{\vR}_1 = \textrm{blkdiag}(\vR_1,\sigma_{\hat{B}_\textrm{UE}}^2)$, in which $\sigma_{\hat{B}_\textrm{UE}}^2$ is variance of the bias which is set higher than variance of the delay estimates. Then, the mean and covariance of the UE prior are given by
\begin{equation}\label{eq:ue_moments}
    \vmu_\vs = \check{\vh}_1(\check{\vz})  \quad \text{and} \quad
    \vSigma_{\vs\vs} = \check{\vH}_1(\check{\vz}) \check{\vR}_1 \check{\vH}_1(\check{\vz})^\top
\end{equation}
where $\check{\vH}_1(\check{\vz})$ denotes the Jacobian of $\check{\vh}_1(\check{\vz})$ evaluated with respect to $\check{\vz}$ and the mean is defined as
\begin{equation}\nonumber
\check{\vh}_1(\tilde{\vz}) = [
    x_\textrm{BS} + \hat{x}_1, \,
    y_\textrm{BS} + \hat{y}_1, \,
    \atan2(-\hat{y}_1, -\hat{x}_1) - \hat{\theta}_1, \,
    \hat{B}_\textrm{UE}
    ]^\top
\end{equation}
in which $\hat{x}_1 = (c \hat{\tau}^b_1 - \hat{B}_\textrm{UE})\cos(\alpha_\textrm{BS} + \hat{\phi}_1)$ and $\hat{y}_1 = (c \hat{\tau}^b_1 - \hat{B}_\textrm{UE})\sin(\alpha_\textrm{BS} + \hat{\phi}_1)$.  After computing the moments using \eqref{eq:ue_moments}, the landmark locations can be computed as presented in Section \ref{sec:landmark_initialization}. Then the trial with lowest cost according to \eqref{eq:robust_rwls_objective_function}, which also involves the landmarks, is selected as the \gls{UE} prior. Due to computational reasons, we solve the described problem using  constrained nonlinear optimization \cite{boyd_vandenberghe_2004} for which the problem can be defined as
\begin{equation}\label{eq:initial_bias_estimate}
    \hat{B}_\textrm{UE} = \underset{B_\textrm{UE,min} \leq B_\textrm{UE} \leq B_\textrm{UE,max}}{\arg \min} \, L(\vx), 
\end{equation} 
with cost $L(\vx)$ as given in \eqref{eq:robust_rwls_objective_function}.}

\subsubsection{Landmark Initialization}\label{sec:landmark_initialization}
The {first order Taylor series based linear approximation of the measurement likelihood} in \eqref{eq:slam_likelihood} around the \gls{UE} prior reads
\begin{equation}
    p(\vz_n \mid \vx_n) \approx \N(\vz_n; \vh_n(\vx) + \vH_n(\vmu_\vs)(\vs-\vmu_\vs),\vR_n),
\end{equation}
where {$\vh_n(\vx)$ is evaluated at $\vx = [\vmu_\vs^\top, \vm_n^\top]^\top$, and} $\vH_n(\vmu_\vs) = \nabla_{\vs} \vh_n(\vx) \vert_{\vs = \vmu_\vs}$ is the Jacobian of $\vh_n(\vx)$ with respect to state $\vs$ computed at $\vmu_\vs$. {Following the derivations in \cite[Eqs. (5.6) - (5.13)]{sarkka2013}}, the likelihood can be approximated as
\begin{equation}\label{eq:affine_likelihood_approximation}
     p(\vz_n \mid \vx_n)  \approx \N(\vz_n; \vh_n(\vx),\vH_n(\vmu_\vs) \vSigma_{\vs\vs} \vH_n^\top(\vmu_\vs) + \vR_n).
\end{equation}
The $n$th map element is initialized by solving a nonlinear optimization problem, defined as
\begin{equation}\label{eq:optimization_problem}
    \vmu_{\vm_n} = \underset{\vm_n}{\textrm{arg\,min}} \, (\vz_n - \vh_n(\vx))^\top \vW_n^{-1} (\vz_n - \vh_n(\vx)),
\end{equation}
where $\vW_n = \vH_n(\vmu_\vs) \vSigma_{\vs\vs} \vH_n^\top(\vmu_\vs) + \vR_n$. The optimization problem is initialized in a similar way as described in \cite{kim2020} and solved using the Gauss-Newton algorithm, with the exact details here omitted, as the algorithm is very similar to the one presented in Algorithm~\ref{alg:gauss_newton_with_line_search}.

%############################################################################
% ENVIRONMENT
% ############################################################################
\vspace{-1mm}
\section{Indoor Environment, Tools and Data}
\label{sec:environments}
\begin{figure*}[!t]
\centering
\vspace{-3mm}
\subfloat[Ray-tracing environment]
{\includegraphics[width=6.1cm]{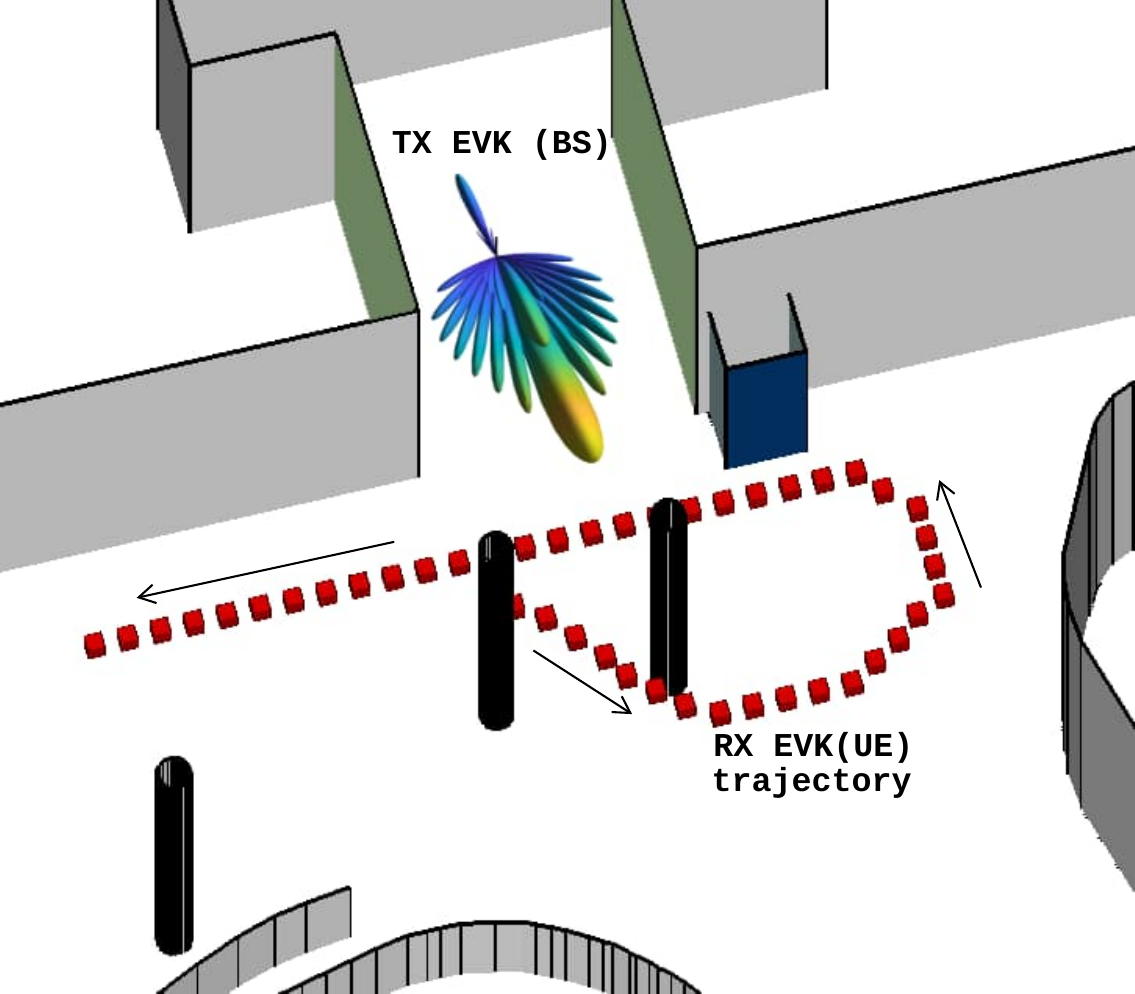}%
}
%\hspace{10mm}
\hspace{6mm}
\hfil
\vspace{1mm}
\subfloat[Photograph of the physical measurement environment]{\includegraphics[width=3.75in]{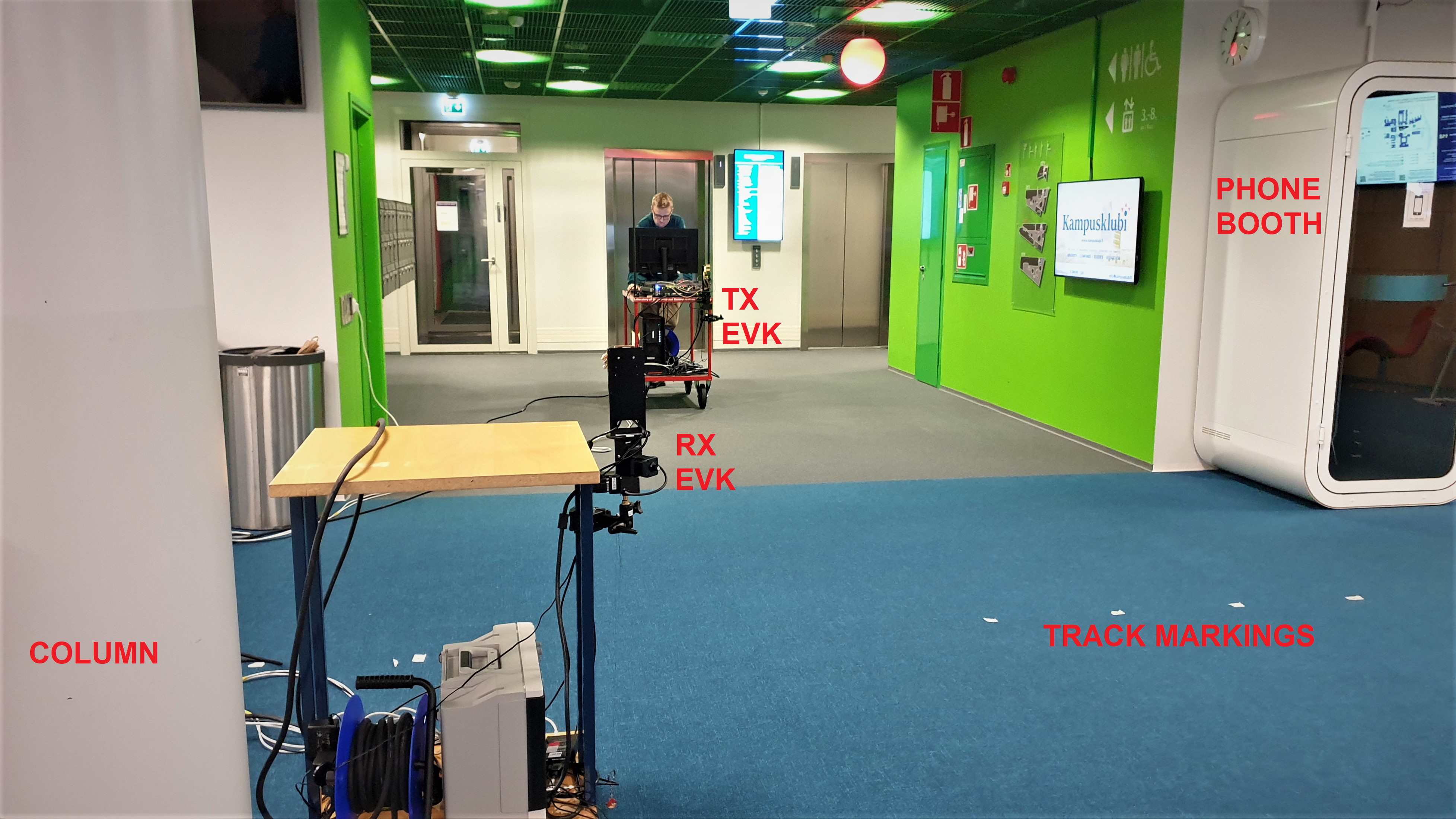}%
}
\vspace{-1mm}
\caption{Illustrations of the considered indoor environment. (a) Ray-tracing model visualization showing the TX/BS position and the RX/UE trajectory; (b) Photograph from the physical premises with the TX and RX trolleys also shown during the measurement campaign. 
}
\vspace{-0mm}
\label{fig:environment}
\end{figure*}

We consider a modern indoor environment at the Hervanta Campus of Tampere University, Finland, located in the so-called Campus Arena building. The environment is illustrated in terms of a floor plan and an actual photograph in {Fig.~\ref{fig:environment}}, consisting of a fairly large partially open space containing a number of different landmarks such as columns, short walls, booths, and so forth. The BS is located in a narrow annex $3$m wide while the UE moves in the area with a trajectory shown in {Fig.~\ref{fig:environment}}. 
Furthermore, $4\times16$ planar antenna arrays are considered at both the transmitting and receiving ends, {with azimuth 3\,dB beamwidth of around 10$^\circ$}. Such antenna system assumption is certainly implementation feasible at BS end while for UEs the current mmWave implementations consider somewhat reduced antenna counts and thus broader beams. 
The considered \gls{PRS} bandwidth is 400\,MHz following \cite{3GPPTS38211}.

\vspace{-3mm}
\subsection{Ray Tracing Tool and Assumptions}
To carry out evaluations with well-defined and known ground-truth, ray tracing was performed with Wireless InSite\textsuperscript{\textregistered} \cite{InSite}. The true physical environment is reproduced using the indoor floor plan editor fitted to the building blueprint, while noting also accurately different movable entities and objects such as an indoor phone booth. The ITU $60$~GHz material models {\cite{ITU_materials}}, namely, layered drywall, wood, glass, floor and ceiling board were used.   
The \gls{BS} position and \gls{UE} trajectory consisting of $45$ points $0.5$m apart are accurately matched to those used in the actual measurements. 
Omnidirectional antennas were used for the ray casting, and the reception with path elevation was limited to $80^{\circ}-100^{\circ}$, { reflecting essentially the azimuth plane.}  
Furthermore, ray casting was limited to $25$ rays per \gls{UE} position, and the number of material interactions to four reflections and one diffraction.  
The ray-tracing model considers \gls{SR} and \gls{D}, such that an unambiguous ground truth for the performance assessment of the proposed and benchmark angle estimation methods can be obtained. 

The ray-tracing model is further combined with I/Q waveform processing in Matlab, such that realistic received I/Q samples can be generated, per UE location. Here, the \gls{TX} and \gls{RX} beampatterns are modelled through classical matched filter type responses where the beamforming weights are matched to the corresponding steering vector of the beamforming angle. 
Additionally, we allow for similar mechanical rotation patterns as in the actual measurements in order to cover $180^{\circ}$ \gls{FOV} in TX and $360^{\circ}$ \gls{FOV} in RX (for further details, see the following subsection). These together facilitate a maximum of $L_\textrm{TX}$ = 126 and $ L_\textrm{RX}$ = 252 beams at TX and RX, respectively.
Moreover, we model accurately the \gls{SNR} characteristics of the environment such that the prevailing \gls{SNR} at each \gls{UE}/\gls{RX} point is adjusted according to the actual \gls{SNR} observed in the corresponding physical RF measurements.

\vspace{-3mm}
\subsection{Measurement Setup and Data}
In the actual mmWave measurements, Sivers Semiconductors Evaluation Kits EVK06002 \cite{EVKinfo} were used as the \gls{TX} and \gls{RX} entities. The overall operational band of EVK06002 is $57-71$~GHz, and it consists of {strip antenna elements} arranged in $4\times16$ arrays for each polarization, integrated with the electrical phase and amplitude control of the individual elements.  
The TX EVK was connected to a PC-controlled  Arbitrary Waveform Generator M8195A, whereas the RX EVK was connected to the Keysight DSOS804A oscilloscope, which serve as data conversion interfaces towards digital signal processing.  
Similar to the ray-tracing case, the transmitted \gls{PRS}-carrying I/Q waveforms were created using the Matlab 5G Toolbox including also embedding of different \gls{PRS} IDs.

The beamforming capabilities of the  Sivers EVK allows for synthesizing electrically controlled beams, building on embedded proprietary codebook.  
{The exact beampatterns are unknown but based on elementary antenna measurements resemble those of the ray tracing model complemented with additional tapering.}  
Since the electrical beamforming \gls{FOV} of the EVK is limited to $[-45^{\circ},45^{\circ}]$, both \gls{TX} and \gls{RX} EVK were complemented with FLIR Pan Tilt PTU-46 for additional mechanical rotation capabilities, in order to cover $180^{\circ}$ and $360^{\circ}$ \glspl{FOV} with maximum of $L_\textrm{TX}$ = 126 and $ L_\textrm{RX}$ = 252 TX and RX beams, respectively. 
Both receiver and transmitter were controlled by the same PC and synchronized to the same clock via a coaxial cable to provide a reference for \gls{ToA} measurements.
Additionally, the true UE locations were recorded accurately for ground-truth purposes.
The measurement setup and a glimpse of the physical environment are shown along the {Fig.~\ref{fig:environment}.}

{The complete 60\,GHz I/Q measurement data set as well as the corresponding processed channel parameter data set, together with supportive scripts are all openly shared along this article -- refer to the first page footnote for further details. 
}

%############################################################################
% RESULTS
% ############################################################################
\vspace{-1mm}
\section{Results}
\label{sec:Results}

In this section, the angle estimation results as well as the complete end-to-end SLAM results are presented and analyzed.  The angle estimator assessment builds on the ray tracing data, as it contains the ground-truth information also for the landmarks. The end-to-end SLAM assessment, in turn, builds on the true measurement data.

\vspace{-3mm}
\subsection{Angle Extraction Results with Ray Tracing Data}
The performance of the proposed SVD-based method is assessed and compared to the available successive cancellation-based reference method described in \cite{Yang2023}. Additionally, the classical \gls{CFAR} detector \cite{Richards2005_book} is also implemented and considered as an additional benchmark solution, as such is widely used, e.g., in radar context for target detection subject to clutter.

\begin{figure*}[t]
    \centering 
\vspace{-2mm}
\includegraphics[width=5.65cm,trim={0.4cm 1.9cm 2.7cm 2.65cm},clip]{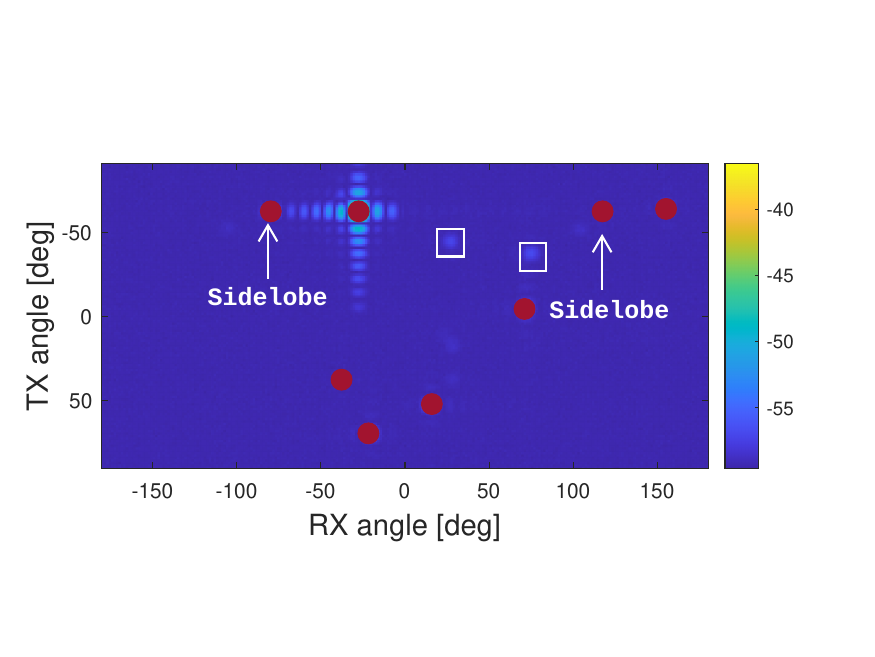}
  \hfil
\includegraphics[width=5.25cm,trim={1.2cm 1.9cm 2.7cm 2.65cm},clip]{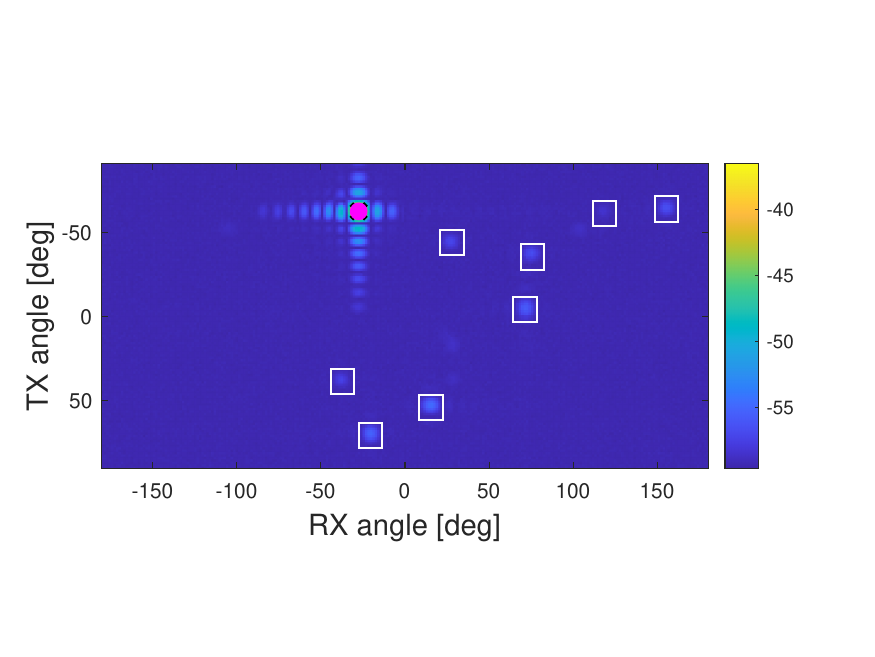}
  \hfil
\includegraphics[width=6.15cm,trim={1.2cm 1.9cm 1.0cm 2.65cm},clip]{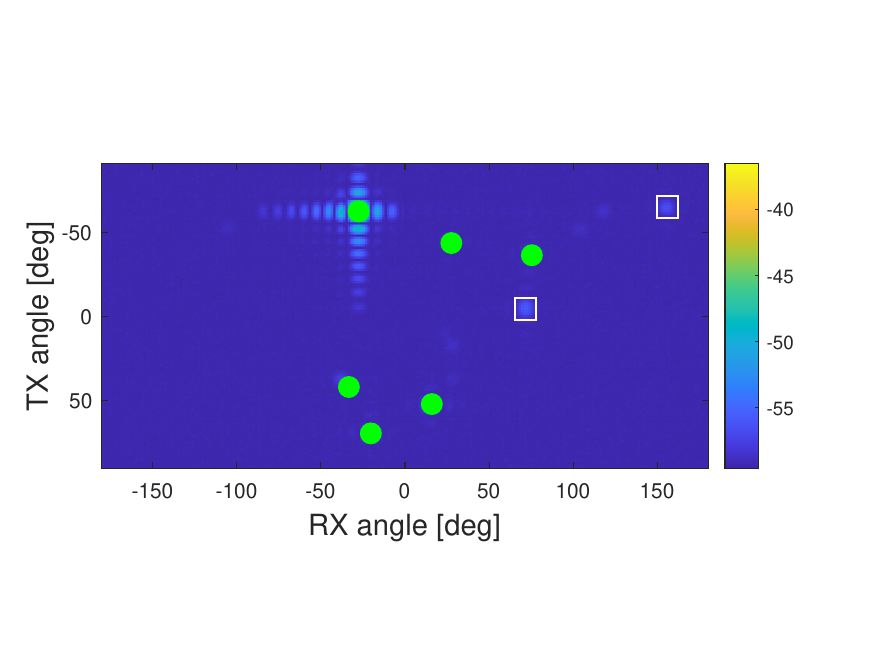}

\vspace{-2mm}
\subfloat[a][Cancellation-based method \cite{Yang2023}]{\includegraphics[width=5.8cm,trim={0.3cm 1.9cm 2.7cm 2.65cm},clip]{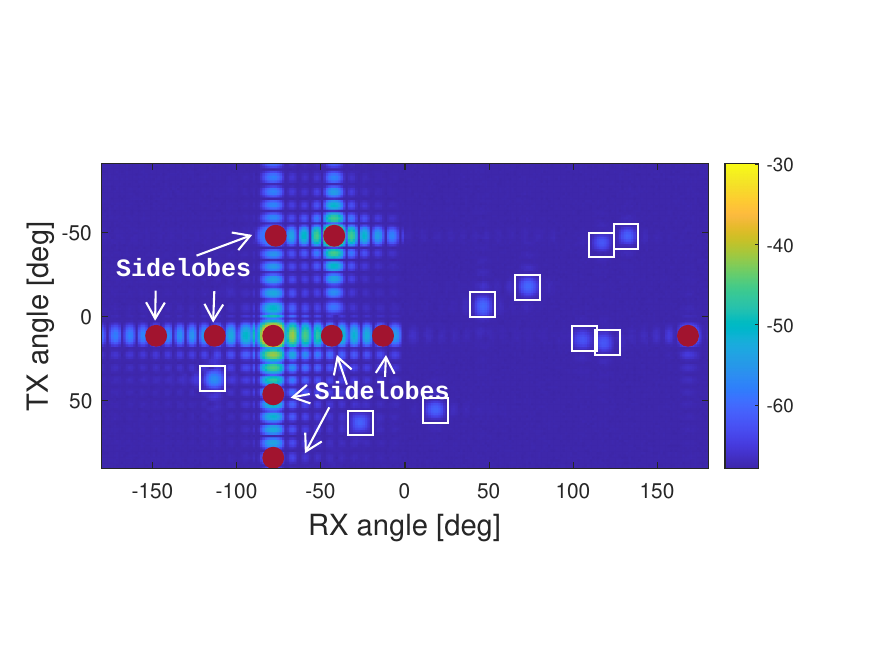}
}
  \hfil
\subfloat[b][CFAR \cite{Richards2005_book}]{\includegraphics[width=5.4cm,trim={1.0cm 1.9cm 2.7cm 2.65cm},clip]{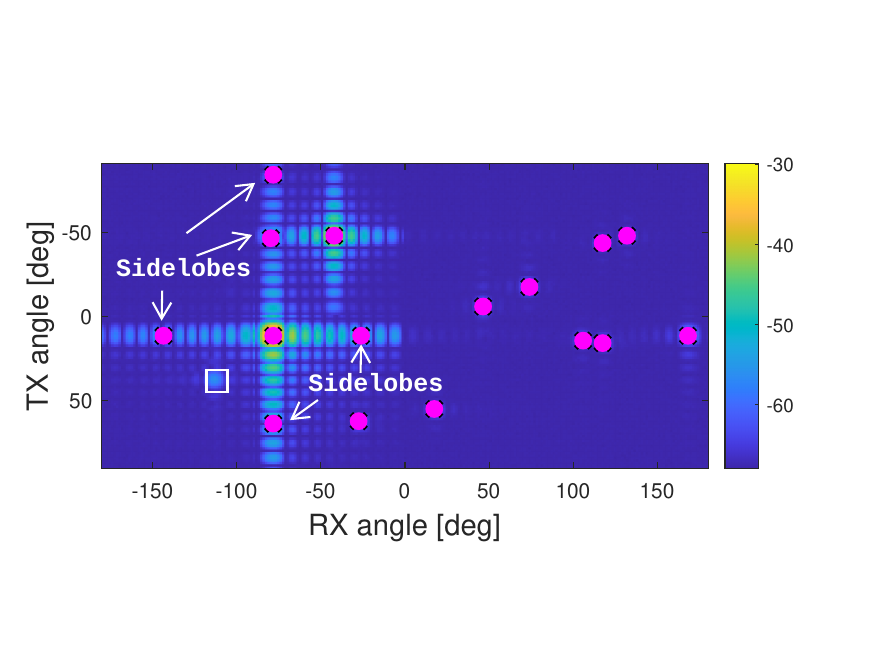}
}
  \hfil
\subfloat[c][{Proposed SVD-based method}]{\includegraphics[width=6.3cm,trim={1.0cm 1.9cm 1.0cm 2.65cm},clip]{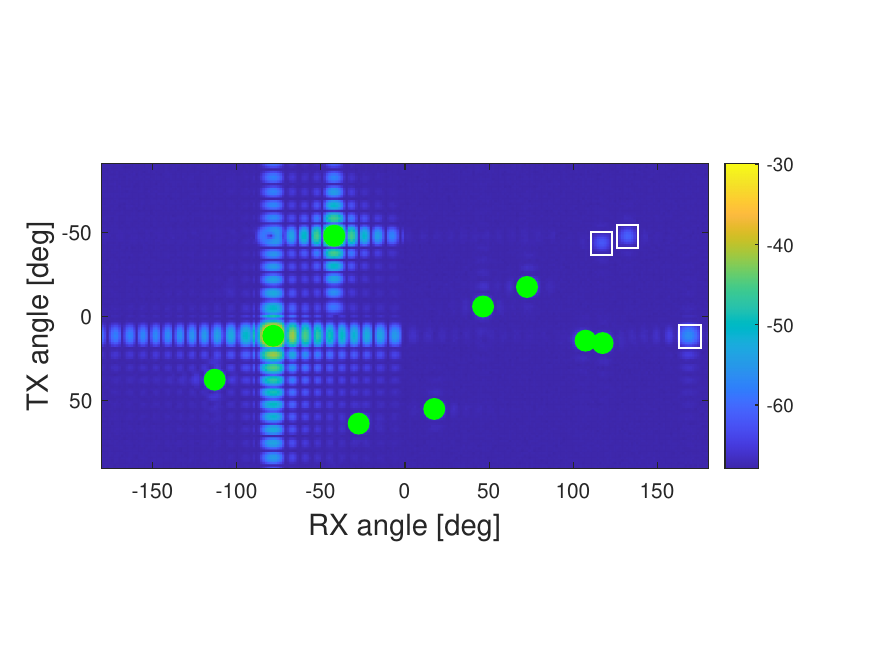}
}

  \label{fig:NLoS}
\vspace{-0mm}
\caption{Illustration of angle extraction methods in NLoS (upper row) and LoS (lower row) UE locations. (a) Cancellation method \cite{Yang2023} for $N_\text{peaks}=10$,  $\epsilon = 0.05$ (NLoS) and $0.1$ (LoS); (b) CFAR \cite{Richards2005_book} for $N_\text{TB}=15$ and $P_\text{FA} = 0.002$ (NLoS) and $0.12$ (LoS); (c) Proposed SVD method for $p$ of $99.9\%$ (NLoS) and $99.99\%$ (LoS){, and with $\beta_\text{th}$ $10\%$ above the noise floor}. Extracted path angles are marked with colored dots, side-lobe false detections with arrows, and missed detections with squares.}
\vspace{-1mm}
\label{fig:images}
\end{figure*}

The method in \cite{Yang2023} 
is parameterised with the so-called support power increment, $\epsilon$, and the maximum number of searched peaks, $N_\text{peaks}$. We additionally apply power thresholding to reduce the amount of false detections, and thereon to have as fair comparison as possible.
The \gls{CFAR} technique, in turn, uses training cells of size $N_\text{TB}$ around the target cell to estimate the local noise level to reach the given false alarm probability $P_\text{FA}$. The Matlab 2D CFAR detector implementation from phased array system toolbox was used, while further clustering was also applied to the results as \gls{CFAR} method yields easily multiple AoD-AoA estimates per one peak. 
{Finally, the proposed SVD-based method is parametrized as follows. For the \emph{power ratio} parameter $p$, we consider the values of $p=99\%, 99.9\%$ and $99.99\%$. The \emph{power thresholding} related parameter $\beta_\text{th}$ is, in turn, set at $10\%$ above the prevailing noise floor. This value is designated with notation $\beta_\text{th}^*$ while additional complementary results where the value of $\beta_\text{th}$ is varied are also provided.}

\subsubsection{Qualitative Comparison}
{The capabilities of the different methods are first visually illustrated in Fig.~\ref{fig:images}, covering both \gls{NLoS} and \gls{LoS} \gls{UE} locations, and with the parametrizations as shown in the caption. As can be observed, the method from \cite{Yang2023} has notable challenges in dealing with antenna sidelobes, while also several true propagation paths are missed especially in \gls{LoS} case. The CFAR has also clear limitations with sidelobes while missing a large number of true landmarks in \gls{NLoS}. The proposed method, in turn, is able to offer enhanced performance in both \gls{NLoS} and \gls{LoS}, while processing efficiently also the antenna sidelobes.}

\begin{figure*}[h!]
    \centering 

\subfloat[a][{ GOSPA values at different measurement locations }]{\includegraphics[width=5.8cm,trim={0.6cm 1.4cm 1.5cm 2.0cm},clip]{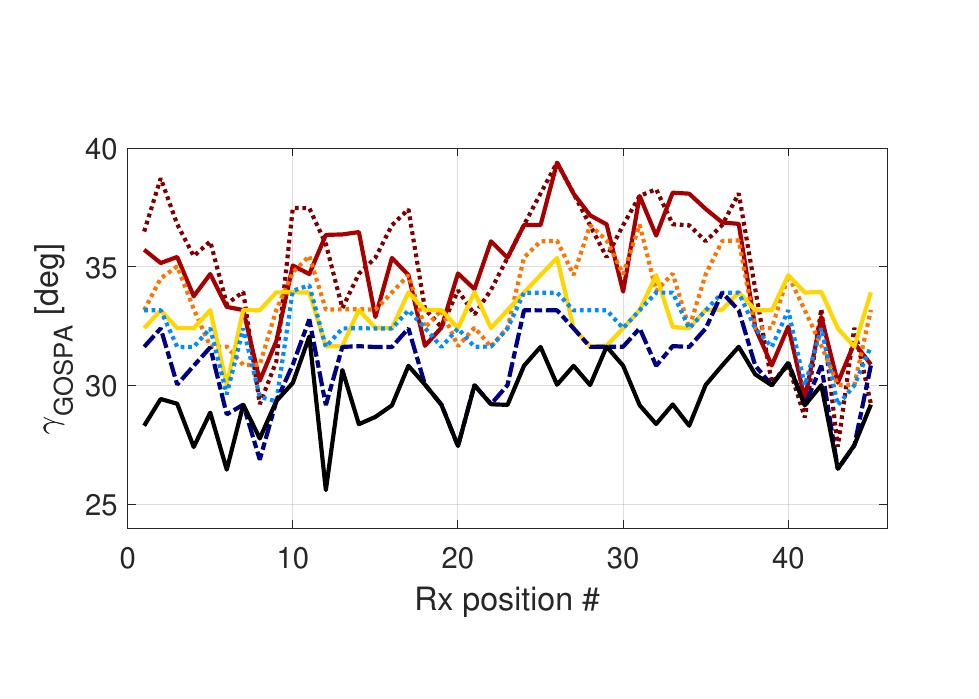}
} %\label{subfig:allgospa}
  \hfil
\subfloat[b][{ Averaged GOSPA values across all locations }]{\includegraphics[width=5.9cm,trim={0.6cm 2cm 1.5cm 2.4cm},clip]{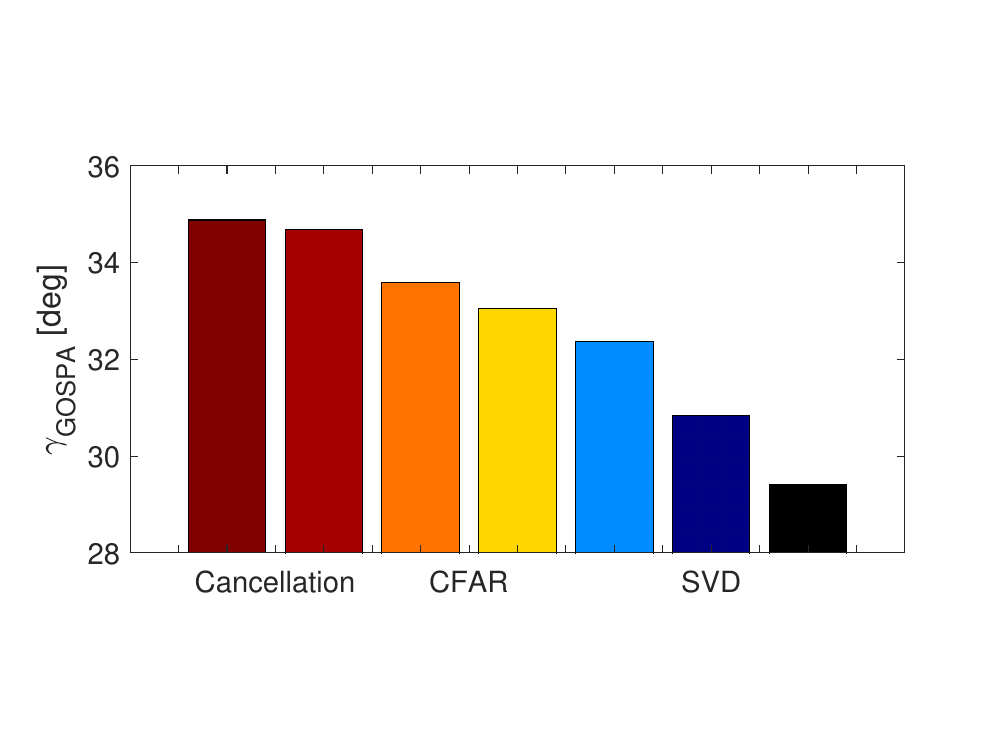}
} %\label{subfig:avergospa}
  \hfil
\subfloat[c][{ Average sidelobe detections across all locations }]{\includegraphics[width=5.9cm,trim={0.6cm 2cm 1.5cm 2.4cmm},clip]{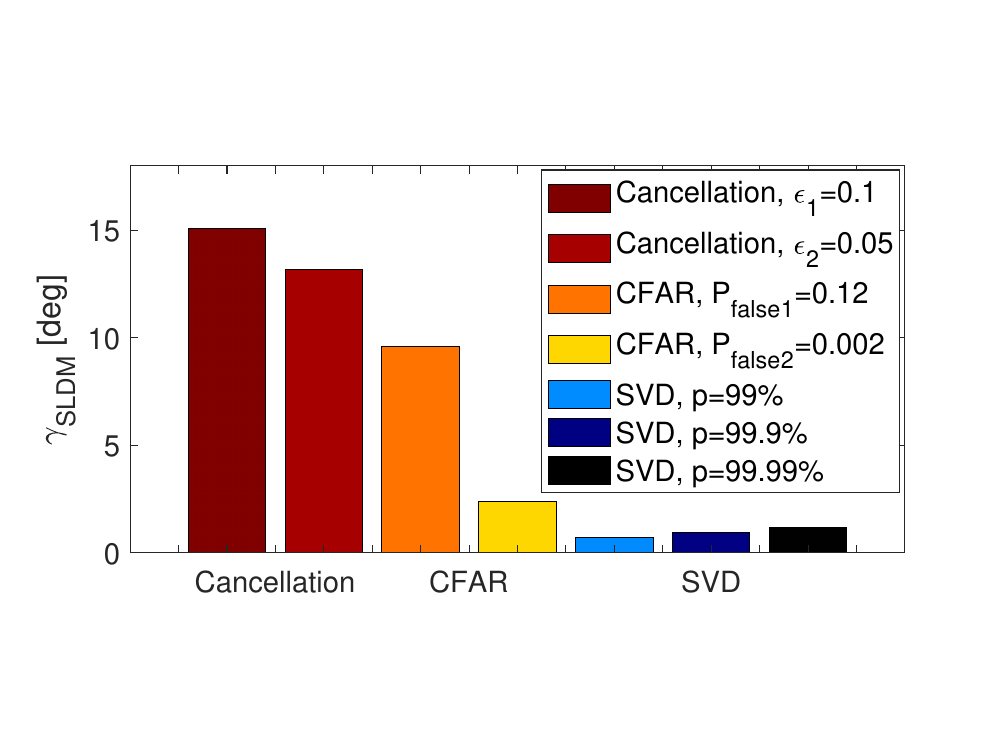}
} %\label{subfig:aversd}
\vspace{-3mm}
\caption{Quantitative assessment of different angle extraction methods. (a) GOSPA metric at different UE/RX locations for  cancellation-based method \cite{Yang2023} with $\epsilon=0.1$ (\protect\suponeline) and $\epsilon=0.05$ (\protect\suptwoline), for CFAR method \cite{Richards2005_book} with $P_\text{FA}=0.12$ (\protect\cfaroneline) and $P_\text{FA}=0.002$ (\protect\cfartwoline), and for the proposed SVD-based method with {$p=99\%$,(\protect\svdoneline) $p=99.9\%$ (\protect\svdtwoline), and $p=99.99\%$ (\protect\svdthreeline), and with $\beta_\text{th}$ $10\%$ above the noise floor}. (b) Corresponding average GOSPA metrics for all methods. (c) Average sidelobe detection metric for all methods. }
\label{fig:allmetric}
\end{figure*}

\subsubsection{Quantitative Comparison}\label{sec:channel_estimation_quantitative_comparison}
As illustrated in Fig.~\ref{fig:images}, the amounts of the detected paths differ from method to another, while the ray-tracing model limits the amount of ground-truth paths to the earlier noted number of 25.  
Such data of different cardinality can be reliably compared and quantitatively assessed using the \gls{GOSPA} metric \cite{Rahmathullah2017}, which takes into account, in addition to \glspl{RMSE} of the quantities of interest, the numbers of false detection, $N_\text{FD}$, and missed detection, $N_\text{MD}$. We thus utilize the \gls{GOSPA} metric for quantitative assessment, and express it in degrees as 
$\gamma_\text{GOSPA} = (\gamma_\text{RMSE} + (N_\text{FD} + N_\text{MD})d_c^{\mathcal{P}} / \alpha)^{(1/\mathcal{P})}$, where the parameters $d_c = 10^\circ$, $\mathcal{P} = 2$ and $\alpha = 2$ are cutoff distance, exponent power and cardinality penalty factor, respectively.

Furthermore, a subset of all false detections are due to the \emph{sidelobes}. 
Such sidelobe effects, unlike those of a noise, cannot be reduced by increasing the number of observations. Thus, we consider \gls{SLFD} a systematic error that is especially detrimental to the performance of the whole end-to-end system. 
To this end, we introduce an additional metric that quantifies the number of \glspl{SLFD}, $N_\text{SLFD}$. {We specifically express the \gls{SLFD} through the following conditions of (\emph{i}) the ToA estimates are within a threshold of $|\hat{\tau}_{n_{1}}-\hat{\tau}_{n_{2}}|\leq \tau_\text{th}$, and (\emph{ii}) the corresponding TX or RX beam indices differ at most by one, i.e., $(|\hat{i}_{n_{1}}-\hat{i}_{n_{2}}| \leq 1) \lor (|\hat{j}_{n_{1}}-\hat{j}_{n_{2}}| \leq 1)$. A tight delay threshold of $\tau_\text{th} = 0.3\,\text{ns}$ is used in the numerical assessment. } Finally, to present the sidelobe detection metric comparable to \gls{GOSPA}, the same penalty factor is used, and thus the final metric is expressed as $\gamma_\text{SLFD} = \left( (N_\text{SLFD} d_c^{\mathcal{P}}) / \alpha\right)^{(1/\mathcal{P})}$.

The quantitative comparison between the three methods in terms of \gls{GOSPA} and sidelobe detection metric are presented in Fig.~\ref{fig:allmetric}, covering the whole UE/RX trajectory shown in Fig.~\ref{fig:environment}, while also noting the options for the different parametrizations. In terms of the \gls{GOSPA} metric, the proposed SVD-based method clearly outperforms the reference methods. Additionally, in terms of the \gls{SLFD} metric, the proposed method outperforms that in \cite{Yang2023} by a large margin. The CFAR with $P_\text{FA} = 0.002$ is closer in performance, however, having clear challenges to detect the actual multipaths as shown in Fig.~\ref{fig:images} already. {It can also be observed that for the proposed method, the \gls{GOSPA} metric decreases with the increase of $p$ due to the increased number of recovered multipath components. At the same time, the sidelobe detection rate slightly increases with the number of detected components, however, the sidelobe detection metric stays generally low for all considered values of $p$ compared to the reference methods.}

\begin{figure}[!t]
\centering
\includegraphics[width=0.38\textwidth,trim={0.cm 0.0cm 0cm 0cm},clip]{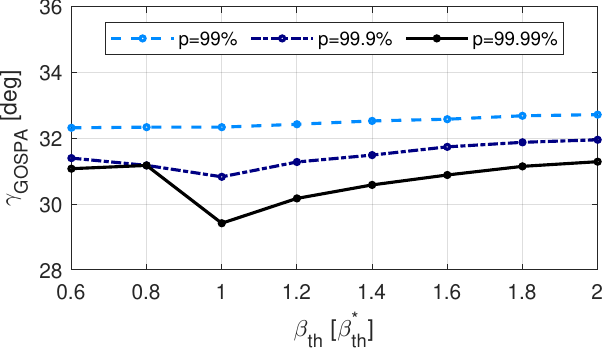}
\vspace{-1mm}
\caption{GOSPA metric for the proposed SVD-based method for $p=99\%$ (\protect\svdoneline), $p=99.9\%$ (\protect\svdtwoline), and $p=99.99\%$ (\protect\svdthreeline) with different values for the power thresholding parameter $\beta_\text{th}$ relative to the baseline value $\beta_\text{th}^*$. 
}
\vspace{-1.75mm}
\label{fig:gospa_beta_th}
\end{figure}

We next provide further complementary assessment of the SVD-based method in terms of the impact of the power thresholding parameter $\beta_\text{th}$. The results are shown in Fig.~\ref{fig:gospa_beta_th}), with $\beta_\text{th}^*$ denoting the baseline value used otherwise in the article. We can observe that the exact value of the power thresholding is impacting the performance only when working under a vary large value of the power ratio parameter $p$ (particularly $p=99.99\%$). This is because in such case, the baseline SVD processing passes through a large number of components, some of which being most likely noise already. Overall, the results in Fig.~\ref{fig:gospa_beta_th} show the robust behavior and performance of the proposed scheme.

\vspace{0mm}
{In summary, we conclude that the proposed SVD approach outperforms the prior-art benchmark methods. As demonstrated by the ray-tracing results, it offers robust performance in both \gls{LoS} and \gls{NLoS} scenarios, while is also having built-in mechanism to suppress the impacts of the unavoidable antenna sidelobes.}

\vspace{-0mm}
\subsection{SLAM Results with Measurement Data}

\begin{figure*}[!t]
\centering
\subfloat[BM1 \cite{wen2021}]{\includegraphics[width=6cm,trim={1.2cm 0 1.6cm 0.8cm},clip]{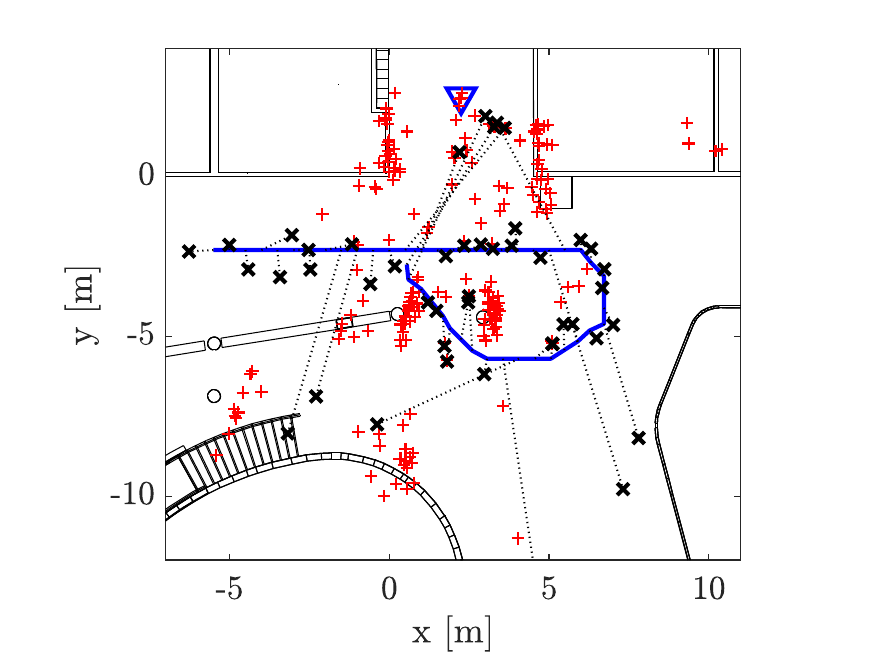}%
%\label{fig:example_performance_BM1}
}
\hfil
\subfloat[BM2 \cite{kaltiokallio2022spawc}]{\includegraphics[width=6cm,trim={1.2cm 0 1.6cm 0.8cm},clip]{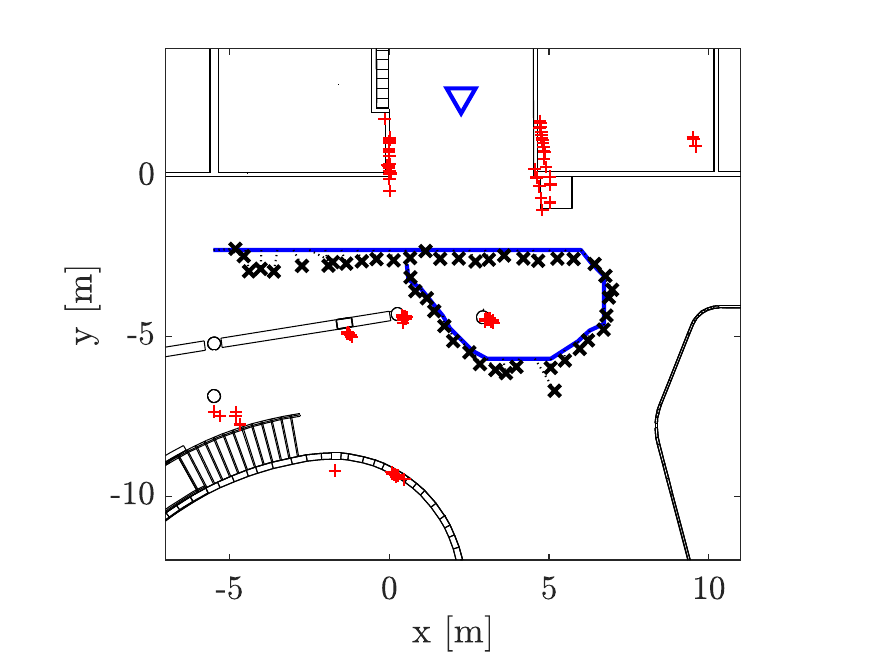}%
%\label{fig:example_performance_BM2}
}
\hfil
\subfloat[Proposed SLAM approach]{\includegraphics[width=6cm,trim={1.2cm 0 1.6cm 0.8cm},clip]{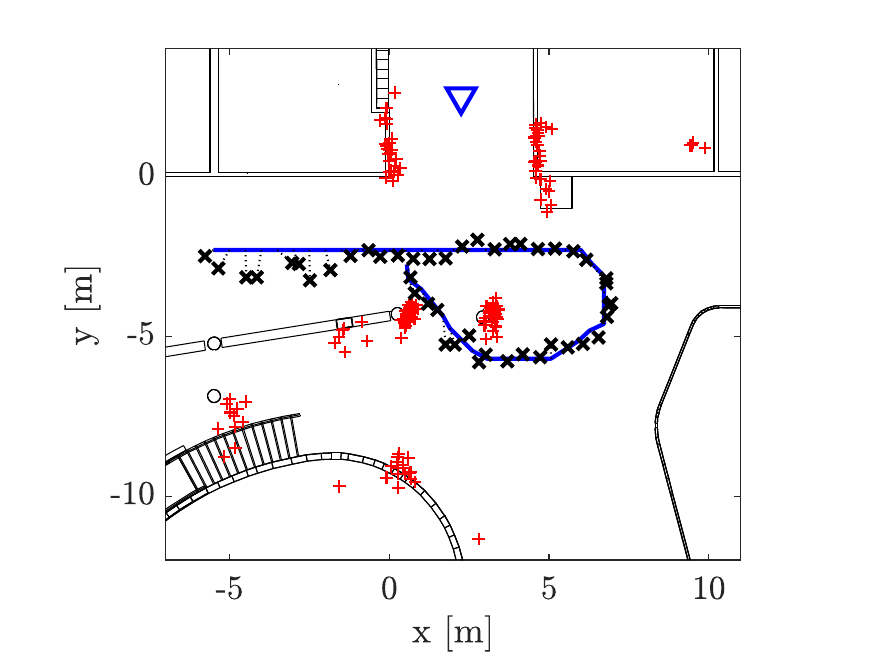}%
%\label{fig:example_performance_proposed}
}
\vspace{-1mm}
\caption{Visual illustration of the performance of the different SLAM algorithms with measurement data including UE clock bias. In the figures, location of the \gls{BS} illustrated using (\protect\bluetriangle), the ground truth \gls{UE} trajectory with (\protect\blueline), the \gls{UE} position estimates using (\protect\blackcross) and the estimated landmark locations with (\protect\redplus).}
\vspace{-1.75mm}
\label{fig:example_performance}
\end{figure*}

\begin{figure*}[!t]
\centering
\subfloat[BM1 \cite{wen2021}]{\includegraphics[width=6cm,trim={1.2cm 0 1.6cm 0.8cm},clip]{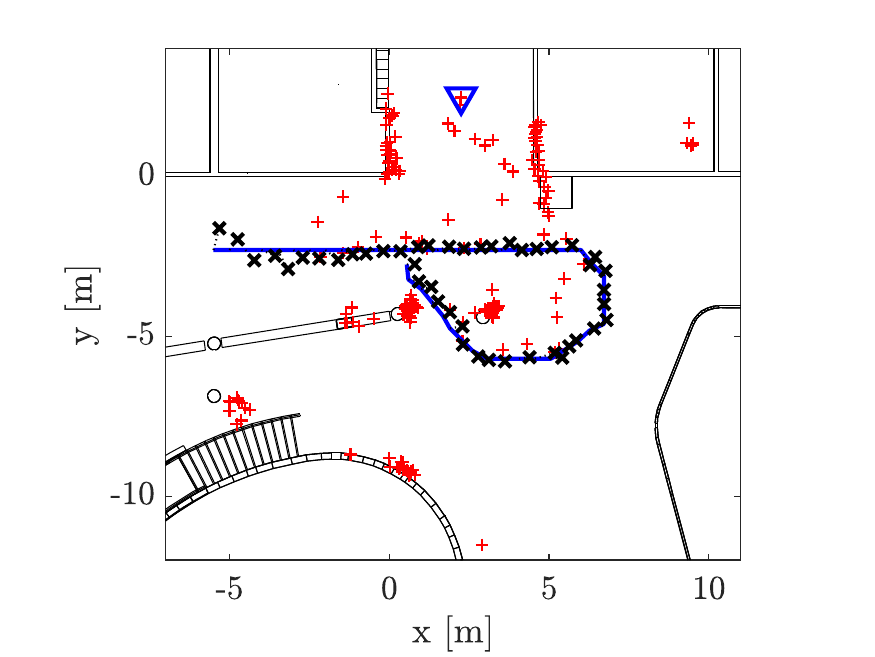}

%\label{fig:example_performance_BM1}
}
\hfil
\subfloat[BM2 \cite{kaltiokallio2022spawc}]{\includegraphics[width=6cm,trim={1.2cm 0 1.6cm 0.8cm},clip]{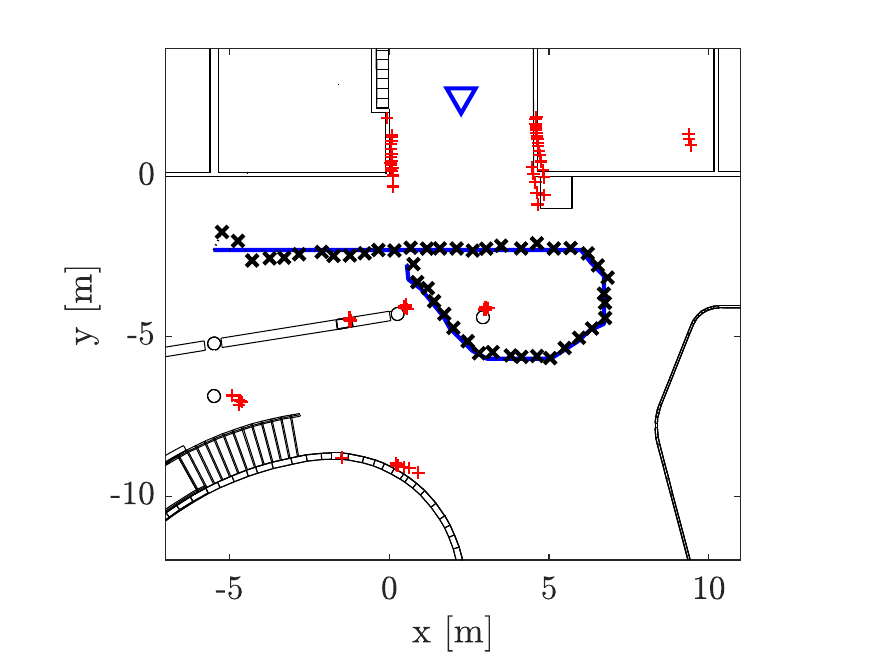}%
%\label{fig:example_performance_BM2}
}
\hfil
\subfloat[Proposed SLAM approach]{\includegraphics[width=6cm,trim={1.2cm 0 1.6cm 0.8cm},clip]{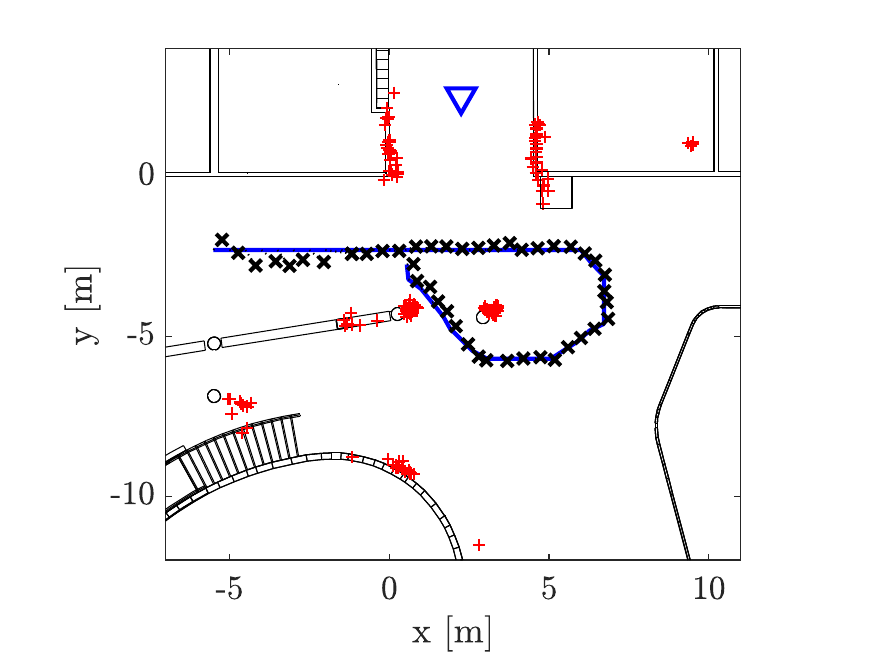}%
%\label{fig:example_performance_proposed}
}
\vspace{-1mm}
\caption{{Visual illustration of the SLAM performance in the reference case where the \gls{UE} and \gls{BS} are mutually synchronized. For other details, refer to the caption of Fig.~\ref{fig:example_performance}).}}
\vspace{-1.75mm}
\label{fig:example_performance_synchronized}
\end{figure*}

Next, the actual end-to-end \gls{SLAM} results are provided and analyzed, utilizing the true measurement data. We use the proposed SVD method for angle estimation, and assume {$p=99\%$}. The \gls{ToA} estimation is carried out as described in \eqref{eq:ToA_coarse}--\eqref{eq:finetoa}, while using brute force search to solve \eqref{eq:fractional_toa}.

The proposed \gls{SLAM} method is benchmarked with respect to two other \gls{SLAM} approaches \cite{wen2021,kaltiokallio2022spawc}. The first benchmark is a geometry based snapshot \gls{SLAM} algorithm \cite{wen2021} referred to as BM1 and the second is a recursive \gls{PHD}-\gls{SLAM} filter \cite{kaltiokallio2022spawc} referred to as BM2. In the following, the time delay and clock bias are converted to distance for convenience, and the clock bias is emulated to evolve according to a random walk model with variance $\sigma_B^2 = 1 \, \textrm{m}^2$. The measurement noise covariance used in the experiments is $\vR_n = \diag([0.3 \, \textrm{m}, \; 3 \, \textrm{deg}, \; 3 \, \textrm{deg} ]^2), \, \forall \, n$. {Recall that no prior knowledge of the map is assumed as discussed in Section~\ref{sec:gn_initialization}.} The mean and covariance of the \gls{UE} prior for the proposed method are $\vmu_\vs = \vmu_{\vs,\textrm{prev}}$ and $\vSigma_{\vs\vs} = \mathbf{I}_4$, in which $\vmu_{\vs,\textrm{prev}}$ denotes the \gls{UE} state estimate at the previous measurement position. In the first measurement position, which is located at $[0.55,\, -2.75]^\top$, there is no prior and estimation is possible because the \gls{LoS} signal exists. {The \gls{UE} initialization algorithm presented in Section \ref{sec:ue_los_initialization} uses $d_\textrm{min}=1 \text{ m}$ and $d_\textrm{max} = 20 \text{ m}$ in the constrained nonlinear optimization problem defined in Equation \eqref{eq:initial_bias_estimate}.} {The benchmark \gls{PHD}-\gls{SLAM} filter considers three types of landmarks \cite{kim2020}: i) \gls{BS} for the \gls{LoS} path; ii) \glspl{VA} for large reflecting surfaces; and iii) \glspl{SP} for small objects. However, the \glspl{VA} are converted to \glspl{SP} using \cite[Eq. (42)]{kim2020} in the following illustrations so that the map of the \gls{PHD}-\gls{SLAM} filter is directly comparable to the maps of the snapshot SLAM algorithms. The benchmark \gls{PHD}-\gls{SLAM} filter uses similar parameter values as in \cite{kaltiokallio2022spawc,kaltiokallio2024tro}, while the values were slightly tuned to maximize the performance in the considered experiment.} The \gls{PHD}-\gls{SLAM} filter is implemented using $1000$ particles, the \gls{UE} state is modeled to evolve according to a random walk model and the process noise is set to $\vQ = \mathbf{I}_4$ which is the same as covariance of the \gls{UE} prior. The \gls{PHD}-\gls{SLAM} filter is initialized using the proposed snapshot \gls{SLAM} algorithm. Since BM1 is sensitive to outliers, measurements labeled as outliers%
\footnote{Under the Gaussian assumption, the quadratic error in \eqref{eq:quadratic_error} follows a $\chi^2$ distribution with three degrees of freedom and if $q_n(\vx) > T_h$, the measurement is labeled an outlier. The quadratic error is evaluated using the ground truth \gls{UE} state and $T_h \approx 16.3$ is computed by choosing tail probability $0.001$ followed by evaluating the inverse cumulative distribution of the $\chi^2$ distribution at $0.999$.} 
are removed from the data for BM1 unless otherwise stated.

\subsubsection{Qualitative Comparison}
The example mapping and localization performance of the algorithms are visualized in Fig.~\ref{fig:example_performance}. Qualitative analysis indicates that the map and \gls{UE} position estimates are more accurate with BM2 and the proposed \gls{SLAM} algorithm than with BM1. For BM1, the estimate is close to the ground truth and the estimated landmark locations are inline with the map in many measurement positions. In several locations however, the estimates are very inaccurate and there are two primary reasons for this. First of all, the method requires four \gls{NLoS} signals so that the system is identifiable and this criterion is not satisfied at every measurement position. For example, when the \gls{UE} is located at {$\vp_{\textrm{UE}} = [4.06, \, -5.70]^\top$}, there are only three propagation paths meaning that the system is underdetermined resulting in an inaccurate estimates as illustrated in Fig.~\ref{fig:example_performance}. The second reason is that in several measurement positions, the cost function of BM1 is multimodal and the global minima is not the one closest to the ground truth. The proposed method and BM2 can operate in mixed \gls{LoS}/\gls{NLoS} conditions and the prior or posterior from the previous time step can be viewed as a regularization term which constrains the posterior update so that the system state is identifiable at every measurement position. Moreover, the resulting estimates for BM2 and the proposed method can be viewed as a weighted average of the evidence provided by the data and the regularization term. Thus, the estimate is expected to be close to the ground truth as long as the prior is not biased and the covariance captures the underlying uncertainties correctly. Lastly, both BM2 and the proposed approach result in sufficient \gls{SLAM} performance despite the measurement data being corrupted by outliers as shown in Fig.~\ref{fig:example_performance}. 

{Since both the \gls{TX} and \gls{RX} were controlled by the same PC, with an opportunity to utilize also a dedicated synchronization cable allowing to synchronize to a common clock, we can also analyze the reference performance of the algorithms with known clock bias $B_\textrm{UE}$. Example performance is illustrated in  Fig.~\ref{fig:example_performance_synchronized} while the results are further elaborated and discussed in Section~\ref{sec:performance_with_known_clock_bias}.}

\subsubsection{Quantitative Comparison}

\begin{table}[!t]
\footnotesize
\renewcommand{\arraystretch}{1.0}
\caption{Performance summary of the different estimators {with and without clock bias}}
\vspace{-.0cm}
\centering
\begin{tabular}{|c|c|c|c|c|}
\hline
Estimator & Position [m] & Heading [deg] & Bias [m] & Time [s] \\ \hline 
BM1 & $2.83 \pm 2.24$ & $32.05 \pm 32.13$ & $2.99 \pm 3.03$ & $41.44$ \\ 
BM2 & $0.56 \pm 0.26$ & $2.85 \pm 2.79$ & $0.55 \pm 0.37$ & $77.41$ \\ 
BM2$^*$ & $2.86 \pm 2.37$ & $35.19 \pm 34.92$ & $1.87 \pm 1.83$ & $1.16$ \\ 
Proposed & $0.56 \pm 0.33$ & $2.30 \pm 2.26$ & $0.54 \pm 0.51$ & $1.44$ \\ \hline
{BM1$^\dagger$} & {$0.34 \pm 0.16$} & {$2.32 \pm 2.33$} & {$-$} & {$0.37$} \\ 
{BM2$^\dagger$} & {$0.26 \pm 0.14$} & {$2.24 \pm 2.21$} & {$-$} & {$73.83$} \\ 
{Proposed$^\dagger$} & {$0.32 \pm 0.16$} & {$1.87 \pm 1.79$} & {$-$} & {$0.15$} \\
\hline
\multicolumn{5}{l}{\scriptsize $^*$BM2 implemented with $10$ particles. {$^\dagger$Known clock bias $B_\textrm{UE}$.}} \\
\end{tabular}
\label{table:performance_summary}
\vspace{-.1cm}
\end{table}

The algorithms are next evaluated quantitatively while since the ground truth landmark locations are unknown, the mapping accuracy is excluded which a common practice in \gls{SLAM} when using experimental data. The performance metrics are tabulated in Table~\ref{table:performance_summary} in which the position, heading and bias errors are computed using the \gls{RMSE} and \gls{STD}. Even without outliers, BM1 results in unsatisfactory performance due to the reason discussed above. The other two methods outperform BM1 and have comparative performance among each other. However, the computational complexity of BM2 is two orders of magnitude higher than with the proposed method. One could decrease the computational complexity by using less particles (see BM2$^*$ in Table~\ref{table:performance_summary}) but at the same time, the accuracy of the filter degrades notably. Thus, a major advantage of the proposed method is that it combines high accuracy together with low computational overhead.

\begin{figure}
  \begin{center}
  \includegraphics[width=0.97\columnwidth]{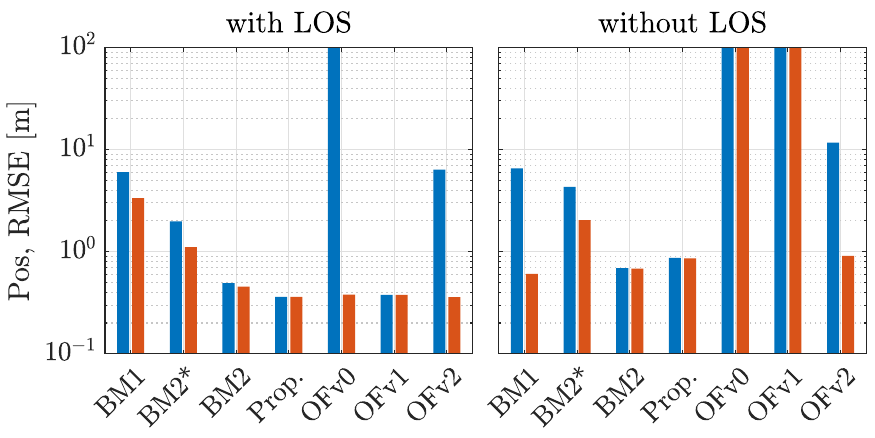}\\
  \vspace{-1mm}
  \caption{Position \glspl{RMSE} of different algorithms in \gls{LoS} and \gls{NLoS} conditions with UE clock bias, and with (\protect\matlabbluebar)  and without (\protect\matlabredbar) outliers. The heading and bias \gls{RMSE} have a similar trend but are omitted from the figure for brevity. The impact of the different objective function variants in \eqref{eq:robust_rwls_objective_function} are also shown, including \emph{i}) OFv0: no prior and quadratic cost function; \emph{ii}) OFv1: no prior and robust cost function; and \emph{iii}) OFv2: prior and quadratic cost function.}
  \vspace{-3mm}
  \label{fig:estimator_evaluation}
  \end{center}
\end{figure}

Next, we evaluate the different algorithms with and without outliers, and in \gls{LoS} and \gls{NLoS} conditions. The results are summarized in Fig.~\ref{fig:estimator_evaluation}. The performance of BM1 degrades significantly if the data contains outliers, whereas the performance of BM2 and the proposed method only degrade slightly. This indicates that both methods are capable of handling noisy measurements that are not inline with the models. The mechanisms how the two methods deal with outliers are quite different. The \gls{PHD}-\gls{SLAM} filter associates the outliers to clutter so that they do not affect the \gls{UE} estimate, whereas the proposed method inflates the covariance according to \eqref{eq:inflated_covariance_matrix} such that outliers are given less trust thus diminishing their impact. Interestingly, the proposed method outperforms BM2 in \gls{LoS} conditions and vice versa in \gls{NLoS} conditions. For the proposed method and when the \gls{LoS} signal exists, the hypothesis that minimizes the cost function is typically the one for which the prior is computed using the \gls{LoS} and the estimate is computed only from the evidence provided by the data. In this particular scenario, in which the prior uncertainty and the process noise of the dynamic model used by BM2 are high, relying solely on the measurements is beneficial. In \gls{NLoS} conditions and when the proposed method cannot be initialized using the measurements, filtering is beneficial and BM2 outperforms the proposed method -- however, at the expense of substantially higher complexity.

Lastly, we decompose the objective function of the proposed method and analyze the performance impact of the prior and the robust cost function. The results are illustrated in Fig.~\ref{fig:estimator_evaluation} from which we can conclude the following: 
\begin{enumerate*}[label=(\roman*)]
  \item Without prior information and using a quadratic cost function (OFv0), the \gls{SLAM} solution is useless in \gls{NLoS} conditions and/or when the data contains outliers. The method outperforms BM1 only if the outliers are removed from the data and if the \gls{LoS} signal exists.
  \item Without prior information and using a robust cost function (OFv1), the estimator yields comparative performance as the proposed method in \gls{LoS} conditions, whereas in \gls{NLoS} conditions the estimates are very innaccurate.
  \item With prior information and using a quadratic cost function (OFv2), the results are comparative to BM1 in both \gls{LoS} and \gls{NLoS} conditions and the performance improves significantly if the outliers are removed from the data.
\end{enumerate*}
Thus to conclude, the robust cost function is a strict requirement of snapshot \gls{SLAM} algorithms that are utilized in realistic scenarios in which the measurements are noisy and contain outliers. Moreover, prior information is required to improve identifiability of the system and enable estimation in mixed \gls{LoS}/\gls{NLoS} conditions.

\subsubsection{SLAM Performance with Known Clock Bias}\label{sec:performance_with_known_clock_bias}
The estimated angles determine the problem geometry up to a scaling which is defined by the unknown clock bias. Thus, the \gls{SLAM} problem is notably easier to solve if the clock bias is known. The experimental arrangements allowed for an additional dedicated synchronization cable between the \gls{TX} and \gls{RX} entities, such that we could also evaluate the reference system performance with perfect synchronization. Example performance with known clock bias was already illustrated in Fig.~\ref{fig:example_performance_synchronized} while the corresponding quantitative performance metrics are tabulated in Table~\ref{table:performance_summary}. As can be observed, the performance improves significantly with all methods when comparing to the practical case with unknown clock bias. Perfect synchronization improves the performance of BM1 the most. Since $B_\textrm{UE}$ is known, only three propagation paths are required to solve the problem and this criterion is satisfied at every measurement position. The cost function is also unimodal and the minima are close to the corresponding ground truth in every measurement position. In addition, the computational complexity of the algorithm is much lower since the problem is solved by finding the minimum cost over $\alpha_\textrm{UE}$, whereas with unknown clock bias, the minimum cost is found by jointly optimizing  $\alpha_\textrm{UE}$ and $B_\textrm{UE}$. With unknown clock bias, the underlying likelihood function that is used to update the particle weights of BM2 is commonly multimodal and the filter can converge to the wrong solution. On the other hand, the likelihood function is generally unimodal if $B_\textrm{UE}$ is known so that the filter more frequently converges to the correct solution. The computational complexity of BM2 only reduces slightly since the only notable difference is that the \gls{UE} state dimension reduces from four to three. Also with the proposed algorithm the performance improves significantly and the underlying cause is the same as with BM1 and BM2, that is, the likelihood function is typically unimodal resulting in an objective function that has a minimum close to the ground truth. Furthermore, the initialization of the proposed algorithm simplifies if the \gls{TX} and \gls{RX} entities are synchronized since mean and covariance of the \gls{UE} prior are directly given by \eqref{eq:ue_moments} and optimization over different trial values of $\hat{B}_\textrm{UE}$ is not required. Thanks to this straightforward initialization procedure and faster convergence of the Gauss-Newton algorithm, the computational complexity of the algorithm is reduced by an order of magnitude. 

\begin{figure}[!t]
  \begin{center}
  \includegraphics[width=0.85\columnwidth]{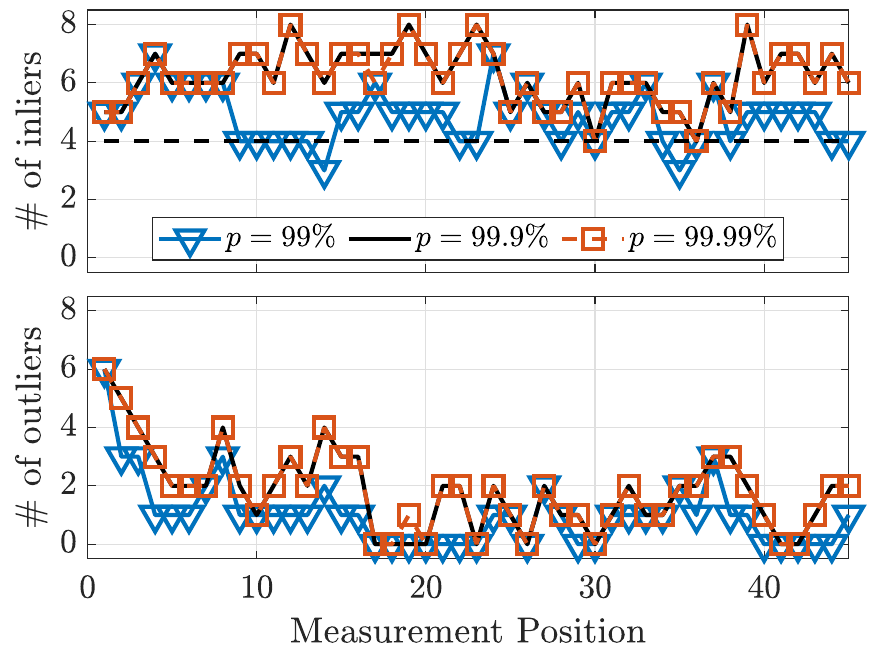}\\
  \caption{{The numbers of inliers and outliers in each measurement position and with different power ratio $p$ values. Minimum number of inliers required by BM1 to solve SLAM is illustrated with the black dashed line.}}
  \label{fig:power_ratio_vs_inliers_and_outliers}
  \end{center}
  \vspace{-1mm}
\end{figure}

\subsubsection{Impact of Power Ratio to SLAM Accuracy}\label{sec:performance_with_different_power_ratio}
Finally, we evaluate and show the \gls{SLAM} performance using channel estimates that are obtained with three different values for the power ratio parameter $p$ in the proposed \gls{SVD} method. As already discussed in Section \ref{sec:channel_estimation_quantitative_comparison}, increasing $p$ grows the rank of the \gls{SVD}-based approximation and as an outcome, the number of estimated paths increases. However, the number of resolvable propagation paths cannot be grown indefinitely since a limited number of physical propagation paths exist and power threshold $\beta_\text{th}$ truncates channel estimates for which the power is close to the noise floor. Higher $p$ values also increase the number of measurements that do not follow the model in \eqref{eq:slam_measurement_model}, such as multi-bounce paths. Such measurements must be treated as unwanted outliers that cannot be utilized by the considered \gls{SLAM} algorithms. The number of inliers and outliers in each of the $45$ measurement positions is illustrated in Fig.~\ref{fig:power_ratio_vs_inliers_and_outliers}. As shown, increasing the power ratio from $99\%$ to $99.9\%$ increases both the number of inliers and outliers. Increasing $p$ even more, the channel estimates of $p=99.9\%$ and $p=99.99\%$ only differ slightly in two measurement positions since already with $p=99.9\%$, the \gls{SVD}-based channel estimator finds all resolvable propagation paths. The power ratio value has the biggest impact on BM1. Recall that four or more propagation paths are required to solve the \gls{SLAM} using BM1 and if $p=99\%$, this criterion is not satisfied at measurement position number $14$ and $35$ as illustrated in Fig.~\ref{fig:power_ratio_vs_inliers_and_outliers}. Now increasing $p$ also increases the number of inliers such that the system is identifiable at every measurement position which improves the accuracy as tabulated in Table~\ref{table:power_ratio_vs_rmse}. The system is always identifiable with the proposed and BM2 methods since the prior constrains the posterior update and the two methods yield similar performance with the tested power ratio values. In general, it is expected that \gls{SLAM} performance improves as the number of propagation paths increases \cite{wymeersch2018}. At the same time, however, the number of outliers increases which can have a negative impact on \gls{SLAM} performance and therefore, the accuracy with BM2 and the proposed approach only improves very little (BM2) or not at all.

\begin{table}[!t]
\footnotesize
\renewcommand{\arraystretch}{1.0}
\caption{{Position RMSE in meters with different power ratio $p$ values}}
\vspace{-.1cm}
\centering
{
\begin{tabular}{|c|c|c|c|}
\hline
Estimator & $p=99\%$ & $p=99.9\%$ & $p=99.99\%$ \\ \hline 
BM1 & $2.83$ & $1.43$ & $1.43$ \\ 
BM2 & $0.56$ & $0.55$ & $0.54$ \\ 
Proposed & $0.56$ & $0.56$ & $0.56$ \\ \hline
\end{tabular}
}
\label{table:power_ratio_vs_rmse}
\vspace{-.0cm}
\end{table}

\subsection{Limitations and Future Directions} 
One of the limitations of the proposed \gls{BRSRP}-based \gls{AoA}/\gls{AoD} extraction algorithm is its inability to decouple two multipath components that have the same \gls{AoA} or \gls{AoD}, which may lead to a missed detection. This type of misdetections, however, do not largely impact the actual SLAM performance, as in majority of cases such undetected components correspond to multi-bounce paths due to the involved geometry. Nevertheless, one future work direction is extending the \gls{SVD}-based \gls{AoA}/\gls{AoD} extraction to support multi-peak detection and selection. One potential way to pursue such can be performing the peak search directly in the singular vector domain, aided by further assumptions about the shape of the beam responses and ToA-based sidelobe rejection. Alternatively, the overall problem may be addressed through a different technical approach -- for example, as a pattern recognition problem solved either through classical methods or potentially through machine learning. On the SLAM and related concepts side, extending the methods to extract information also about the material characteristics of the environment interaction points, on top of their locations, is an interesting future research avenue.

% 
% ############################################################################
% CONCLUSION
% ############################################################################
\vspace{-1mm}
\section{Conclusions}
\label{sec:Conclusions}
In this article, we addressed the timely notion of \gls{mmW} radio \gls{SLAM} from end-to-end processing perspective, under minimal knowledge regarding the array parameters and user mobility. We first proposed a novel \gls{SVD}-based estimation approach for acquiring the \glspl{AoA} and \glspl{AoD} of the involved propagation paths. The method operates on \gls{BRSRP} measurements and does not need information {on the complex antenna patterns or the underlying steering vectors and beamforming weights}, while offering built-in robustness against antenna sidelobes. {The method relies only on known physical correspondence between the beam indices and the beamforming angles, and the sparsity of the beamformed \gls{mmW} channel.} Secondly, a new snapshot \gls{SLAM} method was also proposed, to jointly estimate the locations of the landmarks and the \gls{UE}, offering improved robustness and identifiability compared to prior-art. The performance of the proposed methods was comprehensively assessed through ray tracing and true measurement data at 60\,GHz. The results show that the methods outperform the relevant prior-art, with the end-to-end performance being comparable or even better compared to sequential filtering solutions while offering substantially reduced complexity. Finally, we provide the measured 60\,GHz data openly available for the research community. {Our future work considers extending the proposed angle estimation methods to include multi-bounce detection and estimation with rank-1 singular matrices, thus facilitating further evolved SLAM solutions beyond the common single-bounce approaches.}

% ############################################################################
% REFERENCES
% ############################################################################

% \vspace{-1mm}
% \bibliographystyle{IEEEtran}
% \bibliography{mainReferences}

% ############################################################################
% APPENDIX
% ############################################################################
\vspace{-3mm}
\appendix
\section*{Proof of Lemma~\ref{lemma_approx}}
\label{sec:appendix}

We first denote $\zeta_{k,m,n}^{i,j} = x_{k,m}^{i,j} e^{-\imagunit2\pi k \Delta f \tau_{\mathrm{f},n}} e^{\imagunit 2\pi m T_{\text{sym}} f_{\text{D},n}}$ and $\tilde{G}_{n}^{i,j} = \xi_n G_{\text{TX},i}(\boldsymbol{\psi}_{\text{TX},n}) G_{\text{RX},j}(\boldsymbol{\psi}_{\text{RX},n})$. Now, by substituting \eqref{eq:rx_symbol2} into \eqref{eq:brsrp}, the \gls{BRSRP} measurement 
can be written as
\begin{equation} \label{eq_lemma1_all}
\begin{split}
    \!\!\beta_{i,j}\!&= 
    \!\overbrace{\frac{1}{{N}_\text{RS}} \!\! \sum_{\substack{(k,m) \\ \in \mathcal{M}_\text{RS}}} \! \sum_{n=1}^{N} \Big\vert \tilde{G}_{n}^{i,j} \zeta_{k,m,n}^{i,j} \Big\vert^2}^{=\Sdirect \text{ (direct signal terms)}}\!\!\! +\!\overbrace{\frac{1}{{N}_\text{RS}} \!\! \sum_{\substack{(k,m) \\ \in \mathcal{M}_\text{RS}}} \! \Big\vert \tilde{n}_{k,m}^{i,j} \Big\vert^{2}}^{=\Ndirect \text{ (direct noise terms)}} \\[-3pt]
    &+ \hspace{0cm} \overbrace{\frac{2}{{N}_\text{RS}} \!\!\! \sum_{\substack{(k,m) \\ \in \mathcal{M}_\text{RS}}} \!\!\! \mathfrak{Re} \! \Big[\!\left(\tilde{n}_{k,m}^{i,j}\right)^* \sum_{n=1}^{N}  \tilde{G}_{n}^{i,j} \zeta_{k,m,n}^{i,j}\Big]}^{=\Ncross \text{ (noise cross terms)}} \\[-7pt]
    &+ \overbrace{\frac{2}{{N}_\text{RS}} \!\!\! \sum_{\substack{(k,m) \\ \in \mathcal{M}_\text{RS}}} \!\!\!\mathfrak{Re} \! \Big[ \!\sum_{n_1=1}^{N} \!\!\sum_{\substack{n_2 = 1 \\ n_2 {\neq}  n_1}}^{N} \!\! (\tilde{G}_{n_1}^{i,j} \zeta_{k,m,n_1}^{i,j})^* \tilde{G}_{n_2}^{i,j} \zeta_{k,m,n_2}^{i,j}\Big]}^{=\Scross \text{ (signal cross terms)}}.
\end{split}
\end{equation}
In the following, we elaborate on the limiting behavior of the ratio between the cross and direct terms for both signal and noise components, denoted as $\Scross$, $\Sdirect$, $\Ncross$ and $\Ndirect$, in \eqref{eq_lemma1_all}. 

The ratio between the signal components $\Scross$ and $\Sdirect$ in \eqref{eq_lemma1_all} as $K$ and $M$ tend to infinity can be computed as
\begin{align} \label{eq_lim_frac_s}
    \lim_{\substack{K \to \infty \\ M \to \infty}} \frac{\Scross}{\Sdirect} = \frac{ \lim_{\substack{K \to \infty \\ M \to \infty}} \Scross }{ \lim_{\substack{K \to \infty \\ M \to \infty}} \Sdirect} ~,
\end{align}
provided that the limits on the numerator and the denominator exist. Using \eqref{eq_lemma1_all}, and inserting $\vert \mathcal{M}_\text{RS} \vert = N_\text{RS} = K M$, we can write
\begin{align} \nonumber
    \lim_{\substack{K \to \infty \\ M \to \infty}} \Scross = \mathfrak{Re} \Big[ &\!\sum_{n_1=1}^{N}  \!\! \sum_{\substack{n_2 = 1 \\ n_2 \neq  n_1}}^{N} \!\!\! (\tilde{G}_{n_1}^{i,j})^*\tilde{G}_{n_2}^{i,j} \\ &\times \lim_{\substack{K \to \infty \\ M \to \infty}}  \frac{2}{KM} \sum_{\substack{(k,m) \\ \in \mathcal{M}_\text{RS}}} \!\!\! (\zeta_{k,m,n_1}^{i,j})^* \zeta_{k,m,n_2}^{i,j}\Big] ~. 
\end{align}
Considering $\vert x_{k,m}^{i,j} \vert \!\!=\!\!1$ and the definitions of $\mathbf{b}_{n}[k]$ and $\mathbf{c}_{n}[m]$ in Lemma~1, the numerator and denominator in \eqref{eq_lim_frac_s} yields 
\begin{align} \label{eq_lim_scross}
    \lim_{\substack{K \to \infty \\ M \to \infty}} \Scross &= 0 \text{ and } \lim_{\substack{K \to \infty \\ M \to \infty}} \Sdirect = \sum_{n=1}^{N} \Big\vert \tilde{G}_{n}^{i,j} \Big\vert^2,
\end{align}
respectively. As a result,
\begin{align} \label{eq_lim_frac_s2}
    \lim_{\substack{K \to \infty \\ M \to \infty}} \frac{\Scross}{\Sdirect} = 0 \text{, and thus } \Sdirect + \Scross \approx \Sdirect
\end{align}
for sufficiently large $K$ and $M$.

The ratio between the noise components $\Ncross$ and $\Ndirect$ in \eqref{eq_lemma1_all} as $K$ and $M$ tend to infinity can be computed as
\begin{align} \label{eq_lim_frac_n}
    \lim_{\substack{K \to \infty \\ M \to \infty}} \frac{\Ncross}{\Ndirect} = \frac{ \lim_{\substack{K \to \infty \\ M \to \infty}} \Ncross }{ \lim_{\substack{K \to \infty \\ M \to \infty}} \Ndirect} ~.
\end{align}
Regarding the limiting behavior of the cross term and direct term in \eqref{eq_lim_frac_n}, we note that the summands (over path index $n$) are independent and identically distributed (i.i.d.) random variables. Hence, by invoking the law of large numbers, while exploiting the noise characteristics $\expectation\left\{ \tilde{n}_{k,m}^{i,j} \right\} \!=\! 0$ and $\expectation\left\{   \left\vert \tilde{n}_{k,m}^{i,j} \right\vert^{2}  \right\} \!=\! \sigma_{\mathrm{noise}}^2$, where $\sigma_{\mathrm{noise}}^2$ denotes the variance of $\tilde{n}_{k,m}^{i,j}$, we obtain
\begin{align} \label{eq_lim_frac_n2}
    \lim_{\substack{K \to \infty \\ M \to \infty}} \frac{\Ncross}{\Ndirect} = 0 \text{ and thus, } \Ndirect + \Ncross \approx \Ndirect
\end{align}
for sufficiently large $K$ and $M$.

To conclude, following the \eqref{eq_lim_frac_s2} and \eqref{eq_lim_frac_n2}, the BRSRP measurement in \eqref{eq_lemma1_all} can be approximated for sufficiently large $K$ and $M$ as
\begin{align}\label{eq_betaij2}
     \beta_{i,j} &\approx \lim_{\substack{K \to \infty \\ M \to \infty}} \Sdirect + \lim_{\substack{K \to \infty \\ M \to \infty}} \Ndirect ~,
     \\
     &= \sum_{n=1}^{N} \Big\vert \tilde{G}_{n}^{i,j} \Big\vert^2 + \sigma_{\mathrm{noise}}^2 ~,
\end{align}
which establishes the result \eqref{eq:power_angle} in Lemma~1.

% ############################################################
% BIOGRAPHIES
% ############################################################

\section*{Biographies}
\label{sec:Biographies}

\begin{IEEEbiography}[{\includegraphics[width=1in,height=1.25in,clip,keepaspectratio]{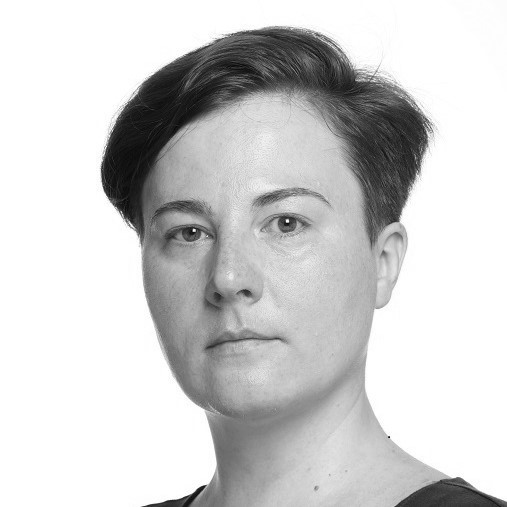}}]{Elizaveta Rastorgueva-Foi}
is a doctoral candidate at the Unit of Electrical Engineering at Tampere University, Finland, where she also received her M.Sc. degree in electrical engineering in 2019. Her research interests include statistical signal processing, positioning and location-aware communications in mobile networks with an emphasis on 5G and beyond systems.
\end{IEEEbiography}

\begin{IEEEbiography}[{\includegraphics[width=1in,height=1.25in,clip,keepaspectratio]{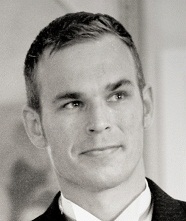}}]{Ossi Kaltiokallio}
(M'23) received the Ph.D. degree in Electrical Engineering in 2017 from Aalto University, Finland. He has held postdoctoral fellow position at the University of Utah, Aalto University and Tampere University. Currently he is a Senior Research Fellow with the Electrical Engineering Unit, Tampere University, Finland. His current research interests lie at the intersection of statistical signal processing and wireless networking for improving radio-based positioning and sensing technologies.
\end{IEEEbiography}

\begin{IEEEbiography}[{\includegraphics[width=1in,height=1.25in,clip,keepaspectratio]{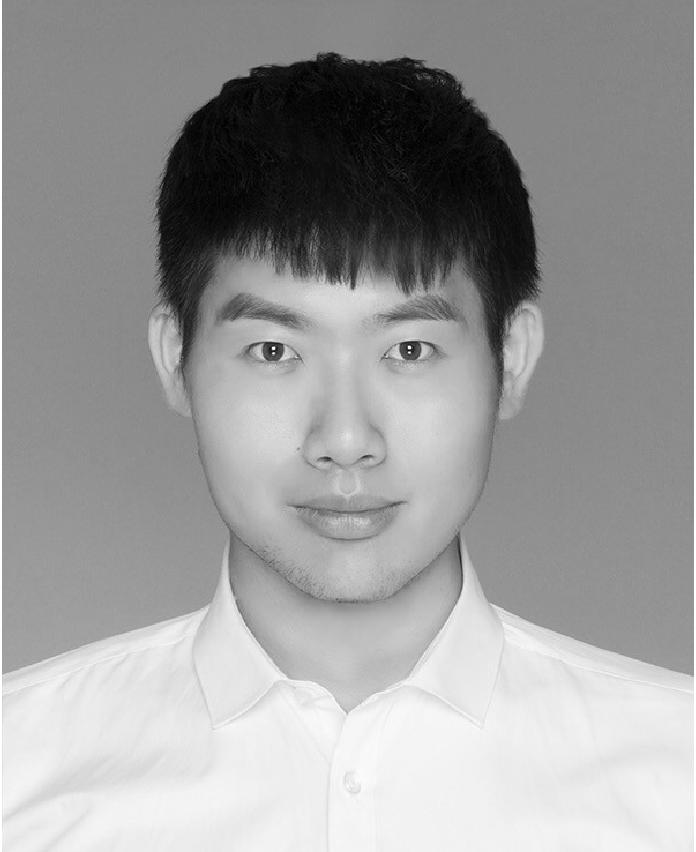}}]{Yu Ge}
(S'20) received his B.E. degree from Zhejiang University, Hangzhou, China, in 2017, and M.Sc. degree from the KTH Royal Institute of Technology, Stockholm, Sweden, in 2019. He is currently a Ph.D. candidate in the Department of Electrical and Engineering at Chalmers University of Technology, Sweden. His research interests include integrated communication and sensing, wireless positioning systems, simultaneous localization and mapping, and multi-object tracking, particularly in 5G and Beyond 5G scenarios.
\end{IEEEbiography}

\begin{IEEEbiography} [{\includegraphics[width=1in,height=1.25in,clip,keepaspectratio]{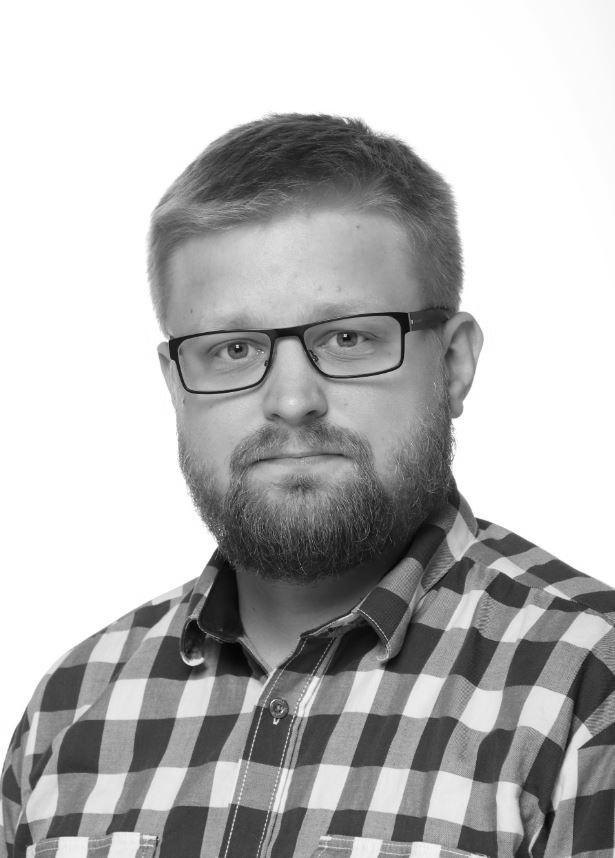}}] {Matias Turunen}
is a researcher and laboratory specialist at the Department of Electrical Engineering, Tampere University (TAU), Finland. His research interests include inband full-duplex radios with an emphasis on analog RF cancellation, OFDM radar, and 5G New Radio systems.
\end{IEEEbiography}

\begin{IEEEbiography}[{\includegraphics[width=1in,height=1.25in,clip,keepaspectratio]{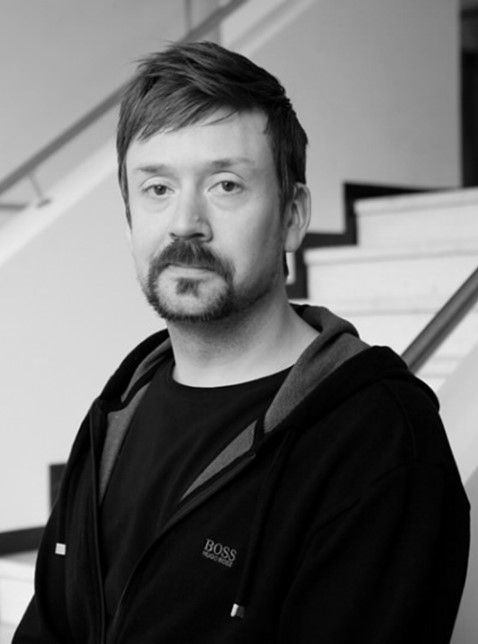}}]{Jukka~Talvitie}
(S’09, M’17) received the M.Sc. and D.Sc. degrees from the Tampere University of Technology, Finland, in 2008 and 2016, respectively. He is currently a University Lecturer with the Unit of Electrical Engineering, Tampere University (TAU), Finland. His research interests include signal processing for wireless communications, radio-based positioning and sensing, radio link waveform design, and radio system design, particularly concerning 5G and beyond mobile technologies. 
\end{IEEEbiography}

\begin{IEEEbiography}[{\includegraphics[width=1in,height=1.25in,clip,keepaspectratio]{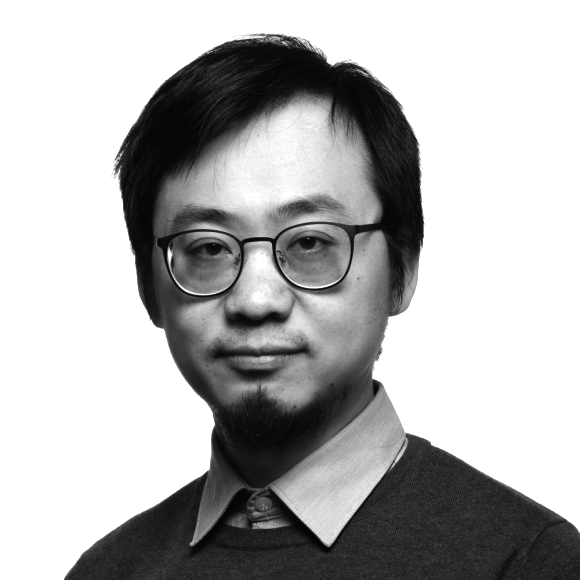}}]{Bo~Tan}
received his PhD from the Institute for Digital Communications (IDCOM), The University of Edinburgh, UK, in Nov. 2013. From 2012 to 2016, he conducted multiple postdoctoral research projects at the University College London and the University of Bristol, UK, contributing to passive radar design and applications in security and healthcare. From 2017 to 2018, he was a lecturer at Coventry University, UK. Since 2023, he has been a tenure-track associate professor at Tampere University, Finland. His research interests include radio sensing and connectivity for intelligent machines.
\end{IEEEbiography}

\begin{IEEEbiography} [{\includegraphics[width=1in,height=1.25in,clip,keepaspectratio]{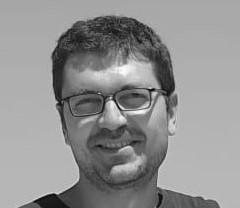}}] {Musa Furkan Keskin}
received the B.S., M.S., and Ph.D. degrees from the Department of Electrical and Electronics Engineering, Bilkent University, Ankara, Turkey, in 2010, 2012, and 2018, respectively. He is currently Research Specialist with the department of Electrical Engineering at Chalmers University of Technology, Gothenburg, Sweden. His current research interests include integrated sensing and communications, RIS-aided localization and sensing, and hardware impairments in beyond 5G/6G systems.
\end{IEEEbiography}

\begin{IEEEbiography}[{\includegraphics[width=1in,height=1.25in,clip,keepaspectratio]{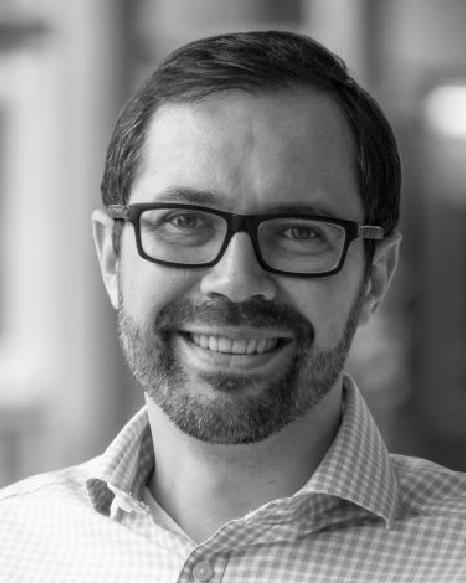}}]{Henk~Wymeersch}
(S'01, M'05, SM'19, F'24) obtained the Ph.D. degree in Electrical Engineering/Applied Sciences in 2005 from Ghent University, Belgium. He is currently a Professor of Communication Systems with the Department of Electrical Engineering at Chalmers University of Technology, Sweden. He is currently a Senior Member of the IEEE Signal Processing Magazine Editorial Board. During 2019-2021, he was an IEEE Distinguished Lecturer with the Vehicular Technology Society. His current research interests include the convergence of communication and sensing, in a 5G and Beyond 5G context. 
\end{IEEEbiography}

\begin{IEEEbiography}[{\includegraphics[width=1in,height=1.25in,clip,keepaspectratio]{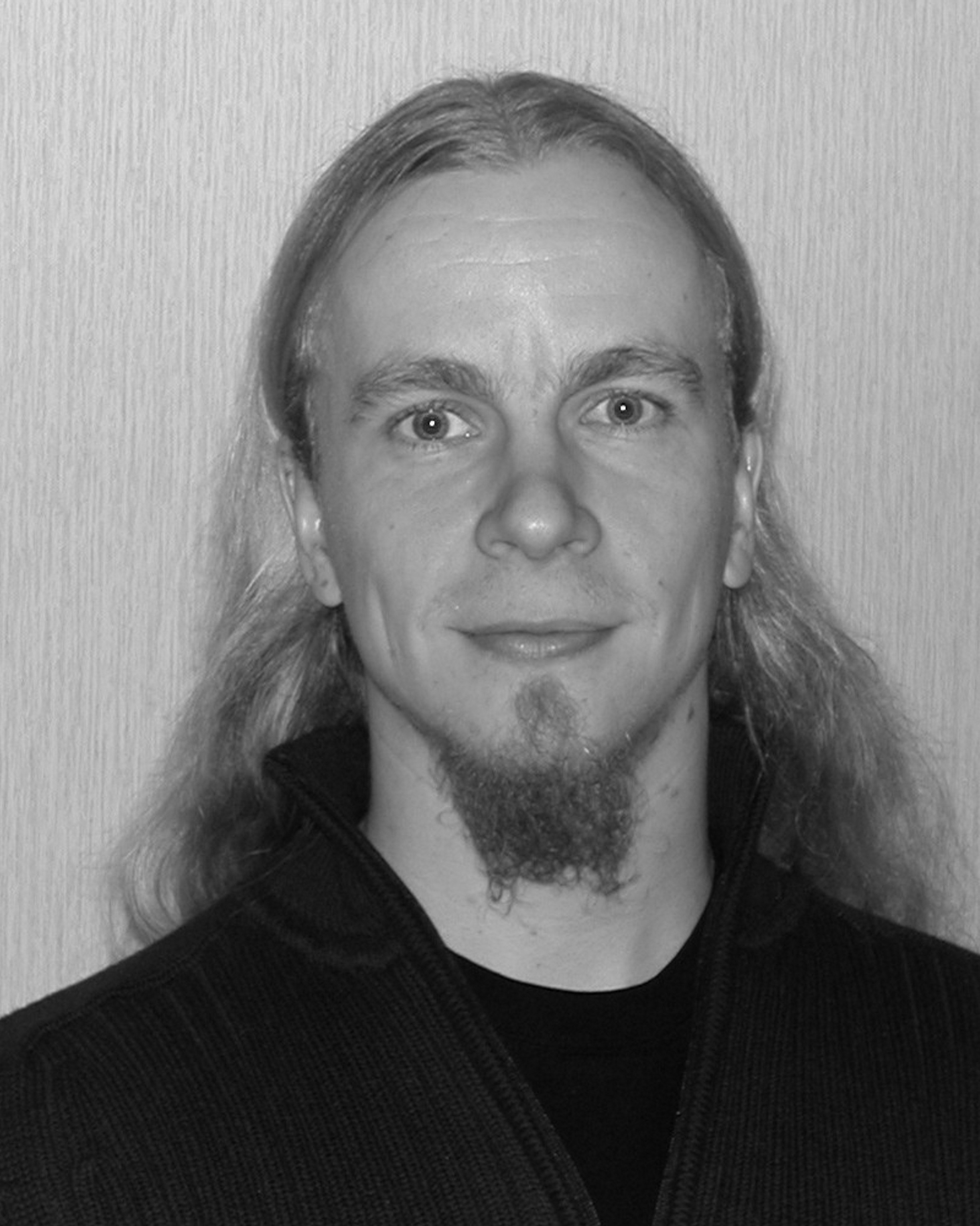}}]{Mikko~Valkama}
[S’00, M’01, SM’15, F’22] received his M.Sc. (Tech.) and D.Sc. (Tech.) degrees (both with honors) from Tampere University of Technology, Finland, in 2000 and 2001, respectively. In 2003, he was with the Communications Systems and Signal Processing Institute at SDSU, San Diego, CA, as a visiting research fellow. Currently, he is a Full Professor and the Head of the Unit of Electrical Engineering at the newly formed Tampere University, Finland. His general research interests include radio communications, radio localization, and radio-based sensing, with particular emphasis on 5G and 6G mobile radio networks. 
\end{IEEEbiography}

\end{document}